\documentclass[twocolumn]{aastex63}

\usepackage[utf8]{inputenc}
\usepackage{tablefootnote}
\usepackage{natbib}
\usepackage{amsmath}
\usepackage{threeparttable}
\usepackage{xspace}
\usepackage{graphicx}
\usepackage{longtable}
\usepackage{multirow}

\newcommand{\tess}{{TESS}\xspace}
\newcommand{\jwst}{{JWST}\xspace}
\newcommand{\kepler}{{\it Kepler}\xspace}
\newcommand{\msun}{M$_\odot$\xspace}
\newcommand{\rsun}{R$_\odot$\xspace}
\newcommand{\rearth}{R$_\oplus$\xspace}
\newcommand{\Teff}{T$_\mathrm{eff}$\xspace}
\newcommand{\teff}{T$_\mathrm{eff}$\xspace}

\newcommand{\target}{TOI-700}

\newcommand{\fbol}{F$_{\mathrm{bol}}$}

\definecolor{pink}{rgb}{1,0.1,.6}

\submitjournal{AAS Journals}

\shorttitle{An Earth-sized Planet in the Habitable Zone of a Nearby Cool Star}
\shortauthors{Gilbert et al.}

\begin{document}

\title{
The First Habitable Zone Earth-sized Planet from TESS. I: Validation of the TOI-700 System 
}

\suppressAffiliations

\author[0000-0002-0388-8004]{Emily A. Gilbert}
\affiliation{Department of Astronomy and Astrophysics, University of
Chicago, 5640 S. Ellis Ave, Chicago, IL 60637, USA}
\affiliation{The Adler Planetarium, 1300 South Lakeshore Drive, Chicago, IL 60605, USA}
\affiliation{NASA Goddard Space Flight Center, Greenbelt, MD 20771, USA}
\affiliation{GSFC Sellers Exoplanet Environments Collaboration}

\author[0000-0001-7139-2724]{Thomas Barclay}
\affiliation{NASA Goddard Space Flight Center, Greenbelt, MD 20771, USA}
\affiliation{University of Maryland, Baltimore County, 1000 Hilltop Cir, Baltimore, MD 21250, USA}

\author[0000-0001-5347-7062]{Joshua E. Schlieder}
\affiliation{NASA Goddard Space Flight Center, Greenbelt, MD 20771, USA}

\author[0000-0003-1309-2904]{Elisa V. Quintana}
\affiliation{NASA Goddard Space Flight Center, Greenbelt, MD 20771, USA}

\author[0000-0001-5084-4269]{Benjamin J. Hord}
\affiliation{University of Maryland, College Park, MD 20742, USA}
\affiliation{NASA Goddard Space Flight Center, Greenbelt, MD 20771, USA}

\author[0000-0001-5347-7062]{Veselin B. Kostov}
\affiliation{NASA Goddard Space Flight Center, Greenbelt, MD 20771, USA}

\author[0000-0002-7727-4603]{Eric D. Lopez}
\affiliation{NASA Goddard Space Flight Center, Greenbelt, MD 20771, USA}

\author[0000-0002-5904-1865]{Jason F. Rowe}
\affiliation{Bishops University, 2600 College St, Sherbrooke, QC J1M 1Z7, Canada}

\author[0000-0001-6541-0754]{Kelsey Hoffman} 
\affiliation{SETI Institute, 189 Bernardo Ave, Suite 200, Mountain View, CA 94043, USA}

\author[0000-0003-2918-8687]{Lucianne M. Walkowicz}
\affiliation{The Adler Planetarium, 1300 South Lakeshore Drive, Chicago, IL 60605, USA}

\author[0000-0003-2565-7909]{Michele L. Silverstein}
\altaffiliation{NASA Postdoctoral Program Fellow}
\affiliation{NASA Goddard Space Flight Center, Greenbelt, MD 20771, USA}
\affiliation{RECONS Institute, Chambersburg, PA 17201, USA}

\author[0000-0001-8812-0565]{Joseph E. Rodriguez}
\affiliation{Center for Astrophysics $\mid$ Harvard \& Smithsonian, 60 Garden St, Cambridge, MA, 02138, USA}

\author[0000-0001-7246-5438]{Andrew Vanderburg}
\altaffiliation{NASA Sagan Fellow}
\affiliation{Department of Astronomy, The University of Texas at Austin, Austin, TX 78712, USA}

\author[0000-0003-4471-1042]{Gabrielle Suissa}
\affiliation{NASA Goddard Space Flight Center, Greenbelt, MD 20771, USA}
\affiliation{GSFC Sellers Exoplanet Environments Collaboration}
\affiliation{Universities Space Research Association (USRA), Columbia, Maryland, USA}

\author[0000-0003-4452-0588]{Vladimir S. Airapetian}
\affiliation{NASA Goddard Space Flight Center, Greenbelt, MD 20771, USA}
\affiliation{GSFC Sellers Exoplanet Environments Collaboration}

\author[0000-0001-8933-6878]{Matthew S. Clement}
\affiliation{Department of Terrestrial Magnetism, Carnegie, 5241 Broad Branch Rd NW, Washington, DC 20015}

\author[0000-0001-8974-0758]{Sean N. Raymond}
\affiliation{Laboratoire d’astrophysique de Bordeaux, Univ. Bordeaux, CNRS, B18N, All\'e Geoffroy Saint-Hilaire, 33615 Pessac, France}

\author[0000-0003-3654-1602]{Andrew W. Mann}
\affiliation{Department of Physics and Astronomy, University of North Carolina at Chapel Hill, Chapel Hill, NC 27599, USA}

\author[0000-0002-0493-1342]{Ethan Kruse}
\affiliation{NASA Goddard Space Flight Center, Greenbelt, MD 20771, USA}

\author[0000-0001-6513-1659]{Jack J. Lissauer}
\affiliation{NASA Ames Research Center, Moffett Field, CA, 94035, USA}

\author[0000-0001-8020-7121]{Knicole D. Col\'on}
\affiliation{NASA Goddard Space Flight Center, Greenbelt, MD 20771, USA}

\author[0000-0002-5893-2471]{Ravi kumar Kopparapu}
\affiliation{NASA Goddard Space Flight Center, Greenbelt, MD 20771, USA}
\affiliation{GSFC Sellers Exoplanet Environments Collaboration}

\author[0000-0003-0514-1147]{Laura Kreidberg}
\affiliation{Center for Astrophysics $\mid$ Harvard \& Smithsonian, 60 Garden St, Cambridge, MA, 02138, USA}

\author[0000-0003-0562-6750]{Sebastian Zieba}
\affiliation{Universit\"at Innsbruck, Institut f\"ur Astro- und Teilchenphysik, Technikerstra\ss e 25, 6020 Innsbruck, Austria}

\author[0000-0001-6588-9574]{Karen A.\ Collins}
\affiliation{Center for Astrophysics $\mid$ Harvard \& Smithsonian, 60 Garden St, Cambridge, MA, 02138, USA}

\author[0000-0002-8964-8377]{Samuel N. Quinn}
\affiliation{Center for Astrophysics $\mid$ Harvard \& Smithsonian, 60 Garden St, Cambridge, MA, 02138, USA}

\author[0000-0002-2532-2853]{Steve B. Howell}
\affiliation{NASA Ames Research Center, Moffett Field, CA, 94035, USA}

\author{Carl Ziegler}
\affil{Dunlap Institute for Astronomy and Astrophysics, University of Toronto, 50 St. George Street, Toronto, Ontario M5S 3H4, Canada}

\author[0000-0002-1864-6120]{Eliot Halley Vrijmoet}
\affiliation{Georgia State University, 33 Gilmer Street SE Atlanta, GA 30303}
\affiliation{RECONS Institute, Chambersburg, PA 17201, USA}


\author[0000-0002-8167-1767]{Fred C. Adams}
\affiliation{University of Michigan, 500 S State St, Ann Arbor, MI 48109, USA}

\author[0000-0001-6285-267X]{Giada N. Arney}
\affiliation{NASA Goddard Space Flight Center, Greenbelt, MD 20771, USA}
\affiliation{GSFC Sellers Exoplanet Environments Collaboration, NASA Goddard Space Flight Center, Greenbelt, MD 20771}

\author[0000-0003-0442-4284]{Patricia T. Boyd}
\affiliation{NASA Goddard Space Flight Center, Greenbelt, MD 20771, USA}

\author[0000-0002-2072-6541]{Jonathan Brande}
\affiliation{NASA Goddard Space Flight Center, Greenbelt, MD 20771, USA}
\affiliation{University of Maryland, College Park, MD 20742, USA}
\affiliation{GSFC Sellers Exoplanet Environments Collaboration}

\author[0000-0002-7754-9486]{Christopher~J.~Burke}
\affiliation{Kavli Institute for Astrophysics and Space Research, Massachusetts Institute of Technology, Cambridge, MA, USA}

\author{Luca Cacciapuoti}
\affiliation{Department of Physics ``Ettore Pancini", Universita di Napoli Federico II, Compl. Univ. Monte S. Angelo, 80126 Napoli, Italy}

\author{Quadry Chance}
\affiliation{University of Florida, Gainesville, FL 32611}

\author[0000-0002-8035-4778]{Jessie L. Christiansen}
\affiliation{Caltech/IPAC, 1200 E. California Blvd. Pasadena, CA 91125}

\author[0000-0002-2553-096X]{Giovanni Covone}
\affiliation{Department of Physics ``Ettore Pancini", Universita di Napoli Federico II, Compl. Univ. Monte S. Angelo, 80126 Napoli, Italy}
\affiliation{INAF - Capodimonte Astronomical Observatory, Salita Moiariello 16, 80131  Napoli, Italy}
\affiliation{INFN, Sezione di Napoli, Compl. Univers. di Monte S. Angelo, via Cinthia, I-80126 Napoli, Italy}

\author{Tansu Daylan}
\altaffiliation{Kavli Fellow}
\affiliation{Department of Physics and Kavli Institute for Astrophysics and Space Research, Massachusetts Institute of Technology, Cambridge, MA, 02139, USA}

\author{Danielle Dineen}
\affiliation{Bishops University, 2600 College St, Sherbrooke, QC J1M 1Z7, Canada}

\author[0000-0001-8189-0233]{Courtney D. Dressing}
\affiliation{Department of Astronomy, University of California at Berkeley, Berkeley, CA 94720, USA}

\author[0000-0002-2482-0180]{Zahra Essack}
\affiliation{Department of Earth, Atmospheric and Planetary Sciences, Massachusetts Institute of Technology, Cambridge, MA 02139, USA}
\affiliation{Kavli Institute for Astrophysics and Space Research, Massachusetts Institute of Technology, 77 Massachusetts Avenue, Cambridge, MA 02139, USA}

\author[0000-0002-5967-9631]{Thomas J. Fauchez}
\affiliation{Universities Space Research Association (USRA), Columbia, Maryland, USA}
\affiliation{GSFC Sellers Exoplanet Environments Collaboration}

\author[0000-0001-5379-4295]{Brianna Galgano}
\affiliation{Fisk University, Department of Life and Physical Sciences, W.E.B. DuBois Hall, 1717 Jackson St, Nashville, TN 37208, USA}

\author[0000-0002-4884-7150]{Alex R. Howe}
\affiliation{NASA Goddard Space Flight Center, Greenbelt, MD 20771, USA}

\author{Lisa Kaltenegger}
\affiliation{Carl Sagan Institute, Cornell University, Space Science Institute 312, 14850 Ithaca, NY, USA}

\author{Stephen R. Kane}
\affiliation{Department of Earth and Planetary Sciences, University of California, Riverside, CA 92521, USA}

\author{Christopher Lam}
\affiliation{NASA Goddard Space Flight Center, Greenbelt, MD 20771, USA}

\author[0000-0002-1228-9820]{Eve J. Lee}
\affiliation{Department of Physics and McGill Space Institute, McGill University, 3550 rue University, Montreal, QC, H3A 2T8, Canada}

\author{Nikole K. Lewis}
\affiliation{Carl Sagan Institute, Cornell University, Space Science Institute 312, 14850 Ithaca, NY, USA}

\author[0000-0002-9632-9382]{Sarah E. Logsdon}
\affiliation{NSF's National Optical-Infrared Astronomy Research Laboratory, 950 North Cherry Avenue, Tucson, AZ 85719, USA}

\author[0000-0002-8119-3355]{Avi M. Mandell}
\affiliation{NASA Goddard Space Flight Center, Greenbelt, MD 20771, USA}
\affiliation{GSFC Sellers Exoplanet Environments Collaboration}

\author[0000-0003-3896-3059]{Teresa Monsue}
\affiliation{NASA Goddard Space Flight Center, Greenbelt, MD 20771, USA}

\author{Fergal Mullally}
\affiliation{SETI Institute, 189 Bernardo Ave, Suite 200, Mountain View, CA 94043, USA}

\author[0000-0001-7106-4683]{Susan E. Mullally}
\affiliation{Space Telescope Science Institute, 3700 San Martin Drive, Baltimore, MD, 21218, USA}

\author[0000-0002-8090-3570]{Rishi R. Paudel}
\affiliation{NASA Goddard Space Flight Center, Greenbelt, MD 20771, USA}
\affiliation{University of Maryland, Baltimore County, 1000 Hilltop Cir, Baltimore, MD 21250, USA}

\author[0000-0001-9771-7953]{Daria Pidhorodetska}
\affiliation{NASA Goddard Space Flight Center, Greenbelt, MD 20771, USA}

\author[0000-0002-8864-1667]{Peter Plavchan}
\affil{Department of Physics \& Astronomy, George Mason University, 4400 University Drive MS 3F3, Fairfax, VA 22030, USA}

\author[0000-0002-1010-3498]{Naylynn Ta\~{n}\'on Reyes}
\affiliation{NASA Goddard Space Flight Center, Greenbelt, MD 20771, USA}
\affiliation{San Diego Mesa College, 7250 Mesa College Dr, San Diego, CA 92111, USA}

\author[0000-0003-2519-3251]{Stephen A. Rinehart}
\affiliation{NASA Goddard Space Flight Center, Greenbelt, MD 20771, USA}

\author[0000-0002-0149-1302]{Bárbara Rojas-Ayala}
\affiliation{Instituto de Alta Investigación, Universidad de Tarapacá, Casilla 7D, Arica, Chile}

\author[0000-0002-6148-7903]{Jeffrey C. Smith}
\affiliation{SETI Institute, 189 Bernardo Ave, Suite 200, Mountain View, CA 94043, USA}
\affiliation{NASA Ames Research Center, Moffett Field, CA, 94035, USA}

\author[0000-0002-3481-9052]{Keivan G.\ Stassun}
\affiliation{Vanderbilt University, Department of Physics \& Astronomy, 6301 Stevenson Center Ln., Nashville, TN 37235, USA}
\affiliation{Fisk University, Department of Physics, 1000 18th Ave. N., Nashville, TN 37208, USA}

\author{Peter Tenenbaum}
\affiliation{SETI Institute, 189 Bernardo Ave, Suite 200, Mountain View, CA 94043, USA}
\affiliation{NASA Ames Research Center, Moffett Field, CA, 94035, USA}


\author[0000-0002-5928-2685]{Laura D. Vega}
\affiliation{NASA Goddard Space Flight Center, Greenbelt, MD 20771, USA}
\affiliation{Vanderbilt University, Department of Physics \& Astronomy, 6301 Stevenson Center Ln., Nashville, TN 37235, USA}

\author[0000-0002-2662-5776]{Geronimo L. Villanueva}
\affiliation{NASA Goddard Space Flight Center, Greenbelt, MD 20771, USA}
\affiliation{GSFC Sellers Exoplanet Environments Collaboration}

\author[0000-0002-7188-1648]{Eric T. Wolf}
\affiliation{University of Colorado, Boulder, CO 80309}
\affiliation{GSFC Sellers Exoplanet Environments Collaboration}

\author[0000-0002-1176-3391]{Allison Youngblood}
\affiliation{Laboratory for Atmospheric and Space Physics, 1234 Innovation Dr, Boulder, CO 80303, USA}


\author{George R. Ricker}
\affiliation{Department of Physics and Kavli Institute for Astrophysics and Space Research, Massachusetts Institute of Technology, Cambridge, MA, 02139, USA}

\author[0000-0001-6763-6562]{Roland K. Vanderspek}
\affiliation{Department of Physics and Kavli Institute for Astrophysics and Space Research, Massachusetts Institute of Technology, Cambridge, MA 02139, USA}

\author[0000-0001-9911-7388]{David W. Latham}
\affiliation{Center for Astrophysics $\mid$ Harvard \& Smithsonian, 60 Garden St, Cambridge, MA, 02138, USA}

\author[0000-0002-6892-6948]{Sara Seager}
\affiliation{Department of Physics and Kavli Institute for Astrophysics and Space Research, Massachusetts Institute of Technology, Cambridge, MA 02139, USA}
\affiliation{Department of Earth, Atmospheric and Planetary Sciences, Massachusetts Institute of Technology, Cambridge, MA 02139, USA}
\affiliation{Department of Aeronautics and Astronautics, MIT, 77 Massachusetts Avenue, Cambridge, MA 02139, USA}

\author{Joshua N. Winn}
\affiliation{Department of Astrophysical Sciences, Princeton University, Princeton, NJ 08544, USA}

\author{Jon M. Jenkins}
\affiliation{NASA Ames Research Center, Moffett Field, CA, 94035, USA}


\author[0000-0001-7204-6727]{G\'asp\'ar \'A. Bakos}
\altaffiliation{Packard Fellow}
\affil{Department of Astrophysical Sciences, Princeton University, NJ 08544, USA}
\affil{MTA Distinguished Guest Fellow, Konkoly Observatory, Hungary}

\author[0000-0001-7124-4094]{C\'{e}sar Brice\~{n}o}
\affiliation{Cerro Tololo Inter-American Observatory, Casilla 603, La Serena, Chile} 

\author[0000-0002-5741-3047]{David R. Ciardi}
\affiliation{Caltech IPAC – NASA Exoplanet Science Institute 1200 E. California Ave, Pasadena, CA 91125, USA}

\author[0000-0001-5383-9393]{Ryan Cloutier}
\affiliation{Center for Astrophysics $\mid$ Harvard \& Smithsonian, 60 Garden St, Cambridge, MA, 02138, USA}

\author[0000-0003-2239-0567]{Dennis M.\ Conti}
\affiliation{American Association of Variable Star Observers, 49 Bay State Road, Cambridge, MA 02138, USA}

\author{Andrew Couperus}
\affiliation{Georgia State University, 33 Gilmer Street SE Atlanta, GA 30303}
\affiliation{RECONS Institute, Chambersburg, PA 17201, USA}

\author{Mario Di Sora}
\affiliation{Campo Catino Astronomical Observatory, Regione Lazio, Guarcino (FR), 03010 Italy}

\author[0000-0002-9138-9028]{Nora L. Eisner}
\affiliation{Department of Physics, University of Oxford, Keble Road, Oxford OX3 9UU, UK}

\author{Mark E. Everett}
\affiliation{NSF's National Optical-Infrared Astronomy Research Laboratory, 950 North Cherry Avenue, Tucson, AZ 85719, USA}

\author[0000-0002-4503-9705]{Tianjun Gan}
\affiliation{Department of Astronomy and Tsinghua Centre for Astrophysics, Tsinghua University, Beijing 100084, China}

\author[0000-0001-8732-6166]{Joel D. Hartman}
\affil{Department of Astrophysical Sciences, Princeton University, NJ 08544, USA}

\author[0000-0002-9061-2865]{Todd Henry}
\affiliation{RECONS Institute, Chambersburg, PA 17201, USA}

\author{Giovanni Isopi}
\affiliation{Campo Catino Astronomical Observatory, Regione Lazio, Guarcino (FR), 03010 Italy}

\author{Wei-Chun Jao}
\affiliation{Georgia State University, 33 Gilmer Street SE Atlanta, GA 30303}

\author[0000-0002-4625-7333]{Eric L. N. Jensen}
\affiliation{Dept.\ of Physics \& Astronomy, Swarthmore College, Swarthmore PA 19081, USA}

\author{Nicholas Law}
\affiliation{Department of Physics and Astronomy, University of North Carolina at Chapel Hill, Chapel Hill, NC 27599, USA}

\author{Franco Mallia}
\affiliation{Campo Catino Astronomical Observatory, Regione Lazio, Guarcino (FR), 03010 Italy}

\author{Rachel A. Matson}
\affiliation{NASA Ames Research Center, Moffett Field, CA, 94035, USA}

\author[0000-0003-4631-1149]{Benjamin J. Shappee}
\affiliation{Institute for Astronomy, University of Hawaii, 2500 Campus Rd, Honolulu, HI 96822}

\author{Mackenna Lee Wood}
\affiliation{Department of Physics and Astronomy, University of North Carolina at Chapel Hill, Chapel Hill, NC 27599, USA}

\author[0000-0001-6031-9513]{Jennifer G. Winters}
\affiliation{Center for Astrophysics $\mid$ Harvard \& Smithsonian, 60 Garden St, Cambridge, MA, 02138, USA}





\correspondingauthor{Emily A. Gilbert}
\email{emilygilbert@uchicago.edu}

\begin{abstract}
 We present the discovery and validation of a three-planet system orbiting the nearby (31.1 pc) M2 dwarf star TOI-700 (TIC 150428135). \object[UCAC3 49-21611]{TOI-700} lies in the \tess continuous viewing zone in the Southern Ecliptic Hemisphere; observations spanning 11 sectors reveal three planets with radii ranging from 1~\rearth to 2.6~\rearth and orbital periods ranging from 9.98 to 37.43 days. Ground-based follow-up combined with diagnostic vetting and validation tests enable us to rule out common astrophysical false-positive scenarios and validate the system of planets. The outermost planet, TOI-700~d, has a radius of $1.19\pm0.11$ \rearth and resides within a conservative estimate of the host star's habitable zone, where it receives a flux from its star that is approximately 86\% of the Earth's insolation. In contrast to some other low-mass stars that host Earth-sized planets in their habitable zones, TOI-700 exhibits low levels of stellar activity, presenting a valuable opportunity to study potentially-rocky planets over a wide range of conditions affecting atmospheric escape. While atmospheric characterization of TOI-700~d with the James Webb Space Telescope (JWST) will be challenging, the larger sub-Neptune, TOI-700~c (R = 2.63~\rearth), will be an excellent target for JWST and future space-based observatories. \tess is scheduled to once again observe the Southern Hemisphere and it will monitor TOI-700 for an additional 11 sectors in its extended mission. These observations should allow further constraints on the known planet parameters and searches for additional planets and transit timing variations in the system.
\end{abstract}


\section{Introduction} \label{sec:intro}

The search for small, rocky planets like Earth orbiting stars outside of our solar system has made rapid progress in the last decade. The \kepler mission \citep{borucki10}, launched in 2009, was designed to explore a specific exoplanet population, Earth-sized planets in Earth-like orbits around Sun-like stars, and aimed to address how common they are. \kepler achieved a number of significant milestones towards this quest, including finding planets within their host stars' habitable zones, the region around a star where liquid water could be exist on the surface of a planet if it has an atmosphere with the appropriate properties \citep{Shapley1953,Strughold1953}.


Among the most important discoveries by \kepler was the high frequency of planets orbiting low-mass M dwarfs \citep{dressing13, dressing15, Gaidos2016, hardegree2019}, particularly small ($<$ 2~\rearth) planets in compact, coplanar, multiplanet systems. The first definitively Earth-sized planet discovered in the habitable zone of its host star, Kepler-186~f (distance = $\sim$ 179 pc), resides in a multiplanet system orbiting an M dwarf about half the mass of the Sun \citep{Quintana2014, torres15}. \kepler's extended mission, \emph{K2}, surveyed substantially more sky than the prime mission and collected data for an order of magnitude more M dwarfs \citep{Dressing2017,Dressing2019} than were observed in \kepler's prime mission \citep[$\sim$ 3000 M dwarfs in the prime mission,][]{Huber2016}. Despite the large number of small planet discoveries, due to the design of the \kepler and \emph{K2} target selections and their limited sky coverage and mission durations, the majority of targets in question are too dim for detailed follow-up observations.


The relative ease of finding small planets orbiting M dwarfs, compared with Sun-like stars, has made them prime targets for exoplanet hunters using both transit photometry and ground-based radial velocity facilities. Radial velocity searches for planets orbiting low-mass stars pre-date \kepler \citep{Delfosse1998, Marcy1998, Rivera2005, plavchan2006, Bonfils2013}, and have led to discoveries of low-mass planets in the habitable zone \citep[e.g.][]{Anglada2013,Anglada2016}. Both ground-based radial velocity and transit photometry surveys searching nearby and bright M dwarfs have discovered systems of planets with the potential for detailed follow-up. The TRAPPIST-1 system, for example, is a late-M dwarf that harbors seven small transiting planets \citep{Gillon2017}, three of which reside in the star's habitable zone. Masses determined via transit timing variations \citep{Luger2017} suggest compositions from rocky terrestrials to more volatile-rich Earth-size planets \citep{grimm18,Dorn2018}.


The Transiting Exoplanet Survey Satellite \citep[\tess,][]{Ricker2015}, launched in April 2018, is performing a near-all-sky photometric survey designed to search for small planets around the Sun's nearest neighbors -- those bright enough for follow-up characterization. The \tess photometric bandpass is redder than \kepler's which provides higher sensitivity to planets orbiting cooler, low-mass stars \citep{sullivan15, Ricker2015, Barclay2018, ballard19}. \tess is now well into its second year of operations and it is delivering on its promise to identify small planets around the closest, brightest M dwarfs. To date, \tess has discovered 17 small planets orbiting 11 M dwarfs with $K_s$-band magnitudes of 6--11. Among these are five compact multiplanet systems: TOI-270 b, c, and d \citep{gunther19}, L 98-59 b, c, and d \citep{kostov2019b}, GJ 357 b (along with non-transiting planets c and d) \citep{luque19}, LP 791-18 b and c \citep{crossfield19}, and TOI-732 b and c \citep{cloutier2020, nowak2020}. As each of the \tess-discovered systems is a new potential benchmark, intensive follow-up is ongoing \citep{cloutier19}, and several planets have been included as targets in Guaranteed Time Observing (GTO) programs for JWST.\footnote{JWST GTO 1201~-~PI D. Lafreni\`ere~-~targets:~GJ 357 b, L 98-59 c and d, LP 791-18 c~-~\url{http://www.stsci.edu/jwst/observing-programs/program-information?id=1201}; JWST GTO 1224~-~PI S. Birkmann~-~target: L 98-59 d~-~\url{http://www.stsci.edu/jwst/observing-programs/program-information?id=1224}}
	
Building on these discoveries from \tess, here we present the discovery and validation of a system of three small planets transiting the nearby (31.1 pc), bright (K = 8.6 mag), M2 dwarf, TOI-700. This system includes a nearly Earth-sized planet in the habitable zone (TOI-700 d). This paper is the first in a series of three papers. In this paper we describe the \tess observations of the system (Section \ref{sec:TESS}), derive precise stellar properties of the host star (Section \ref{sec:stellar}), model planet parameters (Section \ref{sec:lightcurve}), discuss the observational constraints and our vetting and validation of the system (Section \ref{sec:validation}), and explore the dynamics of the system (Section \ref{sec:multiplanet}). In Paper II, \citet{Rodriguez2020} use \emph{Spitzer} observations to provide independent confirmation that TOI-700~d is a transiting planet and refine its parameters, and in Paper III, \citet{Suissa2020} simulate potential climate configurations for TOI-700~d to explore the prospects of both habitable conditions and atmosphere detection.

\section{\tess Observations and Initial Vetting}
\label{sec:TESS}
\begin{figure*}[t]
    \centering
    \includegraphics[width=.8\textwidth]{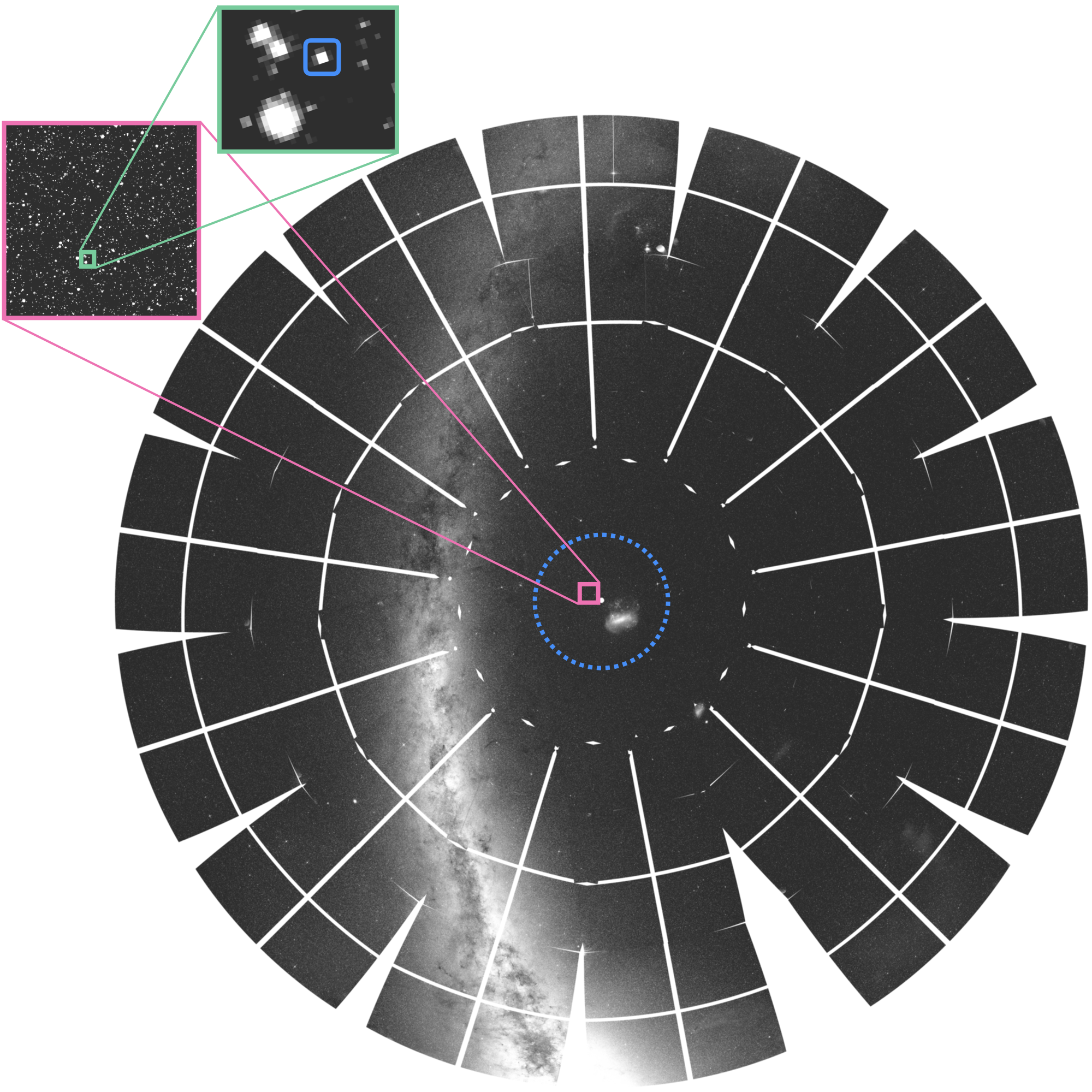}
    \caption{TOI-700 is close to the South Ecliptic Pole and was observed by \tess in 11 of the first 13 sectors of the mission. The field around TOI-700 is relatively uncontaminated, with approximately 1\% of the starlight in the region around TOI-700 coming from other stars. The blue dashed line in the figure is the \tess Continuous Viewing Zone (CVZ). The blue square in the upper-left inset shows TOI-700.}
    \label{fig:sky}
\end{figure*}

TOI-700 (TIC 150428135, 2MASS J06282325-6534456, UCAC4 123-010026) was prioritized for inclusion in the \tess 2-minute cadence mode target list because it was included as a target in the \tess Guest Investigator Program Cycle 1 proposal G011180 - {\it Differential Planet Occurrence Rates for Cool Dwarfs} (PI C. Dressing).\footnote{Details of approved \tess Guest Investigator Programs are available from \url{https://heasarc.gsfc.nasa.gov/docs/tess/approved-programs.html}.} TOI-700 is in a relatively sparsely populated region of the sky only 3$^{\circ}$ away from the South Ecliptic Pole, as shown in Figure~\ref{fig:sky}. This resulted in TOI-700 falling into the field of view of \tess Camera 4 in 11 of the 13 observing sectors that made up the first year of \tess science (sectors 1, 3, 4, 5, 6, 7, 8, 9, 10, 11, and 13), spanning 25 July 2018 to 18 July 2019. During the remaining two sectors, TOI-700 fell into gaps between detectors.

The \tess Science Processing Operation Center (SPOC) pipeline \citep{Jenkins2016} identified three planet candidates transiting TOI-700. These candidate planets had periods of 9.98 (TOI-700.03), 16.05 (TOI-700.01), and 37.42 (TOI-700.02) days, transit depths ranging from 600--3000 ppm, and signal-to-noise ratios of 9.8, 27.4, and 10.0. The pipeline-estimated planet radii were consistent with sub-Neptunes to sub-Saturns, but this was due to missing stellar parameters in the version of the \tess Input Catalog \citep[TIC, ][]{Stassun_CTL_2018,stassun2019} used at the time (TIC V6) and 1 R$_{\odot}$ being adopted by default. The star's broadband colors indicated it was likely an M dwarf. After adopting revised stellar properties based on these colors, the observed transit depths indicated the planets were small, with radii spanning approximately 1--3 R$_{\oplus}$. This early indication of a compact system of small planets transiting a bright M dwarf led to a deeper investigation of the candidate signals, the host star, and subsequently, the planet candidates. 

We performed several initial checks of the \tess data for astrophysical false-positive scenarios that can mimic exoplanet transits. The Data Validation module \citep[DV, ][]{Twicken2018,Li2019} of the TESS SPOC pipeline performs multiple diagnostic vetting tests to investigate such scenarios. The three planet candidates passed all of DV module's diagnostic tests in the multi-sector search of Sectors 1--13. This includes an odd/even depth test; the statistical bootstrap test, which estimates the probability of a false alarm from random noise fluctuations in the light curve and accounts for the non-white nature of the observation noise; the ghost diagnostic test, which compares the detection statistic of the optimal aperture against that of a halo with a 1 pixel buffer ``ring'' around the optimal aperture – this test can identify when transit-like signatures are caused by background scattered light, background eclipsing binaries and background objects such as asteroids; and the difference image centroiding test.

As an additional check, we also used \texttt{DAVE} (Discovery and Vetting of Exoplanets) to perform similar vetting tests on the \tess data. \texttt{DAVE} is an automated pipeline built upon vetting tools developed for \kepler data \citep[e.g.~RoboVetter,][]{coughlin16}, and has been extensively used both for \emph{K2} \citep{hedges19, Kostov2019} and \tess data \citep{crossfield19, kostov2019b}. \texttt{DAVE} performs two sets of vetting tests: light curve-based -- i.e.~odd-even difference between consecutive transits, secondary eclipses, light curve modulations introducing transit-like signals -- and image-based -- i.e.~photocenter motion during transit. Our \texttt{DAVE} analysis confirms that TOI-700 is the transit source for all three planet candidates and rules out simple false-positive features such as odd-even differences or secondary eclipses. Given these results, we moved forward with an investigation of the host star properties.

\section{Determining the Properties of TOI-700}
\label{sec:stellar}
Understanding host stars is an essential component of validating and characterizing exoplanets. Here we use empirically-derived relations based on absolute magnitude (see Section \ref{sec:derived_params}) to estimate \target's fundamental parameters and provide an additional level of characterization using an observed medium resolution spectrum. We then place constraints on the age of TOI-700 using historical photometry (see Section \ref{sec:age}).

\subsection{Empirically Derived Stellar Parameters}
\label{sec:derived_params}

We determined fundamental parameters of TOI-700 using empirical relations for M dwarfs that are based on the variation of mass, radius, luminosity, and temperature with absolute 2MASS $K_s$-band magnitude (M$_{K_s}$). This approach is similar to the methods used in other recent \tess discoveries of small planets transiting M dwarfs \citep[e.g.~L~98-59 and LTT~1445A,][]{kostov2019b, winters2019b}. Specifically, we used the M$_{K_s}$-mass relation of \citet{mann19},\footnote{\url{https://github.com/awmann/M_-M_K-}} calibrated using M dwarf binaries with precise orbital solutions, to estimate the mass of TOI-700. We then used the M$_{K_s}$-radius relationship of \citet{Mann2015}, calibrated using M dwarfs with interferometrically measured radii, to estimate the stellar radius. To calculate the effective temperature (T$_{\rm eff}$), we estimated the $K$-band bolometric correction using the relations of \citet[]{Mann2015} to calculate the stellar luminosity and then combined it with the measured radius estimate using the Stefan-Boltzmann law. The derived parameter estimates are consistent with an M2V~$\pm$~1 dwarf following the color-temperature relations of \citet{Pecaut(2013)}.\footnote{We used the updated stellar parameter table, Version 2019.3.22, available at \url{http://www.pas.rochester.edu/~emamajek/EEM_dwarf_UBVIJHK_colors_Teff.txt}} We estimated parameter uncertainties using Monte Carlo methods assuming Gaussian distributed measurement errors and added the systematic scatter in the parameter relations in quadrature. We found the stellar radius is $0.420 \pm 0.031$ \rsun, mass is $0.416 \pm 0.010$ \msun, effective temperature is $3480 \pm 135$ K, and mean stellar density is $8.0 \pm 1.8$ g cm$^{-3}$. 

We also used the star's photometry to estimate its metallicity via its position on a color-magnitude diagram. Color-magnitude position is mainly sensitive to [Fe/H] for single M dwarfs, unlike for Sun-like stars where color-magnitude diagram position also depends on age (due to main sequence evolution). We interpolated over five different metal-sensitive color-magnitude combinations (using \emph{Gaia}, 2MASS, and APASS photometry) using stars with accurate metallicities from near-infrared spectra \citep{RojasAyala2012,Mann2013a,Newton2014} and parallaxes from \emph{Gaia} DR2. This method yielded a consistent metallicity across all relations, with a final adopted value of [Fe/H] = $-0.07\pm0.11$, and errors limited primarily by the [Fe/H] values applied to the comparison sample.

These stellar properties are adopted as the set we use in the analyses presented in the rest of the paper. They are summarized in Table~\ref{tab:stellarparameters}, along with the star's astrometric and photometric properties. 

\begin{deluxetable}{l l r }[!ht]
\hspace{-1in}\tabletypesize{\scriptsize}
\tablecaption{Stellar Parameters \label{tab:stellarparameters}}
\tablewidth{0pt}
\tablehead{
\colhead{Parameter} & \colhead{Value} & \colhead{Source}
}
\startdata
\multicolumn{3}{c}{\em Identifying Information} \\
Name & TOI-700 & \\
TIC ID & 150428135 & \\
Alt. name & 2MASS J06282325-6534456 & \\
Alt. name & UCAC4 123-010026 & \\  \\
\multicolumn{3}{c}{\em Astrometric Properties} \\
$\alpha$ R.A. (hh:mm:ss) & 06 28 23.229  & $\emph{Gaia}$ DR2 \\
$\delta$ Dec. (dd:mm:ss) & -65 34 45.522  & $\emph{Gaia}$ DR2 \\
$\mu_{\alpha}$ (mas~yr$^{-1}$) & $-102.750 \pm 0.051$ & $\emph{Gaia}$ DR2 \\
$\mu_{\delta}$ (mas~yr$^{-1}$) & $161.805 \pm 0.060$  & $\emph{Gaia}$ DR2 \\
Barycentric RV (km~s$^{-1}$)  & $-4.4 \pm 0.1$ & This work \\
Distance (pc) & $31.127 \pm 0.020$  & $\emph{Gaia}$ DR2\\
\multicolumn{3}{c}{\em Stellar Properties} \\
Spectral Type\dotfill& M2V $\pm$ 1 & This Work \\
\Teff\ (K)\dotfill& $3480 \pm 135$  & This Work\\
$\lbrack $Fe/H$ \rbrack$\dotfill& $-0.07 \pm 0.11$ & This Work \\
M$_{\star}$ (M$_\odot$)\dotfill& $0.416 \pm 0.010$  & This Work\\
R$_{\star}$ (R$_\odot$)\dotfill& $0.420 \pm 0.031$ & This Work\\
L$_{\star}$ (L$_\odot)$\dotfill& $0.0233 \pm 0.0011$ & This Work\\
log(g)\dotfill& $4.81 \pm 0.06$ & This Work\\
$\rho_{\star}$ (g cm$^{-3}$)\dotfill& $8.0 \pm 1.8$  & This Work\\
Rotation period (d)\dotfill&$54.0\pm0.8$&This Work\\
Age (Gyr)\dotfill & $>1.5$ &  This Work\\
\multicolumn{3}{c}{\em Photometric Properties} \\
$B_J$ (mag)\dotfill& $14.550 \pm 0.047$ & APASS DR9 \\
$B_P$ (mag)\dotfill& $13.350 \pm 0.003$ & $\emph{Gaia}$ DR2  \\
$V_J$ (mag)\dotfill&  $13.072 \pm 0.012$ & APASS DR9 \\
$V_{J}$ (mag)\dotfill&  $13.10 \pm 0.01$ & This work \\
$G$ (mag)\dotfill& $12.067 \pm	0.001$  & $\emph{Gaia}$ DR2 \\
$g^{\prime}$ (mag)\dotfill& $13.796 \pm 0.026$  & APASS DR9 \\
$r^{\prime}$ (mag)\dotfill& $12.487 \pm 0.031$  & APASS DR9 \\
$R_{KC}$ (mag)\dotfill& $12.03 \pm 0.01$ & This Work  \\
$R_P$ (mag)\dotfill& $10.960 \pm 0.002$ & $\emph{Gaia}$ DR2  \\
$T$ (mag)\dotfill&  $10.910 \pm 0.007$  & TIC V8 \\
$I_{KC}$ (mag)\dotfill& $10.73 \pm 0.02$ & This Work  \\
$i^{\prime}$ (mag)\dotfill& $11.352 \pm 0.038$  & APASS DR9 \\
$J$ (mag)\dotfill& $9.469 \pm	0.023$	& 2MASS \\
$H$ (mag)\dotfill& $8.893 \pm	0.027$  & 2MASS \\
$K_s$ (mag)\dotfill&  $8.634 \pm 0.023$ & 2MASS \\
$W1$ (mag)\dotfill& $8.523 \pm	0.023$	& AllWISE \\
$W2$ (mag)\dotfill& $8.392 \pm	0.020$  & AllWISE \\
$W3$ (mag)\dotfill& $8.281 \pm	0.019$ & AllWISE \\
$W4$ (mag)\dotfill& $8.234 \pm	0.115$ & AllWISE \\
\enddata
\tablenotetext{}{\emph{Gaia} DR2 - \citep[]{gaia2018,bailerjones2018}, RAVE DR5 - \citep[]{Kunder(2017)}, TIC V8 - \citep[]{stassun2019}, APASS DR9 - \citep[]{Henden2016}, 2MASS - \citep[]{Skrutskie2006}, AllWISE - \citep[]{Cutri2013}}
\end{deluxetable}

For an additional level of stellar characterization, we obtained a spectrum of \target\ with the Goodman High-Throughput Spectrograph \citep{Goodman} on the Southern Astrophysical Research (SOAR) 4.1~m telescope located at Cerro Pachón, Chile. On 2019 September 30 UT and under clear (photometric) conditions, we obtained five spectra of \target, each with an exposure time of 120 seconds. We took all exposures using the red camera, 1200 l/mm grating in the M5 setup, and the 0.46\arcsec\ slit rotated to the parallactic angle, which yielded a resolution of $\simeq$5900 spanning 625--750 nm. For wavelength calibration, we obtained observations of Ne arc lamps taken just before the target, as well as dome flats and biases taken during the afternoon. 

\begin{figure}[htb]
\begin{center}
\includegraphics[width=0.5\textwidth]{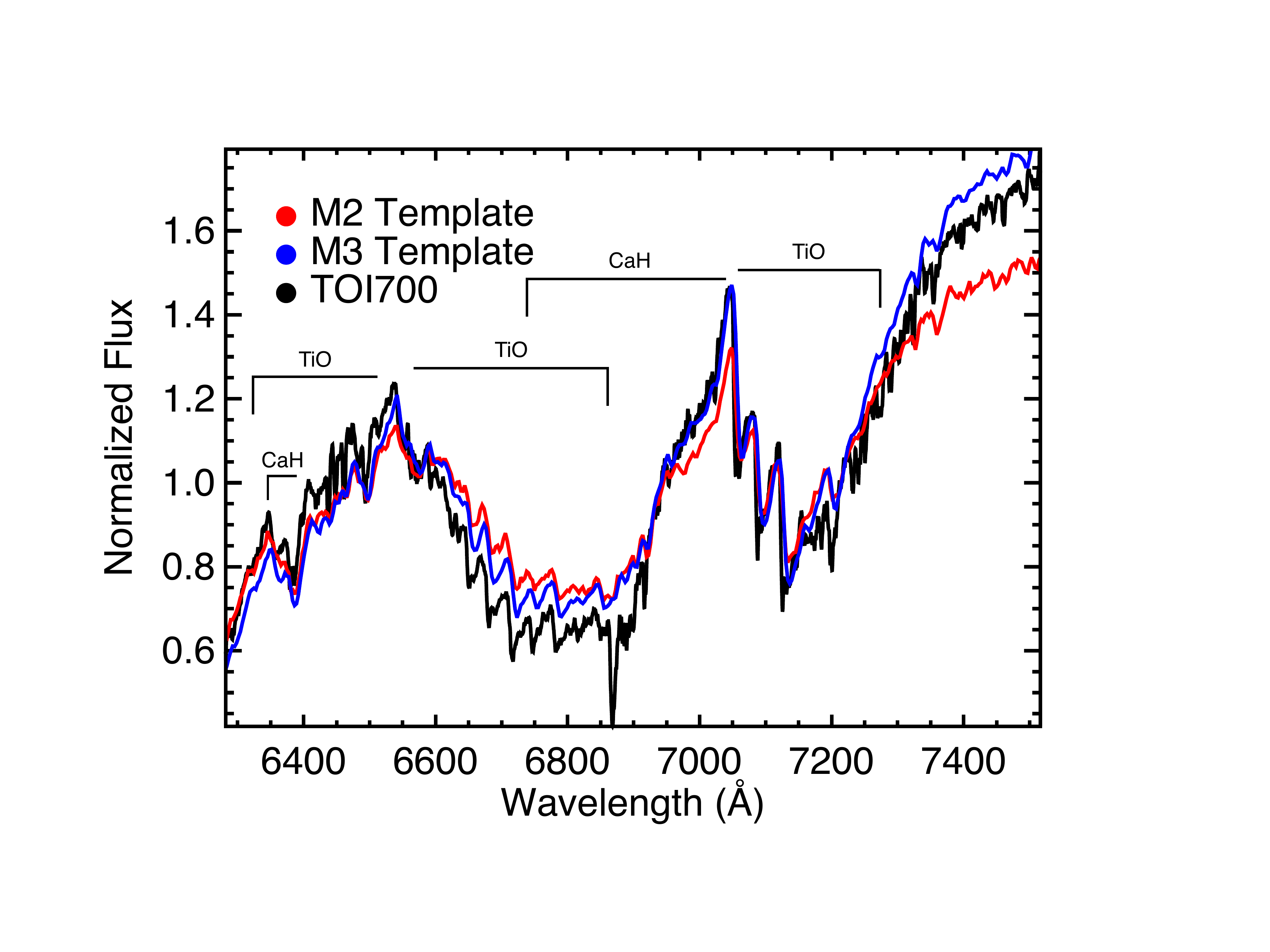}
\caption{SOAR Goodman spectrum of \target\ (black) compared to an M2 (red) and an M3 (blue) template spectrum. The spectrum exhibits a continuum shape and broad TiO and CaH absorption bands that are characteristic of early- to mid-M dwarfs \citep{Kirkpatrick1991}. The good visual match to the M2 and M3 templates is consistent with the M2~$\pm$~1 spectral type estimated from the empirically derived effective temperature in Section \ref{sec:TESS}. Because we used an archival calibration, the flux calibration of our Goodman spectrum is likely only accurate to $\simeq$10\%.}
\label{fig:goodman}
\end{center}
\end{figure}

We performed bias subtraction, flat fielding, optimal extraction of the target spectrum, and mapping pixels to wavelengths using a 4th-order polynomial derived from the Ne lamp data. We then stacked the five extracted spectra using the robust weighted mean (for outlier removal). The stacked spectrum had a signal-to-noise ratio $>100$ over the full wavelength range (excluding areas of strong telluric contamination). While we observed no spectrophotometric standards during the night, we corrected instrument throughput with wavelength using standards from an earlier night. The final spectrum is shown in Figure~\ref{fig:goodman} with M2 and M3 template spectra from \citet{Cushing2005} for comparison. The continuum shape and broad TiO and CaH molecular features \citep[see][]{Kirkpatrick1991} are a good match to these standards and indicate that \target\ is approximately an M2 spectral type, consistent with our fundamental parameter estimates.

As an independent check of the empirically derived stellar parameters presented in this section, we used multiple methods that combine spectral energy distributions (SEDs) and stellar models to estimate parameters in Appendix~\ref{appendix:stellar_params}. We found consistent results regardless of the method used, providing validation of the adopted parameters.

\subsection{Constraints on the Age of TOI-700}
\label{sec:age}
Our stellar parameter analyses indicate that TOI-700 is a main sequence M2 dwarf star. M dwarfs change little over the vast majority of their very long lifespans on the main sequence, therefore precise age determinations for such stars are notoriously difficult \citep[e.g.][]{newton16, Veyette_2018}. Early M dwarfs, like TOI-700, have magnetic dynamos similar to the Sun, and shed angular momentum over time via magnetic braking as the stellar wind interacts with magnetic field lines. This braking results in progressively slower rotation and lower levels of magnetic activity. Stellar magnetic activity manifests in the form of star spots, flares, increased X-ray and UV emission, and emission in activity-sensitive spectral lines (e.g. H$\alpha$, Na I, Ca II), which can provide additional constraints on the age of an M dwarf. In 11 sectors of \tess 2-minute cadence high precision photometry of TOI-700, there are no detectable white-light flares. Additionally, we observed no emission in activity sensitive lines in a high-resolution spectrum (see Section~\ref{sec:spectra}). We also searched for excess UV emission from TOI-700 in the GALEX \citep{morrissey2005} catalog of \citet{bianchi2011}. There is a weak near-UV source near the location of the star, but it is flagged as an image artifact, so we do not attribute this detection to TOI-700.

\begin{figure*}
\includegraphics[width=\textwidth]{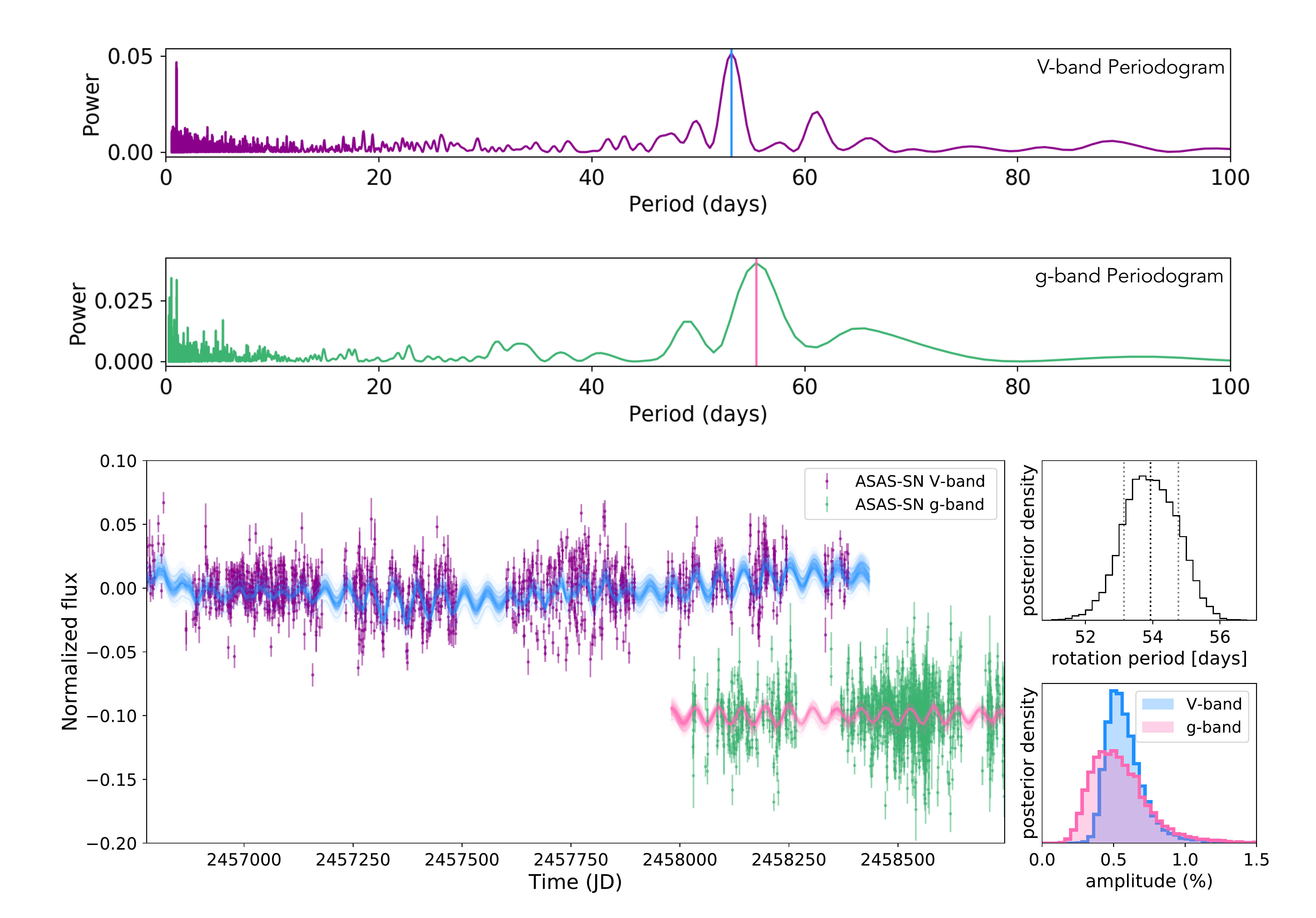}
\caption{Long-term monitoring of TOI-700 by ground-based ASAS-SN telescopes show a $54.0\pm0.8$ day rotation period. \textit{Top} Lomb-Scargle Periodogram of the ASAS-SN $V$-band photometry. \textit{Middle} Lomb-Scargle Periodogram of the ASAS-SN $g$-band photometry. \textit{Bottom} ASAS-SN photometry and GP modeling posteriors. The combined $V$-band data in purple and $g$-band data in green cover five years. The $g$-band data has been offset by -0.1. Fifty posterior draws from a periodic GP kernel model are shown in blue ($V$-band) and pink ($g$-band). The lack of stellar activity and slow rotation period indicate that the star is not young. The posterior distribution of the rotation period and amplitude of the rotation signal are provided (bottom right panels). The amplitude of the rotation is 0.6\% in the $V$-band and 0.4\% in the $g$-band.}
\label{fig:rotation}
\end{figure*}

To estimate the rotation period of TOI-700 we analyze more than five years of archival photometry from the All-Sky Automated Survey for Supernovae (ASAS-SN) \citep{Shappee14, Kochanek17}. We obtained ASAS-SN data (see Figure \ref{fig:rotation}) from the publicly available Sky Patrol database.\footnote{https://asas-sn.osu.edu} The database contained over 2500 photometric observations of TOI-700 in two bands, $V$ and $g$, spanning approximately five years. Both the $V$- and $g$-band long baseline light curves exhibited slowly varying semi-sinusoidal modulation, consistent with periodic brightness variations due to star spots in the photosphere of a rotating star. We calculated the Lomb-Scargle Periodogram of the ASAS-SN data in each band to estimate the stellar rotation period. The power spectra each exhibited one dominant peak: 53 days in the $V$-band and 55 days in the $g$-band (see Figure~\ref{fig:rotation}). Given the consistency of these analyses, we adopt the mean value of 54 days as the initial estimate of the stellar rotation period.

We then used \texttt{exoplanet} \citep{exoplanet:exoplanet} to model the variability in the ASAS-SN data using a periodic Gaussian process kernel \citep{exoplanet:foremanmackey17,exoplanet:foremanmackey18} with the Lomb-Scargle estimated period as a broad Gaussian input prior in the probabilistic model. The particular form of the periodic kernel has two peaks in frequency space: one at the model period and another at half the model period. This kernel is well suited to modeling the signature of stellar rotation (spots coming in and out of view as the star rotates) which often produces two peaks in frequency space owing to multiple spot clusters on the stellar surface. The parameters of the model were the log period, and for each of the two separate data sets the photometric mean, a log amplitude, a log quality factor of the primary frequency, a ratio of the log quality factors between the primary and secondary frequency, a ratio between the amplitude of primary and secondary frequencies, and a log noise parameter that is added in quadrature with the reported uncertainty in the data. All log parameters here are natural logarithms. In addition, for only the $V$-band ASAS-SN data, we included a long term variability term because there appear to be slow changes in the measured brightness of the target in that data. We sampled from this model using the \texttt{PyMC3} \citep{exoplanet:pymc3} implementation of the No U-turn Sampler \citep[NUTS, ][]{NUTS} which is a form of Hamiltonian Monte Carlo. We measured the posterior rotation period to be $54.0\pm0.8$ days. Posterior draws from the model in data space are shown in Figure~\ref{fig:rotation}, along with posteriors for the rotation period and the multi-band amplitudes. This rotation period is typical for inactive early-mid spectral type M dwarfs \citep{Newton2017}. The modeled amplitude of the rotation signal in the $V$-band is $0.6\pm0.1$\% and $0.4\pm0.1$\% in the $g$-band. An independent rotation period analysis using long baseline photometry from the HATSouth telescope network \citep{bakos:2013:hatsouth} that validates the ASAS-SN analysis is presented in Appendix~\ref{appendix:p_rot}.

Stellar galactic kinematics can be combined with the measured rotation period and activity constraints to provide additional age constraints. To calculate the galactic $UVW$ velocities, we followed the prescription of \citet{johnson1987}, updated to epoch J2000. We also adopted a coordinate system where $U$ is positive toward the Galactic center and calculated the $UVW$ velocities corrected to the local standard of rest \citep[LSR,][]{coskunoglu2011}. We used the available astrometry from \emph{Gaia} DR2 and the radial velocity measurement from the CHIRON spectrum presented in this paper (see Section~\ref{sec:spectra}) to calculate ($UVW_{LSR}$) = (-17.83, 20.34, -2.40) $\pm$ (0.29, 0.44, 0.26) km s$^{-1}$, which yielded a total Galactic velocity S$_{LSR}$ = 27.15 km s$^{-1}$ indicating that the star is a likely member of the thin disk population following the kinematic criteria of \citet{bensby2010}. The typical metallicity of stars in the thin disk, -0.7~$<$~$\lbrack $Fe/H$ \rbrack$~$<$~+0.5 dex \citep{bensby14}, is also consistent with the metallicity of TOI-700 estimated in this work. Following the systematic study of M dwarf rotation and kinematics from \citet{Newton2014}, the combined Galactic kinematics and rotation period indicate that TOI-700 is older than $\sim$2 Gyr.
 
As a final check, we used \texttt{stardate} \citep{stardate,Angus2019} to estimate the age of TOI-700 using the photometry listed in Table~\ref{tab:stellarparameters}, the \emph{Gaia} parallax, and the rotation rate from ASAS-SN. This method has been calibrated and tested on stars with \emph{Gaia} $B_P-R_P<2.7$, so is appropriate for TOI-700. The resulting age estimate was $>$1.5 Gyr at 95\% confidence. This result is consistent with the above limit and is adopted as the stellar age in reported in Table~\ref{tab:stellarparameters}.

\section{Measuring the Physical Properties of the Planets Orbiting TOI-700}
\label{sec:lightcurve}

We determined the physical properties of the TOI-700 planets by combining the stellar properties measured previously with an analysis of the \tess time series data. Our \tess data analysis made use of the SPOC-created systematics-corrected light curves from the \tess pipeline \citep{Jenkins2016,Smith2012,Stumpe2014} collected at 2-minute cadence. We first used the \texttt{lightkurve} package to download the datasets from the MAST archive \citep{lightkurve} and used the \texttt{exoplanet} \citep{exoplanet:exoplanet} toolkit to create models of the light curves  and infer the planet properties. Each of the 11 separate sectors of data have different noise properties, so we opted to model these as independent datasets with distinct noise terms. Each sector is modeled with a mean offset, a white noise term parameterized as the natural log variance, and two hyper-parameters, $\ln(S_0)$ and $\ln(\omega_0)$, of a Gaussian Process (GP) that describe a stochastically-driven, damped harmonic oscillator and model residual stellar variability. In addition to the sector-dependent parameters, the model includes two stellar limb darkening parameters, the natural logarithm of stellar density, the stellar radius, and for each planet a natural log orbital period, a natural log planet-to-star radius ratio, impact parameter, eccentricity, periastron angle, and time of first transit.


We used a Normal prior for the stellar radius with mean and standard deviation of 0.42 and 0.03, respectively, in solar units. The natural log mean stellar density, in cgs units, had a Gaussian prior with a mean of $\ln{8.0}$ and standard deviation of 0.3~dex (as per Section 2.1). The limb darkening parameters were estimated following \citet{exoplanet:kipping13} and were sampled uniformly. The impact parameter was uniformly sampled between zero and one plus the planet-to-star radius ratio. The eccentricity had a beta prior \citep[as suggested by ][]{Kipping2013}, with parameters appropriate for systems of small planets \citep{exoplanet:vaneylen19} and was bounded between zero and one. The periastron angle at transit was sampled from an isotropic, two-dimensional normal with the angle given by the arctangent of the ratio of the two coordinates, yielding a uniform prior between -$\pi$ and $\pi$ with no hard boundries \citep{exoplanet:exoplanet}.

We used \texttt{PyMC3} to make draws from the posterior distribution. We used four independent chains and ran 6000 tuning steps and then 5000 draws which we used for inference. The chains were well mixed and the number of effective samples was over 1000 for each model parameter. The Gelman–Rubin diagnostic \citep{Gelman1992} measures convergence between independent chains. All model parameters had a Gelman–Rubin diagnostic within one part in 1000 of unity, providing confidence that the chains had converged. The results of our modeling are shown in Table~\ref{tab:TransitParameters}. The ``Derived Parameters" listed in Table~\ref{tab:TransitParameters} are computed during the sampling as Deterministic parameters in \texttt{PyMC3}.

\begin{figure}
    \centering
    \includegraphics[width=0.43\textwidth]{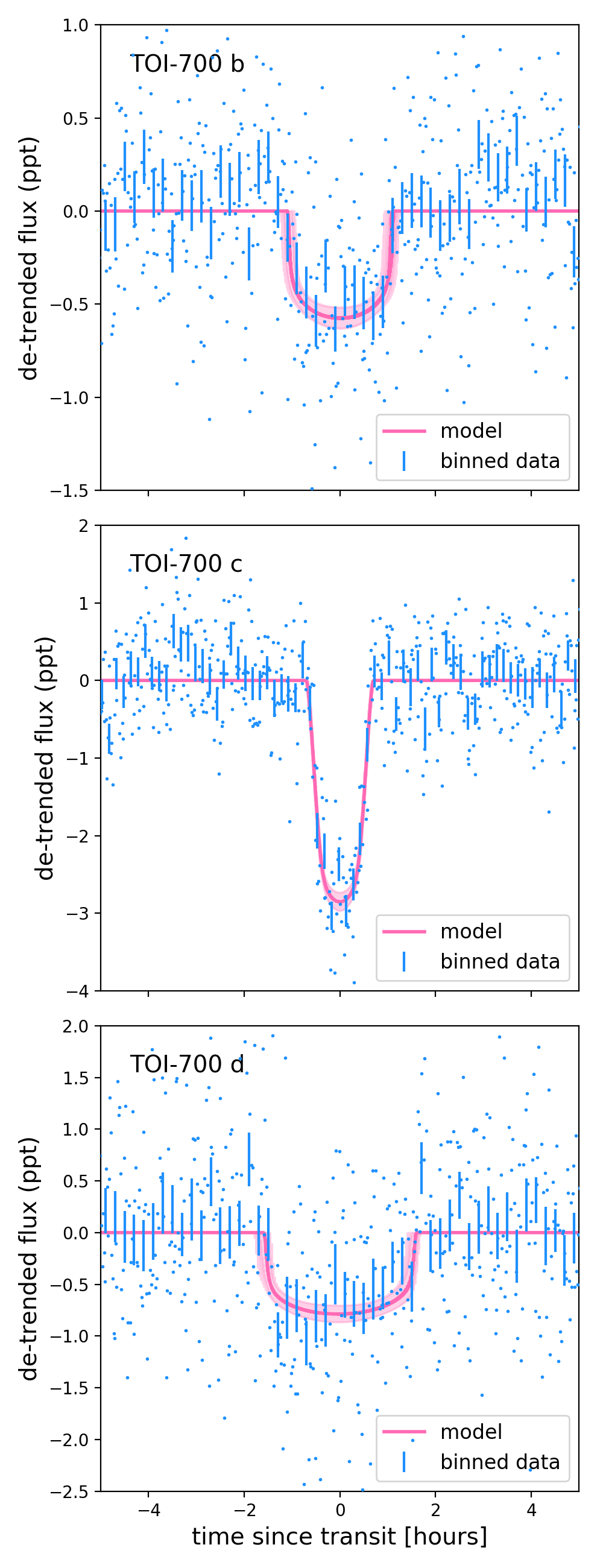}
    \caption{Phase-folded, light curves from 11 sectors of \tess data for planets TOI-700 b (upper panel), TOI-700 c (middle panel), and TOI-700~d (lower panel), along with the respective transit model (pink) showing the 1-sigma range in models consistent with the observed data. The corresponding transit parameters are listed in Table \ref{tab:TransitParameters}.}
    \label{fig:folded_transits}
\end{figure}

The best-fitting transit model for the three planets is shown in Figure~\ref{fig:folded_transits}, along with the 1-$\sigma$ bounds of the transit model shown in the space of the data and binned \tess observations. The radii of the three planets are $1.01\pm0.09$, $2.63\pm0.4$, and $1.19\pm0.11$ R$_\oplus$ from inner to outer planet. TOI-700 b and d are of similar radii to Earth while TOI-700 c is likely a sub-Neptune-type planet \citep{Rogers2015}. TOI-700 d receives an incident flux of $0.86\pm0.2$ that of Earth’s insolation, which places it within the circumstellar habitable zone \citep{kopparapu2013}.

To verify the results of our first \tess light curve model, we repeated this analysis but rather than starting with \tess pipeline generated light curves, we began by using the 2-minute cadence target pixel file (TPF) data products \citep{Jenkins2016}. For each of the 11 TPFs, we manually excluded data with significant stray light. Next, we generated custom apertures for each sector by iteratively adding pixels to the aperture ordered by brightness and then selecting the aperture which minimizes the scatter in the light curve. We then use these apertures to generate light curves for each sector. The light curves were extracted using the \texttt{lightkurve} package. We then masked out transits using the ephemeris generated by the \tess pipeline alerts and subsequently detrended the light curves using pixel-level decorrelation, adapted from the methods of \texttt{everest} \citep{Luger2016}. Once detrended, we combined all 11 sectors into a single light curve. We then used the \texttt{exoplanet} package in a similar manner to that described above, except that we used the entire time series as a single dataset rather than breaking it into 11 separate datasets. The resulting exoplanet parameters were consistent at the $<$0.2$\sigma$ level with the values calculated in our first analysis (see  Table~\ref{tab:TransitParameters}).

\begin{deluxetable}{l c c c}[htb]
\tablecaption{Planet Parameters \label{tab:TransitParameters}}
\tablewidth{0pt}
\tablehead{
\colhead{Parameter} & \colhead{Median} & \colhead{+1$\sigma$}  & \colhead{-1$\sigma$}
}
\startdata
\multicolumn{4}{c}{\em Model Parameters} \\
\textbf{Star} & & & \\
$\ln{\rho}$ [g~cm$^{-3}$] & 2.08 & 0.16 & 0.17 \\
Limb darkening $u_1$ & 0.34 & 0.39 & 0.24 \\
Limb darkening $u_2$ & 0.13 & 0.38 & 0.32\\
{\bf TOI-700~b} & & & \\
${T_0}$ (BJD - 2457000)& 1331.3547 & 0.0048 & 0.0032 \\
$\ln({\textrm{Period}} \textrm{[days]})$ & 2.300284 & 0.000024 & 0.000028 \\
Impact parameter & 0.20 & 0.19 & 0.14 \\
$\ln{R_p/R_*}$ & -3.809& 0.049 & 0.55 \\
eccentricity & 0.032 & 0.050 & 0.024 \\
$\omega$ [radians] & -0.6 & 2.5 & 	1.8 \\
 & & & \\
{\bf TOI-700~c} & & & \\
${T_0}$ (BJD - 2457000)& 1340.0887 & 0.0011 & 0.0010 \\
$\ln{Period}$ [days] & 2.7757773 & 0.0000055 & 0.0000058 \\
Impact parameter & 0.904 & 0.016 & 0.024 \\
$\ln{R_p/R_*}$ & -2.857 & 	0.053 & 	0.046 \\
eccentricity & 0.033 & 0.063 & 0.025 \\
$\omega$ [radians] & 0.4 &	1.8 & 	2.4  \\
 & & & \\
{\bf TOI-700~d} & & & \\
${T_0}$ (BJD - 2457000)& 1330.4737 & 0.0035 & 0.0040\\
$\ln{Period}$ [days] & 3.622365 & 0.000020 & 0.000027\\
Impact parameter & 0.40 & 0.15 & 0.22 \\
$\ln{R_p/R_*}$ & -3.641 &	0.053 & 	0.060 \\
eccentricity & 0.032 & 0.054 & 0.023 \\
$\omega$ [radians] & 0.2 & 	2.0 & 	2.3 \\
\hline
\multicolumn{4}{c}{\em Derived Parameters} \\
{\bf TOI-700~b} & & & \\
Period [days] & 9.97701 & 0.00024 &	0.00028\\
${R_p/R_*}$ & 0.0221 &	0.0011 &	0.0012\\
Radius ${[R_\oplus]}$ & 1.010 & 	0.094 & 	0.087\\
Insolation & 5.0 & 1.1 & 0.9\\
${a/R_*}$ & 34.8 & 1.9 & 1.9 \\
$a$~[AU] & 0.0637& 0.0064& 0.0060\\
Inclination (deg) & 89.67 & 0.23 & 0.32 \\
Duration (hours) & 2.15 & 0.15 & 0.7\\
 & & & \\
{\bf TOI-700~c} & & & \\
Period [days] & 16.051098 &	0.000089 &	0.000092\\
${R_p/R_*}$ & 0.0574 & 	0.0032 & 	0.0026\\
Radius ${[R_\oplus]}$ & 2.63 &  0.24 & 	0.23\\
Insolation & 2.66 & 0.58 & 0.46\\
${a/R_*}$ & 47.8 & 2.7 & 2.6\\
$a$~[AU] & 0.0925 & 0.0088 & 0.0083\\
Inclination (deg) & 88.90 & 0.08 & 0.11\\
Duration (hours) & 1.41 &  0.14 & 0.09\\
 & & & \\
{\bf TOI-700~d} & & & \\
Period [days] & 37.4260 & 0.0007 &	0.0010\\
${R_p/R_*}$ & 0.0262& 	0.0014& 	0.0015\\
Radius ${[R_\oplus]}$ & 1.19 & 	0.11 &	0.11\\
Insolation & 0.86 & 0.19 & 0.15\\
${a/R_*}$ & 84.0 & 4.7 & 4.6\\
$a$~[AU] & 0.163 & 0.015 & 0.015 \\
Inclination (deg) & 89.73 & 0.15 & 0.12 \\
Duration (hours) & 3.21 & 0.27 & 0.26\\
\enddata
\end{deluxetable}

\section{System Validation of TOI-700}
\label{sec:validation}



Here we build upon the TESS pipeline and DAVE vetting analyses and present a validation of TOI-700 b, c, and d. We investigated this system using both observational constraints (Section \ref{sec:observations}) as well as the publicly available software package, \texttt{vespa}, (Section \ref{sec:vetting}) to validate the planetary nature of the signals observed by TESS.

\subsection{Observational Constraints}
\label{sec:observations}
We collected a variety of ground-based observations in order to explore potential false-positive scenarios for the TOI-700 system. The majority of these observational constraints were obtained through the TESS Follow-up Observers Program (TFOP). We utilized archival imaging to place limits on background sources (Section \ref{sec:archival_images}), high-resolution speckle imaging to rule out close-in bound companions (Section \ref{sec:speckle}), high-resolution spectra to place constraints on potential blended sources at even smaller separations (Section \ref{sec:spectra}), and ground-based time series photometry to observe additional planet transits and rule out nearby eclipsing binaries (Section \ref{sec:photometry}).

\subsubsection{Archival Imaging}
\label{sec:archival_images}

\begin{figure*}
    \centering
    \includegraphics[width=\textwidth]{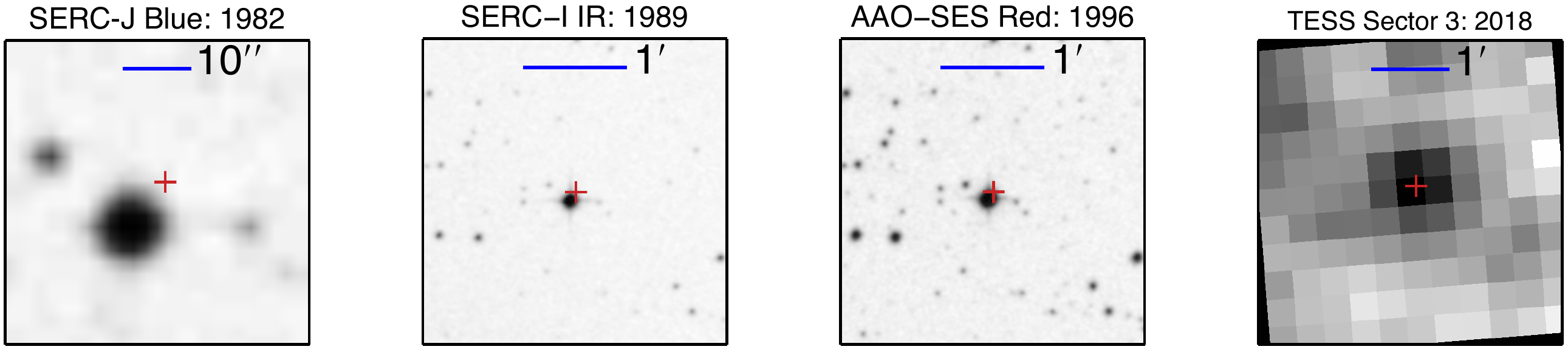}
    \caption{Archival images of TOI-700 from the Digitized Sky Survey showing the location of TOI-700 during the \tess observations (red cross). The star has moved approximately 7$^{\prime\prime}$ since the earliest image in 1982. There are no sources visible at its current location down to a limit of $\approx$~17 mag in the SERC-J ``Blue" band. The faint stars within $\approx$1$^{\prime}$ are $\ge$7.5 mag fainter than TOI-700 in the SERC-I ``IR'' band (the closest available to the \tess bandpass) and do not contribute significant flux to dilute the planet transits.}
    \label{fig:archival}
\end{figure*}

TOI-700 was observed three times in historical large-scale photographic sky surveys \citep{morgan1992} during epochs spanning 1982 to 1996. These Southern Hemisphere observations were obtained using the UK 1.2 m Schmidt Telescope at Siding Spring Observatory and were made available for digital download as part of the Digitized Sky Survey\footnote{\url{https://archive.stsci.edu/cgi-bin/dss_form}} \citep[][ shown in Figure~\ref{fig:archival}]{lasker1990, lasker1994}. TOI-700 was observed on 1982 November 20 during the Science and Engineering Research Council (SERC) J survey using the ``Blue'' photographic emulsion \citep[$\lambda$ = 395-590 nm;][]{monet2003} and 1989 December 18 during the SERC-I survey using the ``IR'' photographic emulsion \citep[$\lambda$ = 715-900 nm;][]{monet2003}. The star was observed again on 1996 February 19 during the Anglo Australian Observatory Second Epoch Survey (AAO-SES or AAO-R) using the ``Red" photographic emulsion \citep[$\lambda$ = 590-690 nm;][]{monet2003}. The relatively large proper motion of TOI-700 allows us to search for background objects at its current position. 

With a total proper motion of 191.673 mas year$^{-1}$, the star has moved approximately 7$^{\prime\prime}$ across the sky to its current location since the SERC-J images were obtained in late 1982. In the archival data, there are no background sources at the star's current position down to $\approx$17 mag in the SERC-J ``Blue'' band as shown in Figure \ref{fig:archival}. We also note there are several faint stars within a separation of $\sim$1$^{\prime}$ of TOI-700 that are within the apertures used to extract the \tess photometry. We compared their photometry in the SERC-I ``IR'' band, the closest available to the \tess bandpass, with TOI-700 as calibrated and presented in the USNO-B1.0 catalog \citep{monet2003}. The brightest star is about 7.5 magnitudes fainter than TOI-700 and we find that none of these stars are bright enough to mimic the transits even if they are totally eclipsing binaries. This is consistent with our ground-based time-series observations that rule out nearby eclipsing binaries at the periods of the TOI-700 planets (see Section~\ref{sec:photometry}).
 
\subsubsection{High-Resolution Imaging}
\label{sec:speckle}
If a star hosting a planet candidate has a close bound companion (or companions), the companion can create a false-positive exoplanet detection if it is an eclipsing binary. Additionally, flux from the additional source(s) can lead to an underestimated planetary radius if not accounted for in the transit model \citep{Ciardi2015, furlanciardi2017, matson2018}. To search for close-in bound companions unresolved in our other follow-up observations, we obtained speckle imaging observations from both Gemini-South's Zorro instrument and the SOAR HRCam. These observations were obtained through TFOP.

\begin{figure}

  \includegraphics[width=0.47\textwidth]{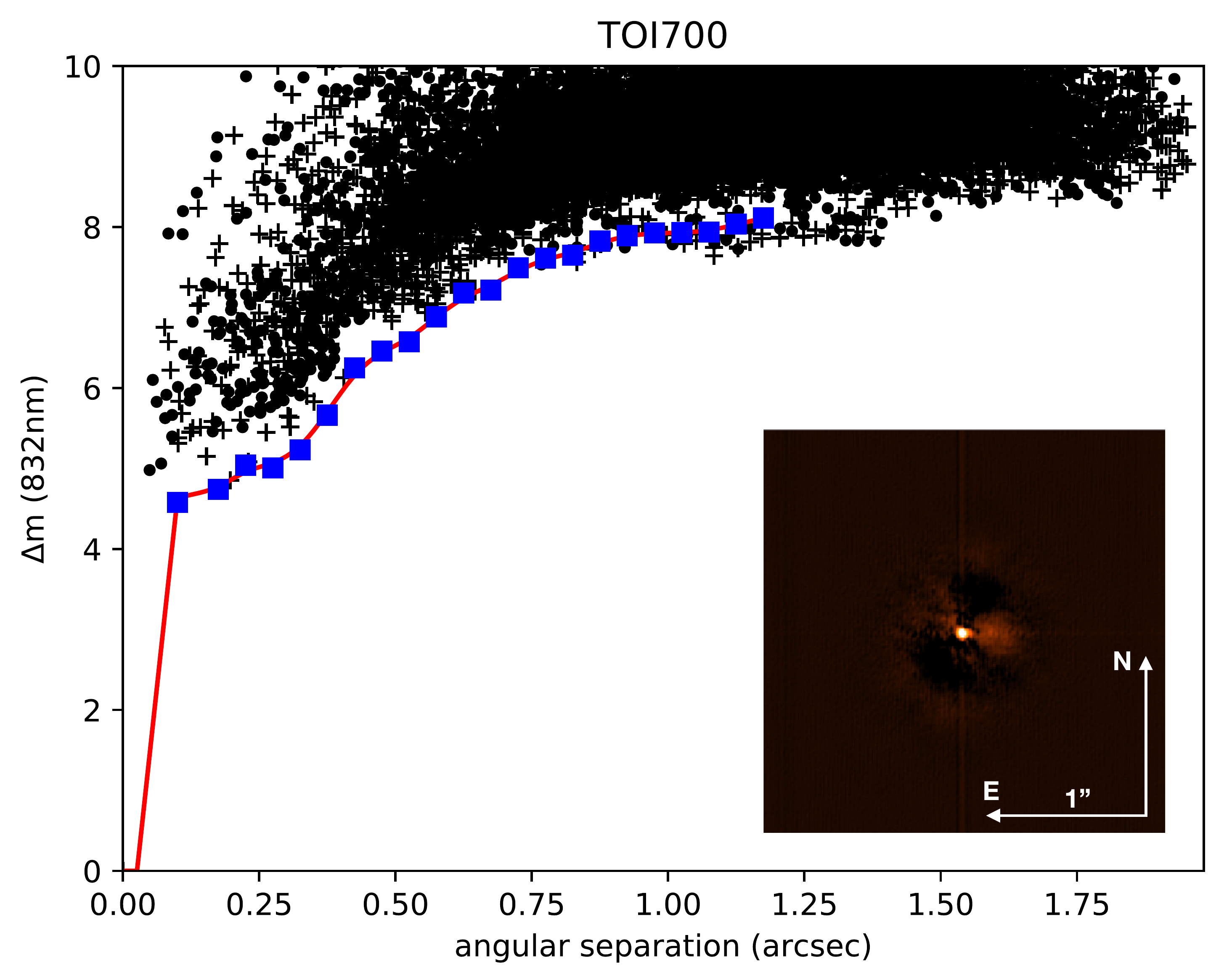}

  \caption{Gemini-South Zorro speckle observations of TOI-700 taken at 832\,nm and the corresponding contrast curve. Our simultaneous 562\,nm observation provides a similar result. The red line fit and blue points in the contrast curve represent the $5\sigma$ fit to the background sky level (black points) revealing that no companion star is detected from the diffraction limit (17\,mas) out to $1.75\arcsec$ within a $\Delta$\,mag of 5 to 8. The reconstructed speckle image (inset) has North up, East to the left, and is $2.5\arcsec$ across.}

  \label{fig:zspeckle}

\end{figure}

TOI-700 was observed on 2019 October 08 UT using the Zorro speckle instrument on Gemini-South. Zorro provides simultaneous speckle imaging in two bands (562\,nm and 832\,nm) with output data products including a reconstructed image and robust contrast limits on companion detections \citep{howell2011, howell2016}. The night had light cirrus, a slight breeze, and very good seeing ($\sim$0.4-0.5\arcsec) during the observations. Figure~\ref{fig:zspeckle} shows our 832\,nm contrast curve result and our reconstructed speckle image. We find that TOI-700 is a single star with no companion brighter than about 5 to 8 magnitudes, respectively, from the diffraction limit out to $1.75\arcsec$. We adopt the Zorro 832\,nm band as approximately equal to the $I$-band and estimate that for TOI-700 these limits correspond to an $I$$\sim$16 mag star at 0.53 AU and $I$$\sim$19 mag star at 54.4 AU.

\begin{figure}[htb]
\begin{center}
\includegraphics[width=0.47\textwidth]{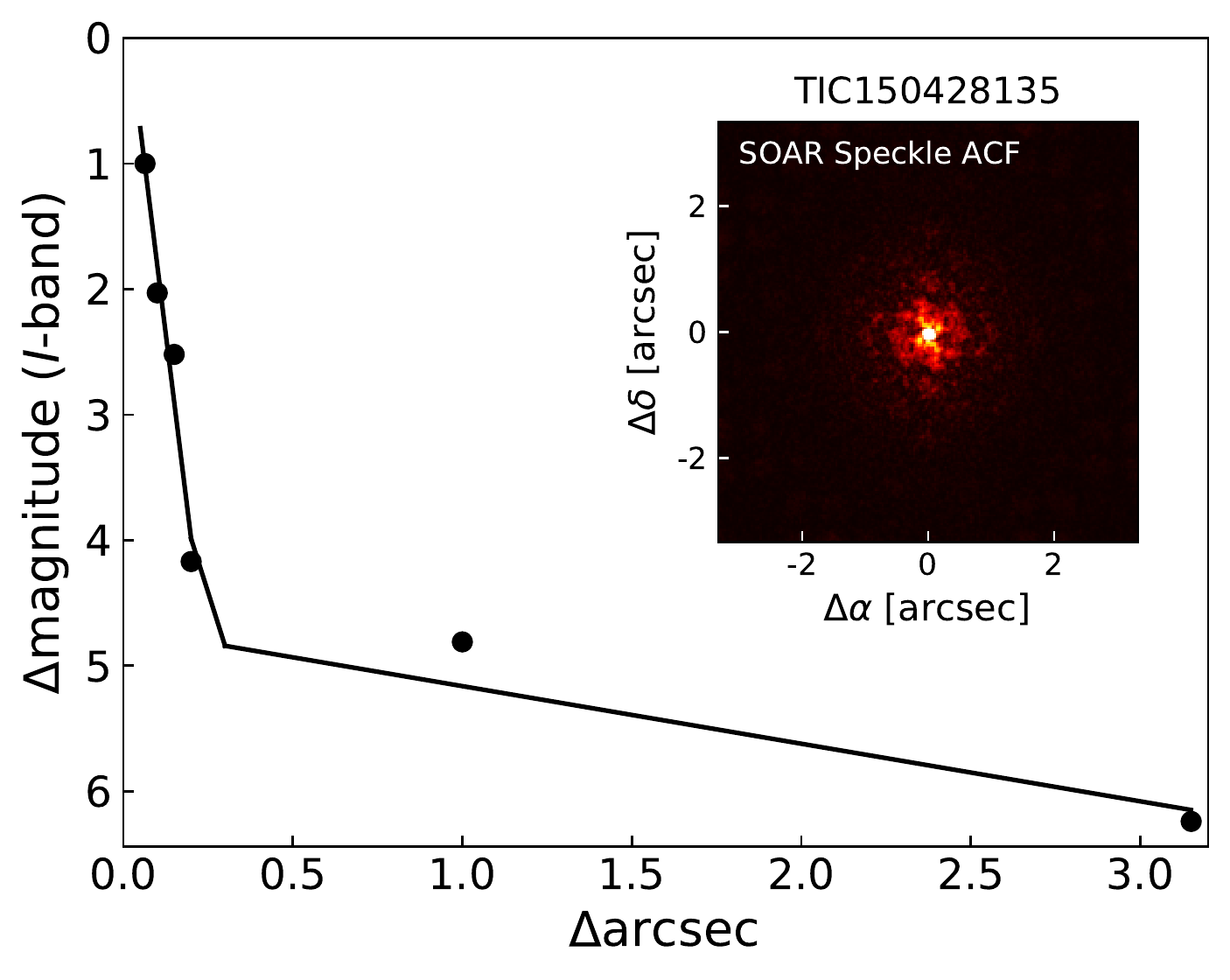}
\caption{SOAR HRCam $I$-band contrast curve and autocorrelation function (inset). The 2-dimensional autocorrelation function is indicative of a single star. The contrast curve shows that TOI-700 hosts no close companions brighter than $\Delta I$ $\approx$~5 mag at separations beyond 0.3$\farcs$}
\label{fig:goodman2}
\end{center}
\end{figure}

We also searched for previously unknown companions to TOI-700 with the SOAR speckle imaging camera \citep[HRCam, see][]{tokovinin18}. Data were taken on 2019 October 16 UT in $I$-band, a similar visible bandpass to \tess. We detected no nearby stars within 3$\arcsec$ (or 93 AU) of TOI-700. The 5$\sigma$ detection sensitivity and the speckle auto-correlation function from the SOAR observation are plotted in Figure \ref{fig:goodman2}.

We also checked for indications of binarity using the Renormalised Unit Weight Error (RUWE) which is calculated for each source in the \emph{Gaia} DR2 catalog. \citet{ziegler19} showed that this measure of fit quality was typically $<$1.4 for single stars. For TOI-700, the RUWE = 1.08, indicating it is comfortably in the single star regime and providing independent verification of the results from the speckle imaging observations.

\subsubsection{High-Resolution Spectroscopy}
\label{sec:spectra}
As part of our TFOP reconnaissance spectroscopy campaign to investigate the activity of the host star and rule out close companions unresolved by speckle imaging, we observed TOI-700 on 2019 October 01 UT using the CTIO high-resolution (CHIRON) spectrograph \citep{Tokovinin2013} in slicer mode on the Cerro Tololo Inter-American Observatory (CTIO) Small and Moderate Aperture Research Telescope System (SMARTS) 1.5 m telescope. CHIRON covers a wavelength range of 410--870 nm and has a resolving power R = 79,000. We obtained three 1200 second exposures, which were then median combined to yield a signal-to-noise ratio per spectral resolution element of roughly 28 at 711.59 nm. Using the TiO molecular bands at 706.5–716.5 nm and an observed template of Barnard’s Star, we calculate a radial velocity of -4.4 $\pm$ 0.1 km s$^{-1}$.\footnote{We note that the total uncertainty on the systemic velocity should include the 0.5 km s$^{-1}$ uncertainty on the Barnard's Star template velocity.} More details on the analysis are described in \citet{Winters(2018)}. We note negligible rotational broadening ($v \ sin \ i$ \textless \ 1.9 km s$^{-1}$) and do not see H$\alpha$ in emission, indicating that the star is inactive. Our analysis of the spectrum reveals no evidence of doubled lines that could originate from unresolved, very close-in, stellar companions. 

\begin{figure}
    \centering
    \includegraphics[width=0.45\textwidth]{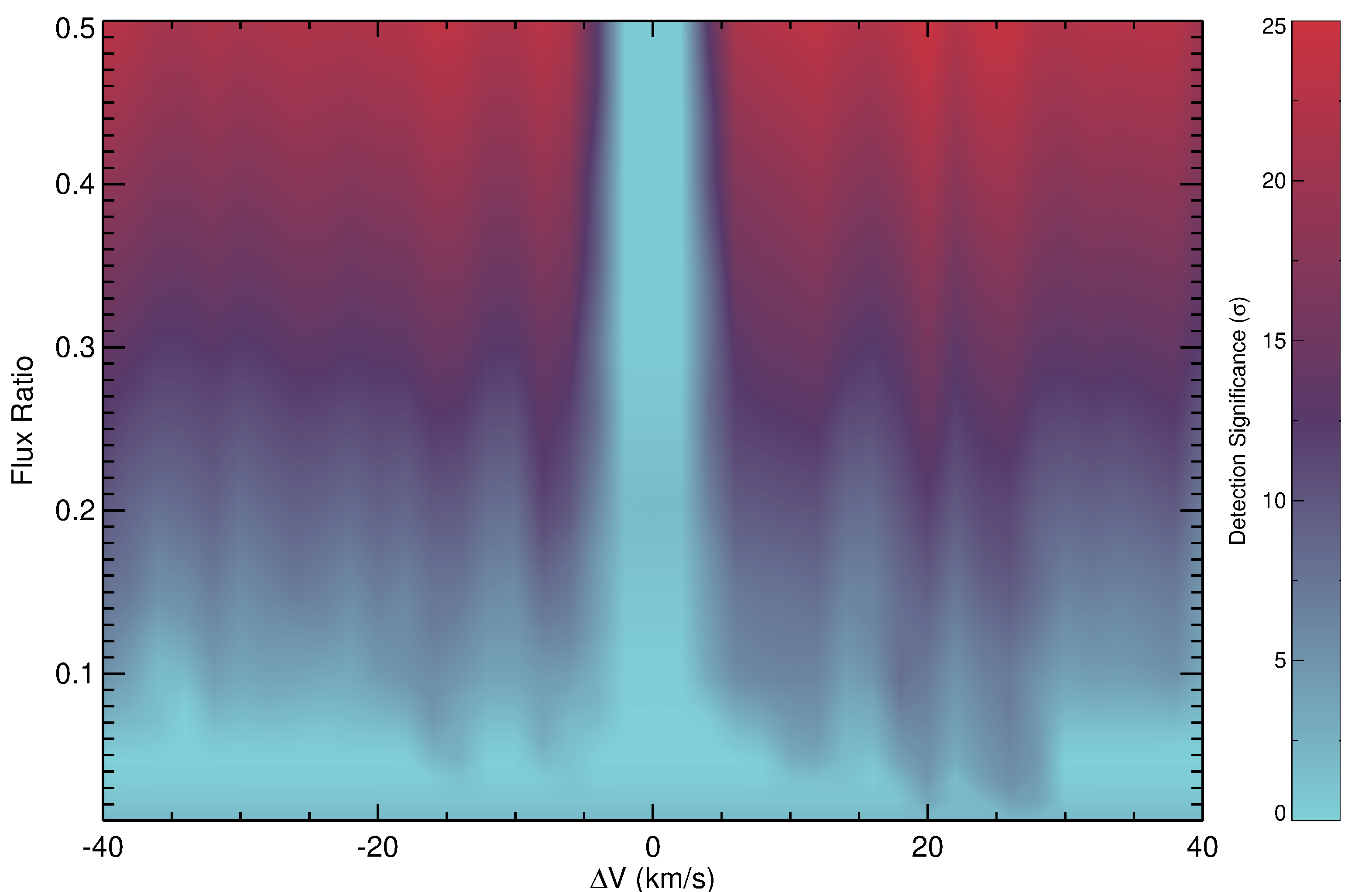}
    \caption{Detection limits for faint companions in the CHIRON spectrum demonstrating that we can rule out the presence of any companion with a flux ratio greater than ~10\% and an RV separation of $>$4 km/s.
    }
    \label{fig:chiron}
\end{figure}

We ran a series of injection and recovery tests to determine how sensitive we are to any remaining unresolved stellar companions. Under the assumption that any bound (M dwarf) companion will have a line profile similar to TOI-700---modulated only by its intensity and rotation---we used the observed least-squares deconvolution profile of TOI-700 as a template. We injected secondary least-squares deconvolution peaks representing companions with properties drawn from grids of flux ratios between 1$\%$ and 50$\%$, radial velocity separations between -100 and 100 km/s, and rotational velocities between 0 and 10 km/s. For each injection, we re-fit the central line profile with a Gaussian and removed it, and performed a search for a second peak in the residuals. We calculated the significance of the best-fitting Gaussian in the residuals, which we plot in Figure \ref{fig:chiron}. We adopt a 5$\sigma$ detection threshold due to the possible additional systematic uncertainty introduced by a mismatch between the line profiles of primary and secondary components. 

We conclude that for radial velocity separations $>$4 km/s, we can rule out all bound companions with flux ratios greater than about 10$\%$. Given the wavelength range of the CHIRON data used in these analyses, this corresponds to companions with $\Delta$ $R$~$\approx$~2.5 mag.
Components with velocity separations $<$4 km/s are blended with the primary peak and difficult to identify. Chance alignments of background stars with different spectral types can also be detected by this analysis, but may suffer from significant template mismatch, and the significance of their detection would therefore tend to be overestimated. For this reason, we limit our quantitative conclusions to hypothetical bound stellar companions of TOI-700.
    
We also placed TOI-700 on an observational Hertzsprung-Russell Diagram and compared it to the 1120 M dwarf primaries within 25 pc, as presented in \citet{Winters(2019a)}. The system is not elevated above the main sequence or among the blended photometry binary sequence, which provides confidence that there are no significantly luminous companions to TOI-700, consistent with the results of our high-resolution imaging and spectroscopy.    
    
\subsubsection{Time-Series Photometry}
\label{sec:photometry}

We conducted ground-based transit observations of the planet candidates associated with TOI-700 through TFOP. To schedule the observations we used the {\tt TESS Transit Finder}, which is a customized version of the {\tt Tapir} software package \citep{Jensen:2013}. These measurements aimed to independently re-detect the transits of the planet candidates to refine the planet and orbital parameters and rule out nearby eclipsing binary contaminants at the relevant periods. The ground based photometric light curves were extracted and analyzed using the {\tt AstroImageJ} ({\tt AIJ}) software package \citep{karen2017}. 

TOI-700~b was observed on 2019 December 02 UT at the Siding Spring Observatory (SSO) using both the Las Cumbres Observatory Global Telescope (LCOGT) network \citep{Brown:2013} 1.0-m telescope and the 0.43-m iTelescope T17.\footnote{\url{https://www.itelescope.net}, \url{https://support.itelescope.net/support/solutions/articles/231915-telescope-17}} The LCOGT time series was obtained in the $z_s$-band\footnote{The $z_s$ or \emph{z}-short filter is similar to a \emph{z}-band filter but with a cutoff at 920 nm.} using exposure times of 50 seconds spanning the event ingress and a partial transit. The images were calibrated by the standard LCOGT \texttt{BANZAI} pipeline. The iTelescope T17 photometric series was obtained using an FLI ProLine E2V CCD in the \emph{Clear} filter with exposure times of 120 seconds. We checked the field for nearby eclipsing binaries at the period of the planet candidate using custom {\tt AIJ} scripts. No transit was definitively detected in either time series, but these observations did allow us to rule out nearby eclipsing binaries (within $2.5\arcmin$) at the period of TOI-700~b.  

TOI-700~c was observed on 2019 November 01 UT at the South African Astronomical Observatory (SAAO) location of the LCOGT. Using the 1.0 m telescope in the $z_s$-band, observations spanning the full transit plus $\sim1$ hour on either side of the transit were obtained with 30 second exposures. We selected a optimal photometric aperture radius of 5.8$^{\prime\prime}$ and an optimal set of seven comparison stars to perform the differential photometry which minimized the 5-minute binned target star model residuals to 0.9 ppt. The planet transit was clearly detected with a transit depth consistent with the TESS data in apertures as small as 2.3$^{\prime\prime}$. The field was also cleared of nearby eclipsing binaries out to 2.5$^{\prime}$ and within $\pm$4$\sigma$ of the SPOC transit ephemeris. A figure showing the LCOGT transit detection and joint transit modeling that includes the \tess data and this ground based transit of TOI-700~c are presented in Paper~II of this series \citep{Rodriguez2020}.

We attempted additional ground based observations of TOI-700~c and TOI-700~d, but the data suffered due to weather and instrumental issues. The resulting light curves were used to independently clear the field of nearby eclipsing binaries at the period of TOI-700~c and to partially clear nearby eclipsing binaries at the period of TOI-700~d.

\subsection{Software Validation Analysis}
\label{sec:vetting}
The suite of follow-up observations presented in the previous sub-sections rule out substantial portions of parameter space where false-positives could exist and mimic planetary transit signals in the \target \ \tess \ data. However, the observational limits are incomplete and not all of the potential parameter space is excluded. Here we statistically analyze the remaining likelihood of false-positive signals. Specifically, we used the publicly available software package \texttt{vespa} \citep{vespa} to calculate the false-positive probabilities for the transit signals in the TOI-700 data. \texttt{vespa} compares transit signals to a number of false-positive scenarios including an unblended eclipsing binary (EB), a blended background EB, a hierarchical companion EB, and the ‘double-period’ EB scenario. Following the prescription described in \citet{schlieder2016}, we ran \texttt{vespa} using the \tess light curves to calculate the false-positive probability independently for each planetary signal. We included observational constraints in our analysis with the addition of the Zorro 832 nm contrast curve (see Section~\ref{sec:speckle}) as well as the radial velocity constraints derived from the CHIRON data (see Section~\ref{sec:spectra}). We also included a constraint on the maximum depth of potential secondary eclipses associated with each candidate. These constraints were estimated from our \texttt{DAVE} analysis. We ran \texttt{vespa} within 1 square degree of TOI-700, but dictate that the maximum aperture radius interior to which the signal must be produced is 21$^{\prime\prime}$, the size of a TESS pixel. Using these inputs, we calculated the false-positive probabilities to be 0.0012, 0.000086, and 0.0019 for planets b, c, and d, respectively. Given our extensive follow-up and the resulting constraints, the only false-positive scenario with any remaining probability was for the case of a background eclipsing binary, but the probability was $\ll$1$\%$ for each planet and is highly disfavored over the true planet scenario. 

With \texttt{vespa} strongly disfavoring astrophysical false positives we statistically validate the planetary nature of the transit signals. Moreover, \texttt{vespa} analysis does not account for any increase in our confidence in a planet scenario based on TOI-700 being a multiplanet system. If we assume that false positives are randomly distributed among stars, then a star with at least one transiting planet is more likely to have a second transiting planet than a false positive \citep{latham2011, Lissauer2012}. For \kepler, this `multiplicity boost' provided approximately a factor of 50 increase in the probability that a planet candidate was a true planet rather than a false positive \citep{Rowe2014,Lissauer2014}. For \tess, that number has been estimated to be 30--60 for small planets like those in the TOI-700 system (Guerrero et al. in preparation). With this in mind, the probability that any of the TOI-700 planet signals are the result of an astrophysical false positive is highly unlikely. 

However, we note that \texttt{vespa} does not take into account potential contamination from instrumental false alarms. \citet{Burke2019} used planet candidates and false positives from \kepler Data Release 25 \citep{Thompson2018} to estimate the instrumental false alarm rate as a function of multiple event statistics \citep[MES, ][]{Jenkins2002} for \kepler data. They recommended a typical threshold for long period planets of MES$>$9 to avoid false alarms. All three planets orbiting TOI-700 have MES statistics above 9. If TOI-700~d were a single \kepler planet, the \citet{Burke2019} estimate of false alarm probability would be 0.18\%, although given TOI-700~d is in a multiplanet system, the \citet{Burke2019} estimate falls to 0.013\% false alarm probability. For TOI-700~b and c these false alarm probability values are vanishing small ($\ll 0.1\%$).

While the instrumental false alarm rate for \tess has not been estimated, \tess detectors have fewer image artifacts than \kepler's \citep{Coughlin2014,Krishnamurthy2019,Vanderspek2018b}, albeit the pointing performance of \tess is less precise than \kepler's and there are background scattered light features in TESS data that were absent from \kepler. If we assume that the \tess instrumental false alarm rate is similar to that seen with the quieter detectors in the \kepler focal plane array, the false alarm rate for TOI-700~d falls to $\ll 1$\%. Therefore, under the assumption that the \tess false alarm rate is similar to or better than \kepler's, the TOI-700 planets are unlikely to be instrumental false alarms. However, this analysis does not independently confirm the planetary nature of the three planets around TOI-700 because \textit{confirmation} of these planets requires detection of a consistent signal with a facility other than \tess. TOI-700~d is a particularly high-interest planet given its size and insolation flux. It is likely to receive a significant amount of follow-up observations from a number of facilities. With this in mind, our group requested, and was awarded, \emph{Spitzer} 4.5 $\mu$m observations to independently confirm a transit of TOI-700~d. We describe these observations and a joint analysis of the \tess and other transit data for each planet in the system in Paper II in this series \citep{Rodriguez2020}.

\section{Gravitational Dynamics}
\label{sec:multiplanet}
Multiplanet systems provide a rich dataset that can reveal information that cannot be obtained from single planet systems. Lacking radial velocity measurements needed to obtain mass measurements, herein we use mass-radius relations to estimate the mass values in order to perform a dynamical stability analysis of the planetary system as shown in Section \ref{sec:stability}. We then present a photodynamics and transit timing variation (TTV) analysis in Section \ref{sec:photodyn} to determine whether we can place mass constraints from the photometry. Finally we conclude with a search for additional planets in the system in Section \ref{sec:planetsearch}. 

\subsection{Stability of the Planetary System}
\label{sec:stability}
Using the planet radii we reported in Table \ref{tab:TransitParameters}, we estimated mass values for each planet using \texttt{Forecaster} \citep{forecaster} to be $1.07^{+0.80}_{-0.43}$, $7.48^{+5.89}_{-3.30}$, and $1.72^{+1.29}_{-0.63}$ M$_{\oplus}$ for planets b, c, and d, respectively. We used these mass values to perform a suite of numerical integrations designed to investigate TOI-700's long term dynamical stability over 1 billion orbits of the outermost planet \citep[note that we choose such long integrations given the lengthy timescales for secular resonance overlaps to develop, see][]{lithwick11}. The \texttt{Forecaster} mass value for TOI-700~c is much higher than the value we constrain using a photodynamic model (see \ref{sec:photodyn} and \ref{tab:photoderivedParameters}), but we explore a range of masses that encompass both in this stability analysis.

\begin{figure}
    \centering
    \includegraphics[width=0.5\textwidth]{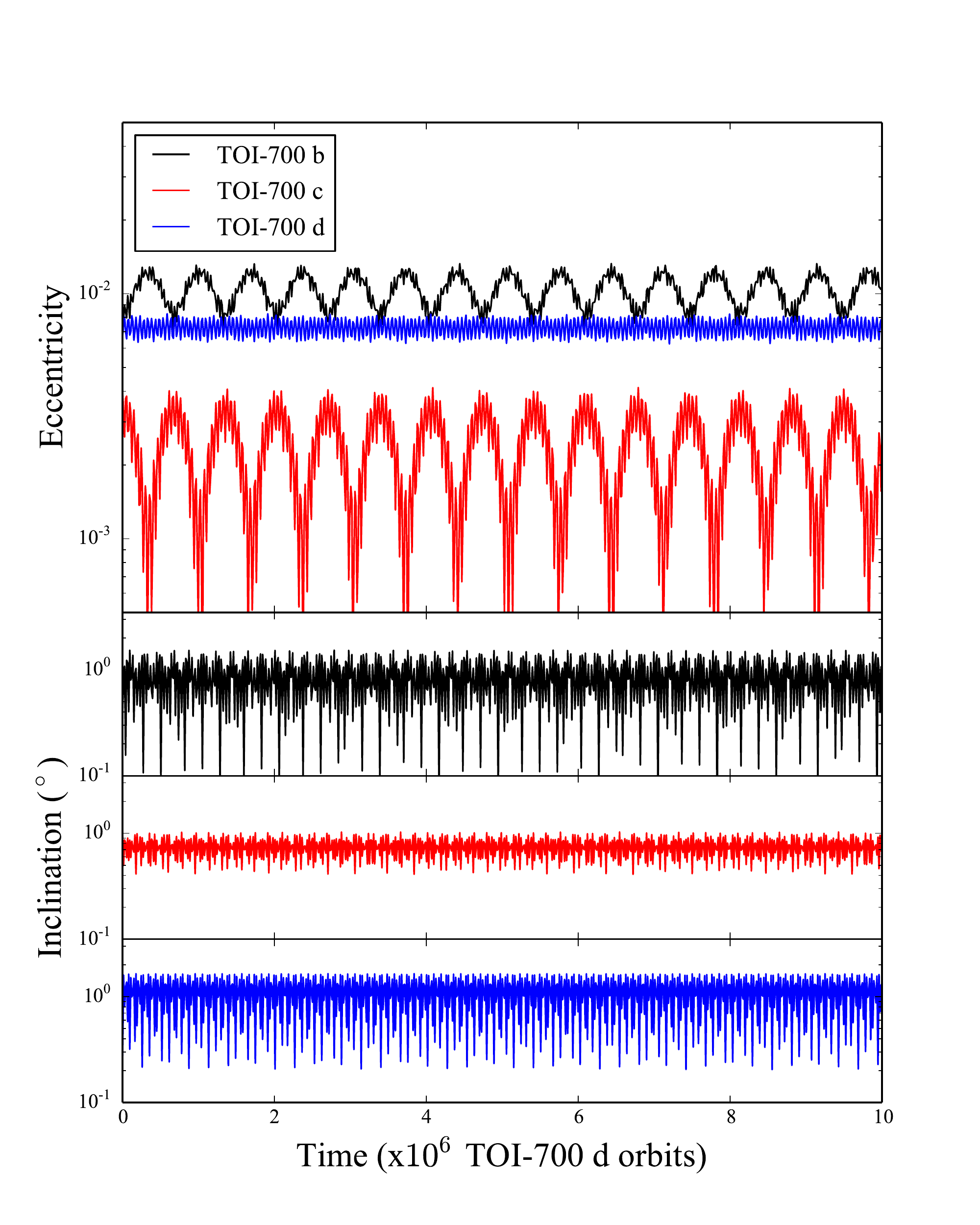}
    \caption{The dynamical evolution of the three planets in TOI-700 was simulated to explore long-term stability of the system. The variations in eccentricity (upper panel) and inclinations (lower panel) are shown here for one sample simulation, illustrating that the system is stable on long timescales.}
    \label{fig:stability}
\end{figure}

Our simulations use the \texttt{Mercury6} integrator \citep{chambers99} and a 10 hour time-step. We selected initial orbits for each planet using the determined nominal semi-major axes and inclinations, and assumed nearly circular initial eccentricities (e$<$0.001). To account for the substantial degeneracy in planet masses given the wide range of possible densities, each simulation varies the respective planets' masses such that the entire density range between 1.0 and 12.0 g cm$^{-3}$ is probed. Note that this range includes the lower density constraints for planet c that are discussed in the following section. In order to briefly investigate the possible existence of external, massive planets, we place an additional Neptune-mass planet at 1.0 AU, on a circular orbit, in half of our simulations. We find that, in each integration, eccentricity variations for all planets are smaller than 0.007 (Figure \ref{fig:stability}). While the moderate inclination of the second planet relative to the other two does drive secular inclination variations within the system (as large as $\sim$1.8$^{\circ}$ for the inner planet in some simulations), this behavior is regular and non-chaotic in all of our integrations. We also check each system for the presence of mean motion resonances and find the planets to be non-resonant within our tested parameter space. A more thorough investigation on the dynamics of the TOI-700 multiplanet system, such as probing the phase space of inclination, eccentricity, and mass of an outer companion \citep{Becker2017}, may provide additional constraints.

\subsection{Photodynamics and Transit Timing Variations}
\label{sec:photodyn}
The ratio of mean orbital periods of TOI-700~b and TOI-700~c (P$_c$/P$_b$=1.609) as observed by \tess is within 1\% of the 8:5 orbital resonance. While this is a weak resonance, this observation motivated a photodynamical analysis to attempt a measurement of the mass of the planets in the system. A photodynamical model can assess the potential for mass measurements from mutual gravitational perturbations of the planetary orbits by combining a transit model with an orbital integrator \citep[e.g.][]{Carter2012}. Gravitational interactions between planets will drive orbital eccentricity to larger values. Thus, constraints on the mean stellar density, $\rho_{\star}$, as derived in Section \ref{sec:stellar}, together with a photodynamical model can, at minimum, place upper limits on the planetary masses. 

Our photodynamical model used positions for each \tess observation calculated using the \texttt{Mercury6} hybrid integrator \citep{chambers99}. We then used these positions in the \texttt{TRANSITFIT5} transit modeling software \citep{Rowe2015,Rowe2016c} to calculate transit photometry of the planetary system. We parameterized the photodynamical model with four global parameters: mean stellar density, $\rho_{\star}$, quadratic limb-darkening, $q_1$, $q_2$ parameterized by \citet{exoplanet:kipping13}, and a factor to scale the photometric uncertainty reported for \tess photometry, d$_{\rm scale}$. For each planet we used seven parameters: the center of transit time, T$_0$, defined as when the projected separation between the star and planet as seen by the observer is minimized, the mean orbital period (P$_{\rm mean}$) as observed by \tess, the impact parameter, $b_{{\rm T}_0}$, observed at T$_0$, the scaled planetary radius, R$_{\rm p}$/R$_{\star}$, the scaled planetary mass, M$_{\rm p}$/M$_{\star}$ and orbital eccentricity parameterized by $\sqrt{e}$\,cos\,$\omega$ and $\sqrt{e}$\,sin\,$\omega$. 

We matched the photodynamical model to \tess photometry using an MCMC analysis. The MCMC routine used an affine-invariant ensemble sampler with 480 walkers \citep{foremanmackey12}. We initialized walkers to sample a wide range of orbital eccentricity and planetary mass to avoid clustering of walkers near a single local minimum. We required initial parameters to be dynamically stable for the duration of the \tess observations. Models were considered to be dynamically unstable if any planet pair came within 3 Hill radii. We adopted a prior on the mean stellar density of $\rho_{\star}$ = 8.0 $\pm$ 1.8 g/cm$^{3}$, as reported in Table \ref{tab:stellarparameters}. We also required masses, radii, and impact parameters to be positive. Orbital inclination is not well constrained by the dynamical portion of photodynamics and negative impact parameters were found to be completely degenerate with positive values in our model. A Markov-chain with a length of 7.68 million was generated. The final 1.68 million entries were examined using the Gelman-Rubin diagnostic \citep{Gelman1992} to assess convergence and adopted to calculate posterior distributions for each model parameter.

\begin{figure}[tp!]
    \centering

    \includegraphics[width=0.5\textwidth]{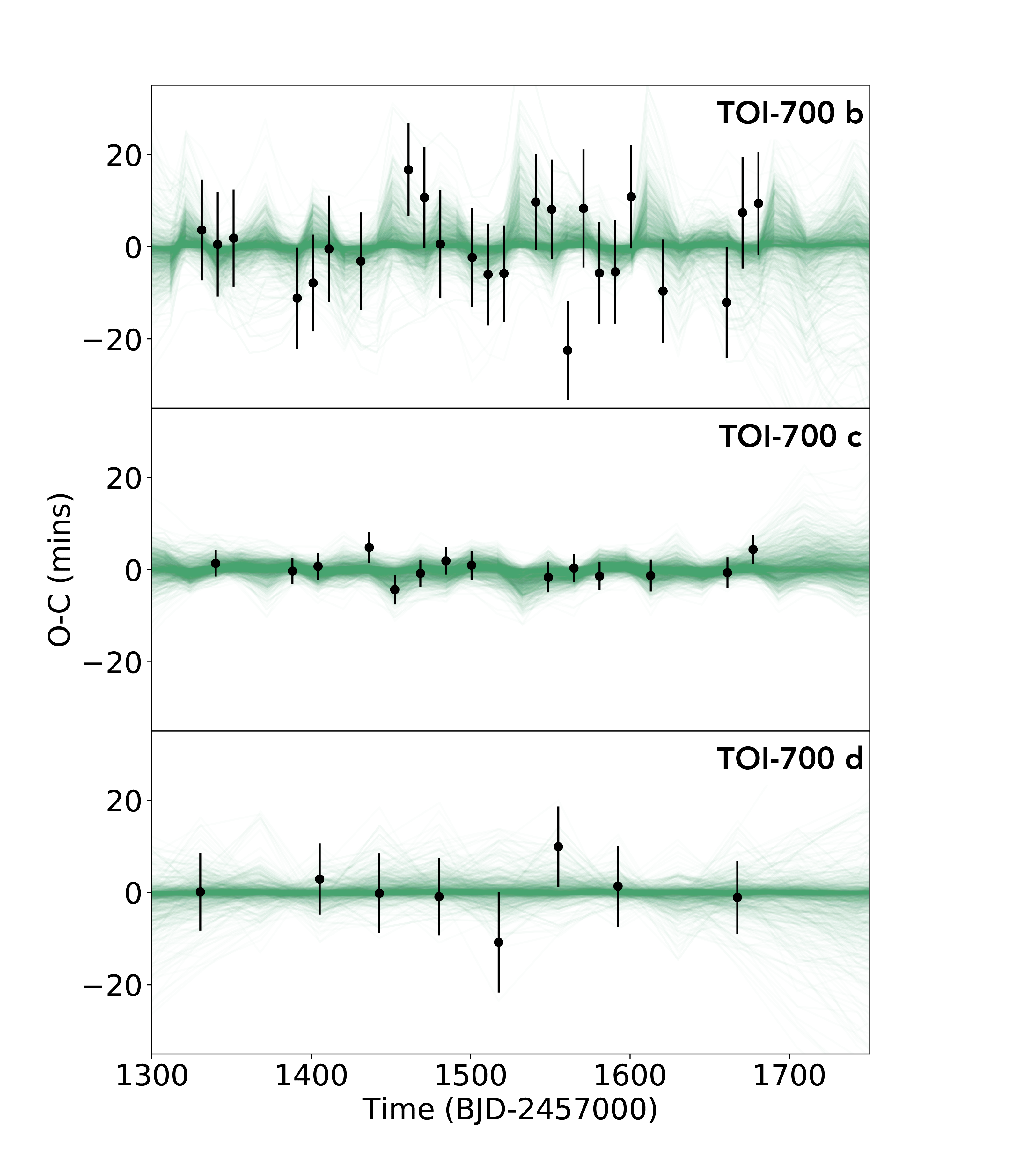}
    
\caption{The Observed minus Calculated (O-C) transit times for TOI-700~b, c, and d are presented; comparing measured transit times (black markers) and photodynamical models (green lines). For each \tess transit, the Observed transit time from photodynamics is compared to the Calculated transit time based on the modelled mean orbital period and displayed as green lines. There are 2000 green lines that present models randomly sampled from MCMC analysis of \tess photometry with our photodynamical model. Thus, the density of the lines indicated the probability of deviations from a strictly periodic orbit. The black markers are measured transit times based on a template analysis of \tess photometry and are presented to visualize the expected timing for each observed transit. The green lines are not fit to the black timing measurements but represent the range of TTVs allowed by \tess photometry which can be visually compared to timing measurements of each individual transit. 
\label{fig:TTVandPhotodyn}}
\end{figure}

Table \ref{tab:photoParameters} presents the adopted photodynamical model parameters based on our MCMC analysis and includes the mode and 68.27\% interval centered on the mode. The mode and interval for each parameter were calculated using a Kernel Density Estimator from \texttt{scipy} \citep{scipy}. In Figure \ref{fig:TTVandPhotodyn}, we compare the transit timing predictions from our photodynamical analysis (green lines) with TTVs measured using a best-fit transit model template from \texttt{TRANSITFIT5} (black dots with 1$\sigma$ uncertainty). The photodynamical model was not fit to the template extracted TTVs displayed in Figure \ref{fig:TTVandPhotodyn}, but was fit directly to \tess photometry.

Using stellar parameters reported in Table \ref{tab:stellarparameters} the posterior distribution in the planet mass (M$_{\rm p}$), planetary radius (R$_{\rm p}$) and planet density ($\rho_{\rm p}$) are provided in Table~\ref{tab:photoderivedParameters}. The results show that TTVs for TOI-700~b and c are allowed with potential changes in the orbital period of a few minutes per orbit and provide constraints on the mass and density of the planets. The density of TOI-700~c is fairly well constrained with a 1$\sigma$ upper limit of 1.9 g/cm$^3$. With this constraint, TOI-700~c could potentially have a significant H/He envelope with a density that is significantly lower than what would be expected for a rocky planet. This density limit may also allow a water world (although that would require an unexpectedly large water/rock ratio). The orbit of TOI-700~d was not found to be strongly perturbed by TOI-700~b or TOI-700~c in our analysis on the timescale of \tess observations. However, additional transit timing measurements of the TOI-700 system are needed to reach strong conclusions for planets b and c as the models diverge very quickly.

\begin{deluxetable}{l c c c}[htb]
\tablecaption{Photodynamic Model Parameters \label{tab:photoParameters}}
\tablewidth{0pt}
\tablehead{
\colhead{Parameter} & \colhead{Mode} & \colhead{+1$\sigma$}  & \colhead{-1$\sigma$}
}
\startdata
$\rho_{\star}$ (g cm$^{-3}$) & 8.1 & +1.9 & -1.0 \\
$q_1$ & 0.052  & +0.263 & -0.052 \\
$q_2$ & 0.122 & +0.478 & -0.114 \\
d$_{\rm scale}$ & 0.8841 & +0.0020 & -0.0011  \\
 & & & \\
{\bf TOI-700~b} & & & \\
T$_0$ (BJD - 2457000) & 1331.3568 & +0.0059 & -0.0053  \\
P$_{\rm mean}$ (days) & 9.97681 & +0.00033 & -0.00021 \\
$b_{{\rm T}_0}$ & 0.0586 & +0.234 & -0.047  \\
$R_{\rm p}/R_{\star}$ & 0.0227 & +0.0011 & -0.0011  \\
$M_{\rm p}/M_{\star} \times 10^6$ & 3.1 & +17.9 & -3.1 \\
$\sqrt{e}\,$cos$\,\omega$ & -0.03 & +0.18 & -0.19  \\
$\sqrt{e}\,$sin$\,\omega$ & -0.14 & +0.23 & -0.11  \\
 & & & \\
{\bf TOI-700~c} & & & \\
T$_0$ (BJD - 2457000) & 1340.0898 & +0.0020 &-0.0016 \\
P$_{\rm mean}$ (days) & 16.050989 & +0.000130 & -0.000083 \\
$b_{{\rm T}_0}$ & 0.920 & +0.030 & -0.035 \\
$R_{\rm p}/R_{\star}$ & 0.0575 & +0.0035 & -0.0022 \\
$M_{\rm p}/M_{\star} \times 10^6$ & 7.7 & +39.3 & -7.7 \\
$\sqrt{e}\,$cos$\,\omega$ & 0.131 & +0.099 & -0.232 \\
$\sqrt{e}\,$sin$\,\omega$ & 0.117 & +0.089 & -0.220 \\
 & & & \\
{\bf TOI-700~d} & & & \\
T$_0$ (BJD - 2457000) & 1330.4698 & +0.0072 & -0.0077 \\
P$_{\rm mean}$ (days) & 37.4260 & +0.0011 & -0.0014  \\
$b_{{\rm T}_0}$ & 0.53 & +0.12 & -0.28  \\
$R_{\rm p}/R_{\star}$ & 0.0277 & +0.0010 & -0.0023  \\
$M_{\rm p}/M_{\star} \times 10^6$  & 7.5 & +30.1 & -7.5  \\
$\sqrt{e}\,$cos$\,\omega$ & 0.217 & +0.078 & -0.388  \\
$\sqrt{e}\,$sin$\,\omega$ & 0.19 & +0.12 & -0.28  \\
\enddata
\end{deluxetable}

\begin{deluxetable}{l c c c}[htb]

\tablecaption{Photodynamic Derived Parameters \label{tab:photoderivedParameters}}
\tablewidth{0pt}
\tablehead{
\colhead{Parameter} & \colhead{Mode} & \colhead{+1$\sigma$}  & \colhead{-1$\sigma$}
}
\startdata
{\bf TOI-700~b} & & & \\
$R_{\rm p}$ ($R_{\oplus}$) & 1.041 & +0.088 & -0.097  \\
$M_{\rm p}$ ($M_{\oplus})$ & 0.42 & +2.5 & -0.42 \\
$\rho_{\rm p}$ (g cm$^{-3}$)  & 2.2 & +12.1 & -2.2  \\
 & & & \\
{\bf TOI-700~c} & & & \\
$R_{\rm p}$ ($R_{\oplus}$) & \ \  \ \ \ 2.66 \ \ \ \ \ & \  \ \ \ \ +0.26 \ \ \ \ \ & \ \ \ \ \ -0.24 \ \ \ \ \ \\
$M_{\rm p}$ ($M_{\oplus})$ & 1.1 & +5.4 & -1.1 \\
$\rho_{\rm p}$ (g cm$^{-3}$)  & 0.3 & +1.6 & -0.3  \\
 & & & \\
{\bf TOI-700~d} & & & \\
$R_{\rm p}$ ($R_{\oplus}$) & 1.22 & +0.14 & -0.10  \\
$M_{\rm p}$ ($M_{\oplus}$) & 1.0 & +4.1 & -1.0 \\
$\rho_{\rm p}$ (g cm$^{-3}$)  & 3.1 & +13.1 & -3.1  \\
\enddata
\end{deluxetable}

\subsection{Search for Additional Planets}
\label{sec:planetsearch}
To complement and reinforce the SPOC pipeline planet detections, we ran our own independent planet search on the light curve. Using \texttt{QATS} \citep[Quasi-periodic Automated Transit Search,][]{kruse19}, we recovered the three planet candidates but found no evidence for further transiting planets in the system; the \texttt{QATS} search also allowed for planets exhibiting TTVs, but no additional candidates hidden by strong TTVs were found.

\section{Discussion}
\label{sec:discussion}
TOI-700 is an exciting three-planet system orbiting a nearby M dwarf star. In this section we aim to put TOI-700 into context with other planetary systems, and consider the value of this system for habitability and atmospheric studies and the prospects for future follow-up characterization. 

\subsection{Comparison to other Multiplanet Systems}
\label{sec:comparison}

\begin{figure}
    \centering
    \includegraphics[width=0.5\textwidth]{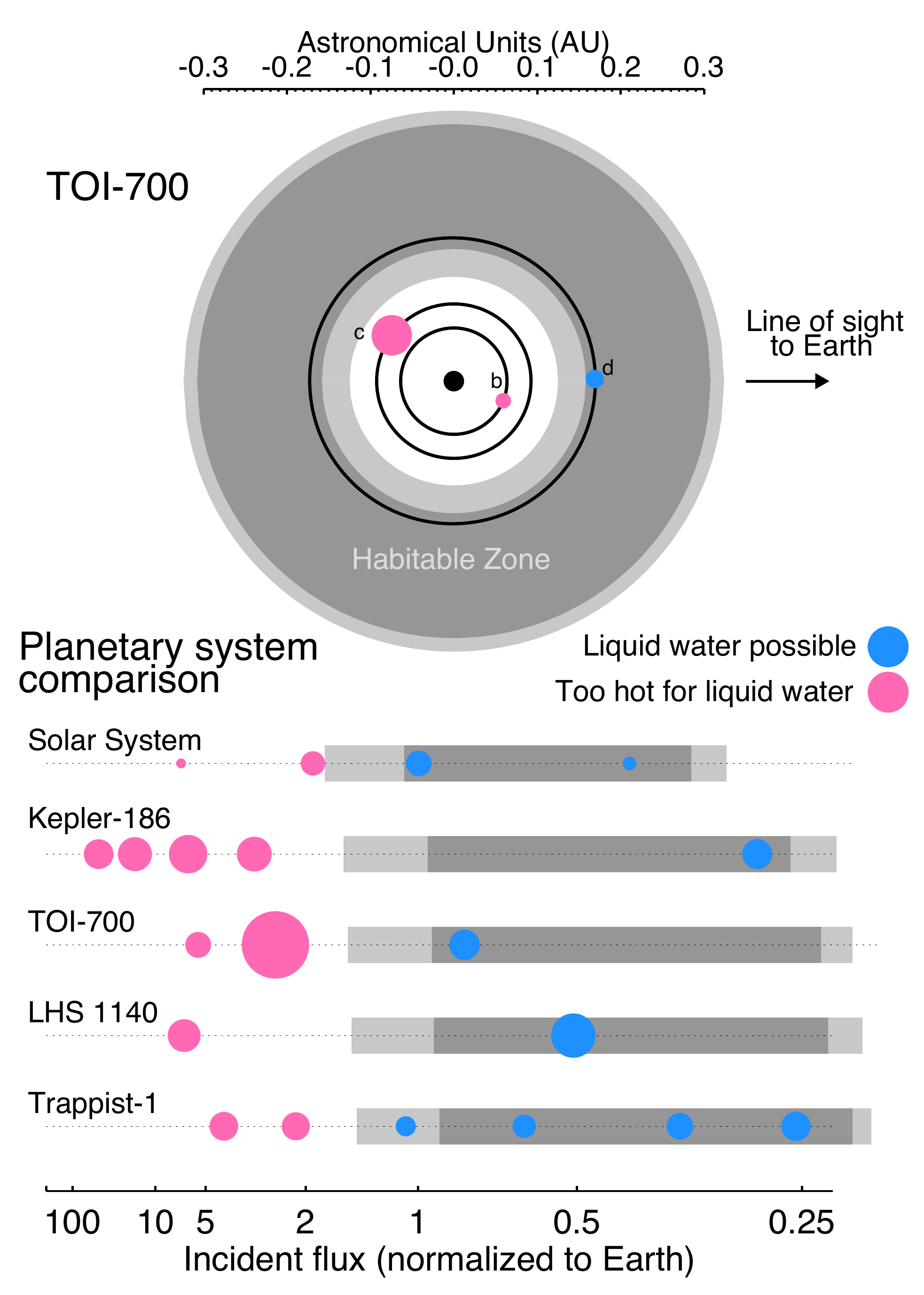}
    \caption{A top-down view of the orbits of the TOI-700 planets (upper panel). The relative sizes of the planets are to scale, but are not on the same scale as the orbits. The conservative habitable zone is shown in dark gray, and the optimistic habitable zone in light gray \citep{kopparapu2013}. We also compare the TOI-700 system to the Solar System and other benchmark exoplanet systems with low-mass host stars and small habitable-zone planets (lower panel).}
    \label{fig:habitability}
\end{figure}

The TOI-700 planetary system consists of three planets, with two approximately Earth-sized planets and a larger planet (2.6 times the size of Earth) orbiting in-between. This architecture is unusual compared to other multiplanet systems with small habitable zone planets (Figure \ref{fig:habitability}). Studies of the \kepler multiplanet population have found that planets within a given multiplanet system tend to have similar sizes, regular orbital spacings, and circular and coplanar orbits (if measureable) \citep{Millholland2017,Weiss2018}. The TOI-700 system architecture breaks this trend.

Planetary embryos which grow by accreting planetesimals tend to end up at similar sizes \citep{Lissauer1987,Kokubo1998}.  This is also true for pebble accretion \citep{Lambrechts2014,Ormel2017}.  While one might expect gas accretion to proceed at a similar rate for neighboring planets \citep{Ikoma2001,Millholland2017}, small differences in the planets' formation times or the local gas opacity could easily change this.  

What formation scenarios might explain the origin of a system like TOI-700 containing a low-density planet bracketed on either side by higher-density planets with similar masses?  Perhaps the two inner planets formed faster and accreted significant gaseous envelopes but the outer planet formed more slowly and accreted less gas.  Photo-evaporation is extremely sensitive to the orbital separation \citep{Lopez2013}, so the inner planet may have lost its envelope later.  Alternately, long-range orbital migration causes large diversity in planetary feeding zones, and therefore, compositions \citep{Raymond2018}.  One could imagine that planet c migrated inward from the outer parts of the disk and thus formed under different conditions (and perhaps faster) than planets b and d. Given that the masses of the planets are not tightly constrained (see Table \ref{tab:photoderivedParameters}), this second scenario would become more plausible if future studies indicate that the mass of planet c is significantly larger than that of planets b and d. It could be that planet c has more rock and was thus able to accrete and retain a much larger atmosphere.

The sizes of the planets orbiting TOI-700 span the observed gap in the transiting planet radius distribution \citep{Fulton2017,vaneylen2018,Cloutier2019b}. The inner and outer planets are likely to be rocky, whereas the middle planet likely has a gaseous envelope and is more akin to Neptune \citep{Rogers2015,Lopez2014}. This system is therefore a great laboratory to explore the formation mechanisms of compact multiplanet systems and for future atmospheric studies.

\subsection{Atmospheric Stability}
\label{sec:atmstability}

One of the key questions for the exoplanet community is, ``under what conditions are rocky exoplanets able to retain an atmosphere?" Recent observations of thermal emission from the rocky exoplanet LHS 3844~b indicate that it is likely airless \citep{Kreidberg2019}. Furthermore, a large and growing body of literature indicates that most of the rocky exoplanets found by \kepler have likely been heavily sculpted by extreme atmospheric escape \citep[e.g.,][]{Lopez2012, Owen2013,Owen2017, Lopez2017, Zahnle2017, McDonald2019, Neil2019}. This is a particular concern for planets around M dwarfs, where the host stars' long pre-main sequence lifetimes and frequently high activity levels mean that even rocky planets with heavier secondary atmospheres in or near the habitable zone are highly vulnerable to extreme atmospheric escape driven by space weather in the form of ionizing radiation (x-ray and Extreme UV [XUV], 1-1240 \AA ) and stellar wind particles \citep[e.g.,][]{Lissauer2007,Lammer2007,Cohen2014, Owen2016, Airapetian2017,Bolmont2017, Garcia-Sage2017, Dong2017, Garraffo2017,  Cohen2018, Airapetian2020}.

In this context, the TOI-700 system presents an exceptional opportunity since it contains three planets well-suited to detailed characterization around a bright, nearby M dwarf with low levels of stellar activity. As discussed in Section \ref{sec:age}, over the 11 sectors observed with \tess we do not observe a single white-light flare, and its slow rotation rate of 54 days places it firmly into the low-activity sample of M dwarfs identified by \citet{Newton2017}. Stars with rotation rates this slow are observed to have low x-ray luminosities with L$_{\mathrm{X}}$/L$_{\mathrm{bol}}$ $\approx 10^{-5}$ \citep[e.g.][]{Kiraga2007}, whereas more active M dwarfs like Proxima Centauri and TRAPPIST-1 have L$_{\mathrm{X}}$/L$_{\mathrm{bol}}$ in the range of $\approx 2\times10^{-4}-10^{-3}$ \citep{Wheatley2017}. 

TOI-700 has been observed in the soft x-ray band (0.1-2.5 keV) by the NICER mission, but there was no detection of x-ray emission. This suggests the upper bound for the star’s x-ray luminosity \textbf{at $L_x < 10^{27}$ erg,} which is comparable to the x-ray luminosity of the Sun at solar maximum \citep{Aschwanden1994,Peres2000}. This lower x-ray luminosity is critically important to atmospheric survival as it also strongly correlates with other key drivers of atmospheric escape including EUV irradiation and stellar wind particle flux \citep[e.g.,][]{Lammer2003,Owen2012, Khodachenko2007, Cohen2015, Airapetian2017, Dong2018}.

TOI-700~d is of particular interest as a likely rocky planet in the habitable zone. An empirical relationship between EUV and x-ray fluxes for G, K, and early M dwarfs \citep{Sanz-Forcada2011} implies a total XUV incident flux at TOI-700~d of approximately 65 ergs s$^{-1}$ cm$^{-2}$, approximately 35 times greater than the XUV flux at present-day Earth and 50 times lower than that received by TRAPPIST-1~e \citep{Wheatley2017}. 

Much work still needs to be done to understand the processes that drive atmospheric escape from rocky exoplanets. However, to get an initial idea we used an escape rate scaling law for an Earth-like planet, to estimate the possible rate of $\text O^+$ and $\text N^+$ ion escape \citep{Airapetian2017}. Assuming Earth’s surface gravity, atmospheric composition, and magnetic moment, along with quiescent conditions from the host star, with no observed flares and associated coronal mass ejections, gives a total ion mass loss rate of $1\times10^5$ g/s. At this escape rate, a planet with a 1 bar Earth-like atmosphere would survive for longer than $\gtrsim$1 Gyr even if there was no atmospheric replenishment due to volcanic activity. Assuming that the XUV emission at the early phase of the stellar evolution was about 10 times higher, the corresponding escape rate would be comparable to the outgassing rate of $1\times10^6$ g/s via volcanic activity on the early Earth-like planet \citep[e.g][]{Claire2008,Schaefer2007}, suggesting that this planet may have been able to retain an Earth-like secondary atmosphere. Recent studies of the interaction of stellar wind with TOI-700~d also suggest that the planet should retain a thick atmosphere over a few billion tears \citep{cohen2020, dong20}. Along with other recent discoveries of potentially rocky transiting planets like those in the TRAPPIST-1 system, we believe that TOI-700 presents a valuable opportunity to compare the atmospheres of rocky planets in the habitable zone over a wide range of conditions affecting atmospheric escape.

\subsection{Prospects for Follow-up}
\label{sec:future}
Prior to the launch of \kepler, it was unknown whether Earth-sized planets in the habitable zones of other stars existed. Particularly for M dwarfs, the Galaxy's most common type of star, this question has been of great interest due to the implications for the abundance of habitable planets in our galaxy. Of the more than 4,000 exoplanets discovered to date, only about a dozen are Earth-sized and reside in their stars' habitable zones. However, we now know that Earth-sized, habitable zone planets orbit stars that span the full range of M dwarf masses: from the ultra-cool M8 dwarf TRAPPIST-1 (0.08 \msun), to M3 dwarf K2-72 (0.3 \msun), to M0 dwarf Kepler-186 (0.5 \msun). We can now add the M2 dwarf, TOI-700 (0.42 \msun), to this growing list.

For detecting and characterizing planetary atmospheres, TRAPPIST-1 is a prime target since the planet-to-star size ratio is extremely high due to the diminutive size of the star (approximately the size of the planet Jupiter). TRAPPIST-1 also resides at 12 pc and has a $K$-band magnitude of 10.3. TOI-700 also has the small star advantage, but another advantage over \kepler and \emph{K2} targets is the star's proximity to observers (31 pc, versus 70 and 179 pc for K2-72 and Kepler-186, respectively), and its $K$ magnitude of 8.6. The TRAPPIST-1 and TOI-700 systems provide an opportunity to compare planets within the same system which formed in the same stellar environment to those that formed in very different M dwarf stellar environments. While TRAPPIST-1 and TOI-700 are both M dwarfs, the difference in mass between the two is more than a factor of four, whereas the masses of TOI-700 and the Sun differ by less than a factor of three. Moreover, TOI-700 is relatively old and quiet, whereas TRAPPIST-1 is fairly active \citep{Vida2017}, providing the opportunity to explore how activity affects atmospheric escape.

Following the methods of \citet{Kempton2018}, we took an initial look at the potential for future atmospheric follow-up with \jwst by calculating the transmission spectroscopy metric (TSM) of each planet. The TSMs for planets TOI-700~b, TOI-700~c, and TOI-700~d are 5.40, 73.64, and 3.49, respectively. While a TSM of 3.49 is the highest of any habitable zone planet smaller than 1.5 \rearth outside of the TRAPPIST-1 system, it is still relatively low. 

Achieving a $\sim$5$\sigma$ detection of biosignatures or other molecules in the atmosphere of TOI-700 d would likely require over 100 transits using JWST (See Paper II in the series, \citet{Rodriguez2020}. Paper III in this series, \citet{Suissa2020}, provides detailed modeling of plausible atmospheres of TOI-700 d and the resulting detectability using future observing facilities.

TOI-700~c, on the other hand, is a sub-Neptune-sized planet around a bright M-dwarf with a high TSM value, making it an excellent candidate for further investigation. A TSM of 74 is amongst the highest of planets in the `Venus Zone' \citep{Kane2014}, and may provide an excellent opportunity to characterize this sub-Neptune with the Hubble Space Telescope and JWST.

\subsubsection{Radial Velocity Follow-up}
For radial velocity observations, we estimated the signals needed to constrain the masses of the TOI-700 planets. The three planets in the system, from inner to outer, TOI-700~b, c, and d, have expected Doppler semi-amplitudes of 0.57, 3.4, and 0.59 m/s, respectively, with uncertainties around 20\% (using the \texttt{Forecaster} mass-radius relation). 
While the velocity semi-amplitude of planet c is well within the capabilities of current Southern Hemisphere instruments such as HARPS and PFS \citep{mayor03,teske16}, the orbital period of TOI-700 c of 16.05 days is close to one-third of the $\sim$54-day stellar-rotation period. The rotational modulation of stellar activity introduces apparent velocity changes of a few m/s for quiet, main sequence dwarfs. The strongest of these changes occur at time-scales equal to one-third, one-half and one times the stellar rotation period for intensely sampled cadences \citep{vanderburg16}, and also at other spurious periods both longer and shorter than the rotation period that can persist for multiple observing seasons for less well-sampled cadences \citep{nava19}. This will confound the interpretation of the radial velocity signal for all of the TOI-700 planets without novel methods for mitigating stellar activity in radial velocities such as recently probed with line-by-line analysis and chromatic radial velocities \citep{Cretignier19, Dumusque18, lanza19, talor18}. 

Planets b and d will be challenging because of the relatively low expected amplitudes (under 1 m/s), and will require excellent instrument stability. ESPRESSO is currently the only Southern Hemisphere facility with demonstrated instrument single measurement precision of less than 0.5 m/s on sky that can access TOI-700 \citep{pepe14, faria19}. Recent work shows promising ESPRESSO stability in the mass measurement of Proxima Centauri b with a typical photon noise-limited radial velocity semiamplitude precision of 27 cm s$^{-1}$ \citep{proxcen_espresso}. Moreover, they did not find that stellar jitter noise was detectable above the photon noise limit of the observations. TOI-700 provides an excellent benchmark case for ESPRESSO to explore the limits of techniques for stellar activity correction in radial velocity spectra time-series for early M dwarfs with multiplanet systems.

While planet c is well within the capabilities of current instruments, planet b, and particularly planet d, will be challenging because of the length of the orbital periods and low expected radial velocity amplitudes. Mass measurements of these two planets will require excellent instrument stability.

\begin{figure}
    \centering
    \includegraphics[width=0.5\textwidth]{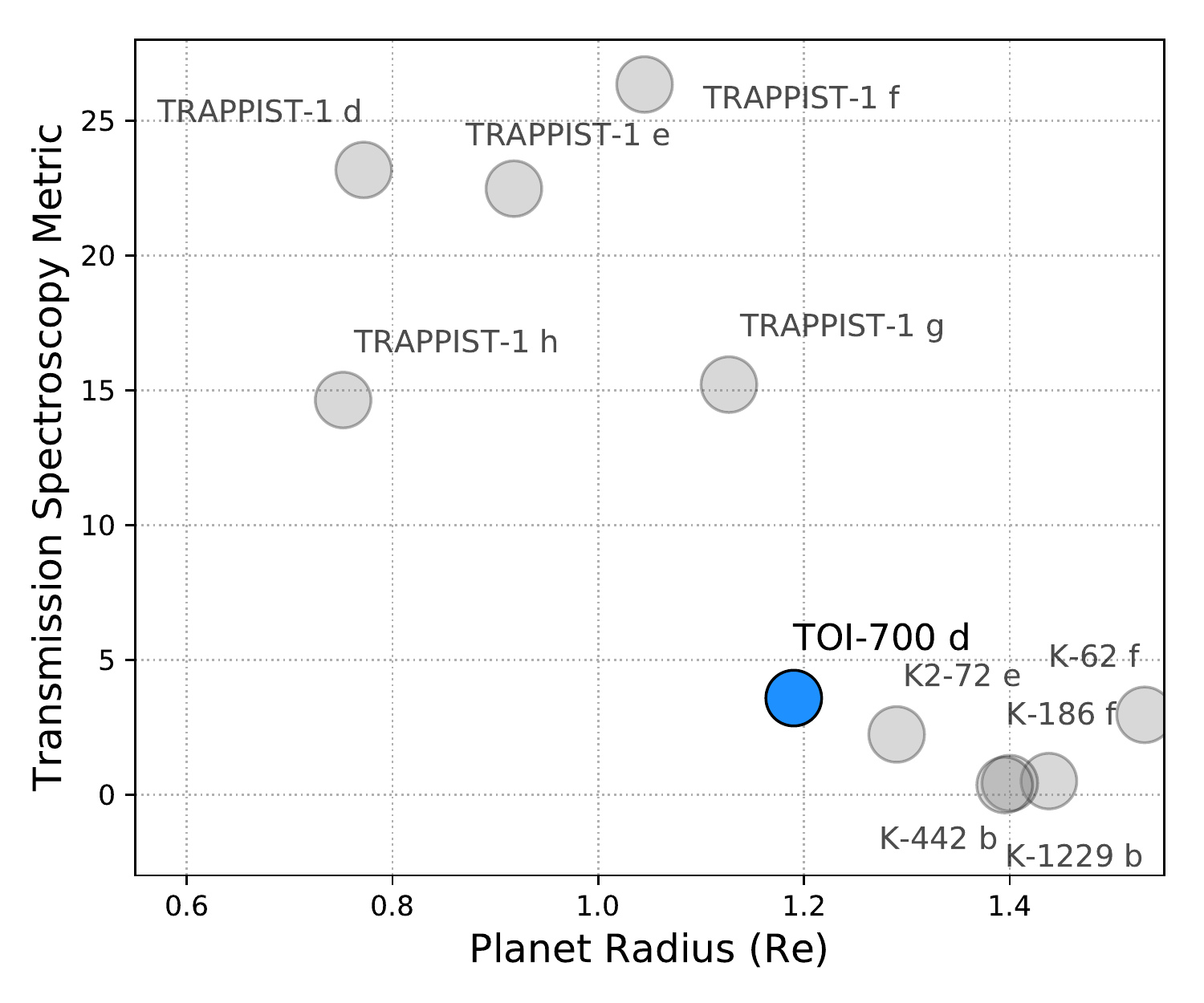}
    \caption{There are now 11 known exoplanets that have radii less than 1.5\rearth and orbit within their star's optimistic habitable zone \citep{kopparapu2013}. Plotted are these planets' TSM values. The top candidates for atmospheric characterization orbit TRAPPIST-1. Beyond these, TOI-700~d has the highest TSM, although characterizing this planet will be challenging.}
    \label{fig:tsm}
\end{figure}

\subsubsection{Additional Photometry from \tess's Extended Mission}
The \tess extended mission is scheduled to begin July 4, 2020. \tess will return to the Southern Hemisphere where it will re-observe TOI-700 for 11 of the 13 sectors in \tess Guest Investigator Program Cycle 3. The full-frame image data will be collected at 10-minute cadence in the extended mission, and targets can be proposed for both 2-minute cadence observations and a new 20-second cadence mode. Additional photometry, combined with the data presented herein, will allow for better constraints on planet parameters, enable searches for additional planets, and collect more transit time measurements to improve our TTV analysis.

\section{Conclusions}
\label{sec:end}
We present the discovery and validation of three small planets (R$_p$ = 1.01, 2.63, 1.19 R$_\oplus$) orbiting TOI-700, a bright, nearby (distance = 31.1 pc) M2 dwarf (0.416 \msun, 0.42 \rsun, with a temperature of 3480 K). The outermost planet, TOI-700~d, is approximately Earth-sized and resides in the star's habitable zone.

After initial vetting and extensive ground-based follow-up observations, we found no evidence of binarity or contamination of the light from the host star. We then validated the system using the \texttt{vespa} software package, and showed that the signals in \tess data are planetary in nature and highly unlikely to be false positives. 

TOI-700~d affords us the exciting opportunity to study an Earth-sized, habitable zone planet. TOI-700~c is also an excellent target for detailed follow-up. The sizes of the planets in the system span the observed gap in the transiting planet radius distribution, therefore, this system is an intriguing target for studies of planet formation and comparative planetology. TOI-700 is a quiet star, with no detectable flares in the optical \tess data, making it an optimal target for habitability studies of planets orbiting M dwarfs.

\tess will return to the Southern Hemisphere observe TOI-700 for an additional 11 sectors in \tess's extended mission, which is scheduled to begin in July 2020. This will enable studies for additional evidence of transit timing variations, place further constraints on planet parameters, and searches for additional planets in the system.

\acknowledgments
This paper includes data collected by the \tess mission, which are publicly available from the Mikulski Archive for Space Telescopes (MAST). Funding for the \tess mission is provided by NASA's Science Mission directorate. We acknowledge the use of public \tess Alert data from pipelines at the \tess Science Office and at the \tess Science Processing Operations Center. 

This research has made use of the Exoplanet Follow-up Observation Program website, which is operated by the California Institute of Technology, under contract with the National Aeronautics and Space Administration under the Exoplanet Exploration Program. 

This work has made use of data from the European Space Agency (ESA) mission {\it Gaia} (\url{https://www.cosmos.esa.int/gaia}), processed by the {\it Gaia} Data Processing and Analysis Consortium (DPAC, \url{https://www.cosmos.esa.int/web/gaia/dpac/consortium}). Funding for the DPAC has been provided by national institutions, in particular the institutions participating in the {\it Gaia} Multilateral Agreement.

Some of the observations in the paper made use of the High-Resolution Imaging instrument Zorro at Gemini-South. Zorro was funded by the NASA Exoplanet Exploration Program and built at the NASA Ames Research Center by Steve B. Howell, Nic Scott, Elliott P. Horch, and Emmett Quigley.

Resources supporting this work were provided by the NASA High-End Computing (HEC) Program through the NASA Advanced Supercomputing (NAS) Division at Ames Research Center for the production of the SPOC data products.

This work makes use of observations from the LCOGT network.

E.A.G. thanks the LSSTC Data Science Fellowship Program, which is funded by LSSTC, NSF Cybertraining Grant \#1829740, the Brinson Foundation, and the Moore Foundation; her participation in the program has benefited this work. E.A.G. and E.V.Q. are thankful for support from GSFC Sellers Exoplanet Environments Collaboration (SEEC), which is funded by the NASA Planetary Science Division’s Internal Scientist Funding Model.

J.F.R. acknowledges research funding support from the Canada Research Chairs program and NSERC Discovery Program. This research was enabled, in part, by support provided by Calcul Québec (\url{www.calculquebec.ca}) and ComputeCanada (\url{www.computecanada.ca})

A.V.'s work was performed under contract with the California Institute of Technology (Caltech)/Jet Propulsion Laboratory (JPL) funded by NASA through the Sagan Fellowship Program executed by the NASA Exoplanet Science Institute.

R.C. is supported by a NASA grant in support of the TESS science mission.

C.D.D acknowledges support from the NASA TESS Guest Investigator Program through Grant 80NSSC18K1583.

B.J.S. is supported by NASA grant 80NSSC19K1717 and NSF grants AST-1908952,
AST-1920392, and AST-1911074.

B.R-A acknowledges the funding support from FONDECYT through grant 11181295.

J.G.W. is supported by a grant from the John Templeton Foundation. The opinions expressed in this publication are those of the authors and do not necessarily reflect the views of the John Templeton Foundation.

T.D. acknowledges support from MIT’s Kavli Institute as a Kavli postdoctoral fellow.

V.S.A. was supported by Sellers Exoplanetary Environments Collaboration (SEEC) Internal Scientist Funding Model (ISFM) at NASA GSFC and NICER Cycle 1 GO program.

\facilities{ASAS-SN,
CTIO:0.9m (2048x2046 Tek2K CCD),
CTIO:1.5m (CHIRON),
Exoplanet Archive,
\emph{Gaia},
Gemini:South (Zorro),
MAST,
LCOGT,
SOAR (Goodman Spectrograph, HRcam),
TESS,
WISE}

\software{
AstroImageJ \citep{karen2017}, 
astropy \citep{exoplanet:astropy13,exoplanet:astropy18}, 
celerite \citep{exoplanet:foremanmackey17,exoplanet:foremanmackey18}, 
emcee \citep{foremanmackey12}, 
exoplanet \citep{exoplanet:exoplanet}, 
DAVE \citep{Kostov2019}, 
Forecaster \citep{forecaster}, 
IPython \citep{ipython}, 
Jupyter \citep{jupyer}, 
Lightkurve \citep{lightkurve}, 
M\_-M\_K- \citep{mann19}, 
Matplotlib \citep{matplotlib},
Mercury6 \citep{chambers99},
NumPy \citep{numpy}, 
Pandas \citep{pandas}, 
PyMC3 \citep{exoplanet:pymc3}, 
SciPy \citep{scipy}, 
stardate \citep{stardate,Angus2019}, 
STARRY \citep{exoplanet:luger18, exoplanet:agol19}, 
Tapir  \citep{Jensen:2013}, 
TRANSITFIT5 \citep{Rowe2015,Rowe2016c},
Theano \citep{exoplanet:theano}, 
TTVFast \citep{Deck2014}, 
TTV2Fast2Furious \citep{Hadden2018}, 
vespa \citep{Morton2012,vespa}
}

\pagebreak
\appendix
\section{Validation of the Stellar Parameters with Alternative SED Based Methods}
\label{appendix:stellar_params}

As an independent check of the stellar parameters derived in Section~\ref{sec:derived_params}, we used multiple SED-based methods to derive stellar parameters to validate the previous analysis. The first check employed the methods and procedures described in \citet{kostov2019b} and combined the stellar SED with the {\it Gaia\/} DR2 parallax to determine an empirical measurement of the stellar radius. We used the $B_T V_T$ magnitudes from {\it Tycho-2}, the $BVgri$ magnitudes from APASS, the $JHK_S$ magnitudes from {\it 2MASS}, the W1--W4 magnitudes from {\it WISE}, the $G$ magnitude from {\it Gaia}, and the NUV magnitude from {\it GALEX}. Together, the available photometry spans the full stellar SED over the wavelength range 0.2--22~$\mu$m. 

We performed a fit using NextGen stellar atmosphere models, with the priors on effective temperature ($T_{\rm eff}$), surface gravity ($\log g$), and metallicity ([Fe/H]) from the values provided in the TIC \citep{stassun2019}. The remaining free parameter is the extinction ($A_V$), which we set to zero because of the star's proximity. Integrating the model SED gives the bolometric flux at Earth, F$_{\rm bol} = 7.15 \pm 0.34 \times 10^{-10}$ erg~s~cm$^{-2}$. Taking the $F_{\rm bol}$ and T$_{\rm eff}$ together with the {\it Gaia\/} DR2 parallax\footnote{Adjusted by $+0.08$~mas to account for the systematic offset reported by \citet{Stassun2018}.} provides a stellar radius $R = 0.404 \pm 0.023$~R$_\odot$. Finally, estimating the stellar mass from the empirical relations of \citet{Torres2010}, assuming solar metallicity, gives M~$ = 0.44 \pm 0.03 \ $ M$_\odot$, which when combined with the radius results in a mean stellar density $\rho = 9.52 \pm 0.12$ g~cm$^{-3}$. These results are consistent with those from the empirically driven parameter analysis.

As a second independent check on the stellar parameters, we employ the SED fitting method of Silverstein et al.~(in preparation), which is based upon the method described by \cite{Dieterich2014}. In this analysis, we compared the star's Johnson $V$ ($V_J$), Kron-Cousins $RI$ ($R_{KC}I_{KC}$), 2MASS $JHK_s$,~and \textit{WISE} AllWISE Release $W1W2W3$ to those extracted from the BT-Settl 2011 photospheric model spectra \citep{Allard2011}. We obtained $V_JR_{KC}I_{KC}$ photometry observations at the SMARTS/CTIO 0.9 m telescope in Chile on 2019 August 20 UT using the 2048$\times$2048 Tektronix CCD camera. Following standard RECONS SMARTS/CTIO 0.9m photometry procedures \citep{Jao2003,Jao2005,Winters2011}, we took observations, reduced the data, and performed aperture photometry. 

In Silverstein et al.~(in preparation), we found nine photometric colors to be effective probes of temperature for early M dwarfs. Here we compared the colors of \target \ to colors extracted from the BT-Settl 2011 model photospheres. Each color yielded a best-matching spectrum and corresponding effective temperature. The resulting value for TOI-700, \teff~=~$3480\pm50$~K, is the mean of these temperatures. We estimated the temperature uncertainty by adding their standard deviation in quadrature with a systematic error based on the discrete nature of the model grid. We then calculated the flux within the full wavelength range covered by the available photometry using an iterative procedure that scaled a 3500 K model spectrum, the closest grid point to our results, until all model magnitudes were within 0.03 mag of their observed counterparts. Next, we integrated the scaled spectrum within the wavelength range of the $V_J$ to $W3$ photometry, and we performed a correction to bolometric flux by calculating the flux that would be missing from a blackbody of the same effective temperature. We calculated the bolometric luminosity, L$_{bol}$~=~0.0235~$\pm$~0.0004~L$_\odot$, by scaling the resultant bolometric flux, \fbol~=~7.73~$\pm$~0.12~$\times 10^{-10}$ erg~s~cm$^{-2}$, by the inverse square of the \textit{Gaia} DR2 parallax. We then derived a radius of $R=0.421~\pm~0.025$~\rsun using the Stefan-Boltzmann law. We also calculated the mass of the star using the \citet{Benedict(2016)} absolute $V$- and $K$-band mass-luminosity relations for main sequence M dwarfs. We determined the weighted mean of the masses from each relation and found M~$=0.42~\pm~0.02$~\msun. These parameters are also consistent with those estimated in Section~\ref{sec:derived_params}.

\begin{figure}[htb]
\begin{center}
\includegraphics[width=0.5\textwidth]{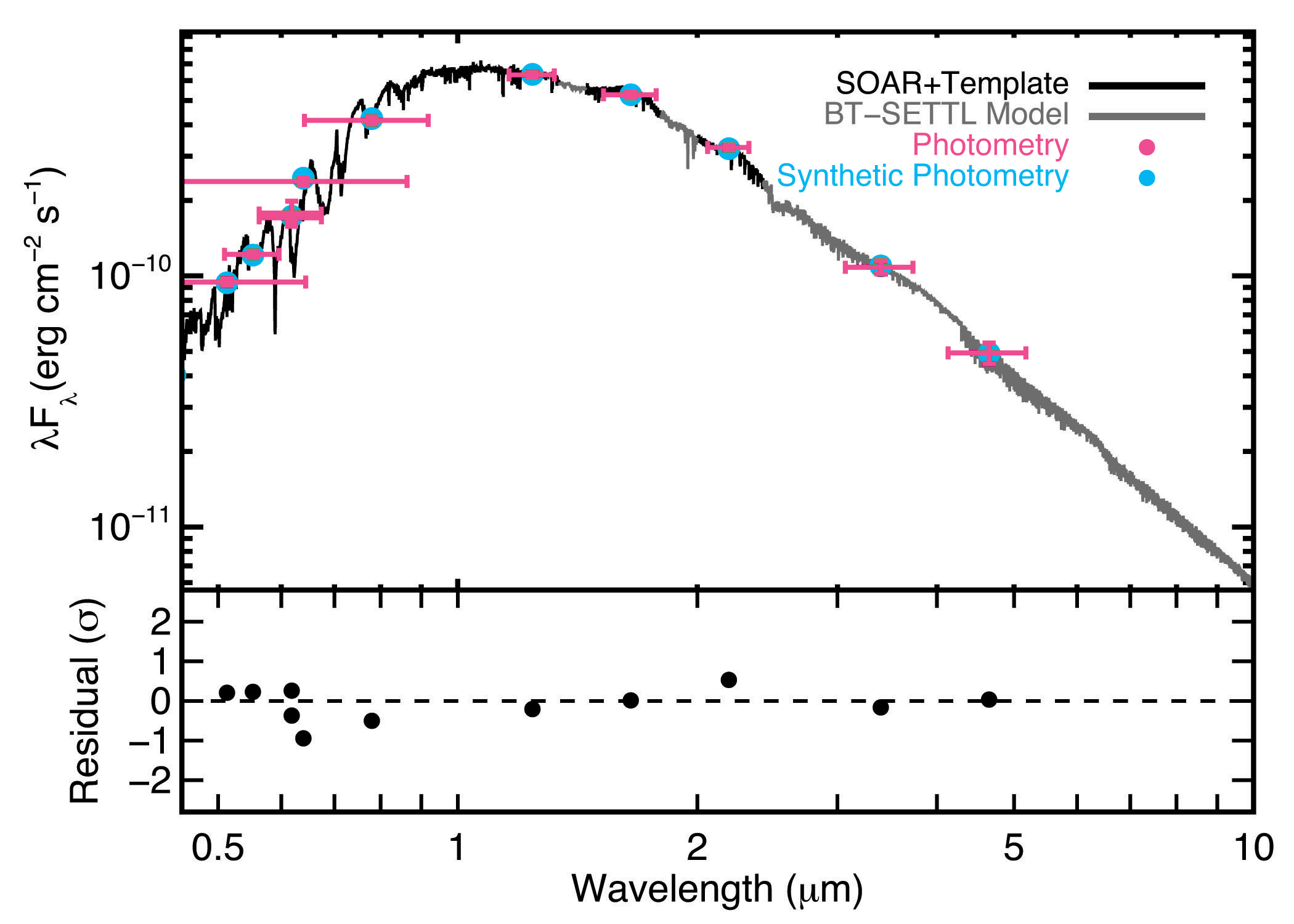}
\caption{Best-fit spectral template and Goodman spectrum (black) compared to the photometry of \target. Gray regions are BT-Settl models, used to fill in gaps or regions of high telluric contamination. Literature photometry is shown in pink, with horizontal errors corresponding to the filter width and vertical errors the measurement errors. Corresponding synthetic photometry is shown as blue points. The bottom panel shows the residuals in terms of standard deviations from the fit.}
\label{fig:sed}
\end{center}
\end{figure}

As a final alternative, we also estimated the stellar parameters using the SOAR Goodman spectrum described in Section~\ref{sec:derived_params}. We constructed and fit an SED using available photometry, the spectrum, and M dwarf templates from \citet{Gaidos2014}. More details of our method can be found in \citet{Mann2015}, which we summarize here. We first downloaded literature optical and NIR photometry from the 2MASS \citep{Skrutskie2006}, the Wide-field Infrared Survey Explorer \citep[WISE,][]{Wright2010}, \emph{Gaia} data release 2 \citep[DR2,][]{Evans2018,GaiaDr2}, and AAVSO All-Sky Photometric Survey \citep[APASS,][]{Henden2012}. We compared this photometry to synthetic magnitudes computed from the combination of our SOAR spectrum, a grid of template M dwarf spectra, and PHOENIX BT-Settl models \citep{Allard2011} to cover gaps in the spectra. The Goodman spectrum was not as precisely flux-calibrated as the data used in \citet{Mann2015}, so we included two additional free parameters to fit out wavelength-dependent flux variations (so the major constraint comes from the molecular band shape and depth). This joint fitting procedure yielded a \teff\ of $3460 \pm 65$~K and a L$_*$ of $0.0236 \pm 0.0005$ L$_\odot$. Using the Stefan-Boltzmann Law, this yielded a radius value consistent at $<1\sigma$ with the value derived from the $M_{K_S}-$R$_*$ relation described in Section~\ref{sec:derived_params}. The final calibrated and combined spectrum along with archival and synthetic photometry used to construct the SED is shown in Figure~\ref{fig:sed}.

\section{Validation of the Stellar Rotation Period with HATSouth}
\label{appendix:p_rot}

Long-baseline photometry of TOI-700 was also obtained using the HATSouth telescope network \citep{bakos:2013:hatsouth} from 2017 Feb 15 through 2017 May 09. A total of 1137 $r^{\prime}$-band exposures of 4 minute duration were obtained containing TOI~700 as a point source. The median FWHM of the point-spread-function was $7\arcsec$ at the location of TOI-700. The observations were reduced to an ensemble-corrected light curve via aperture photometry following the method described by \citep{penev:2013:hats1}. The light curve shows a clear quasi-sinusoidal variation that phases up at a period of $53.1 \pm 1.2$\,days and a peak-to-peak amplitude of $12.6 \pm 0.7$\,ppt. If this is the rotation period of the star, the observations span 1.6 cycles. After fitting and subtracting a sinusoid model from the light curve, we find that the residuals have a point-to-point r.m.s.\ scatter of 6.4\,ppt. The sinusoidal variation persists after applying the standard de-trending techniques used by HATSouth, indicating an astrophysical origin. The scatter in the HATSouth light curve is too large to permit detection of any of the three transiting planet signals identified by TESS, and the time-coverage is such that no transit events were observed for TOI-700.01 or TOI~700.03. The observations do cover a predicted transit for TOI~700.02, though the transit is too shallow to be detected. While no obvious flare events are seen in the HATSouth light curve we do find a slight imbalance between the number of bright outliers in the light curve compared to faint outliers, with six total $3\sigma$ bright outliers and two $3\sigma$ faint outliers. These HATSouth observations are consistent with those from ASAS-SN and confirm the estimated rotation period of TOI-700.

\bibliography{bibliography}{}

\begin{thebibliography}{}
\expandafter\ifx\csname natexlab\endcsname\relax\def\natexlab#1{#1}\fi
\providecommand{\url}[1]{\href{#1}{#1}}
\providecommand{\dodoi}[1]{doi:~\href{http://doi.org/#1}{\nolinkurl{#1}}}
\providecommand{\doeprint}[1]{\href{http://ascl.net/#1}{\nolinkurl{http://ascl.net/#1}}}
\providecommand{\doarXiv}[1]{\href{https://arxiv.org/abs/#1}{\nolinkurl{https://arxiv.org/abs/#1}}}

\bibitem[{{Agol} {et~al.}(2019){Agol}, {Luger}, \&
  {Foreman-Mackey}}]{exoplanet:agol19}
{Agol}, E., {Luger}, R., \& {Foreman-Mackey}, D. 2019, arXiv e-prints

\bibitem[{{Airapetian} {et~al.}(2017){Airapetian}, {Glocer}, {Khazanov},
  {Loyd}, {France}, {Sojka}, {Danchi}, \& {Liemohn}}]{Airapetian2017}
{Airapetian}, V.~S., {Glocer}, A., {Khazanov}, G.~V., {et~al.} 2017, \apjl,
  836, L3, \dodoi{10.3847/2041-8213/836/1/L3}

\bibitem[{{Airapetian} {et~al.}(2020){Airapetian}, {Barnes}, {Cohen},
  {Collinson}, {Danchi}, {Dong}, {Del Genio}, {France}, {Garcia-Sage},
  {Glocer}, {Gopalswamy}, {Grenfell}, {Gronoff}, {G{\"u}del}, {Herbst},
  {Henning}, {Jackman}, {Jin}, {Johnstone}, {Kaltenegger}, {Kay}, {Kobayashi},
  {Kuang}, {Li}, {Lynch}, {L{\"u}ftinger}, {Luhmann}, {Maehara}, {Mlynczak},
  {Notsu}, {Osten}, {Ramirez}, {Rugheimer}, {Scheucher}, {Schlieder},
  {Shibata}, {Sousa-Silva}, {Stamenkovi{\'c}}, {Strangeway}, {Usmanov},
  {Vergados}, {Verkhoglyadova}, {Vidotto}, {Voytek}, {Way}, {Zank}, \&
  {Yamashiki}}]{Airapetian2020}
{Airapetian}, V.~S., {Barnes}, R., {Cohen}, O., {et~al.} 2020, International
  Journal of Astrobiology, 19, 136, \dodoi{10.1017/S1473550419000132}

\bibitem[{{Allard} {et~al.}(2011){Allard}, {Homeier}, \&
  {Freytag}}]{Allard2011}
{Allard}, F., {Homeier}, D., \& {Freytag}, B. 2011, in Astronomical Society of
  the Pacific Conference Series, Vol. 448, 16th Cambridge Workshop on Cool
  Stars, Stellar Systems, and the Sun, ed. C.~{Johns-Krull}, M.~K. {Browning},
  \& A.~A. {West}, 91

\bibitem[{{Anglada-Escud{\'e}} {et~al.}(2013){Anglada-Escud{\'e}}, {Tuomi},
  {Gerlach}, {Barnes}, {Heller}, {Jenkins}, {Wende}, {Vogt}, {Butler},
  {Reiners}, \& {Jones}}]{Anglada2013}
{Anglada-Escud{\'e}}, G., {Tuomi}, M., {Gerlach}, E., {et~al.} 2013, \aap, 556,
  A126, \dodoi{10.1051/0004-6361/201321331}

\bibitem[{{Anglada-Escud{\'e}} {et~al.}(2016){Anglada-Escud{\'e}}, {Amado},
  {Barnes}, {Berdi{\~n}as}, {Butler}, {Coleman}, {de La Cueva}, {Dreizler},
  {Endl}, {Giesers}, {Jeffers}, {Jenkins}, {Jones}, {Kiraga}, {K{\"u}rster},
  {L{\'o}pez-Gonz{\'a}lez}, {Marvin}, {Morales}, {Morin}, {Nelson}, {Ortiz},
  {Ofir}, {Paardekooper}, {Reiners}, {Rodr{\'\i}guez},
  {Rodr{\'\i}guez-L{\'o}pez}, {Sarmiento}, {Strachan}, {Tsapras}, {Tuomi}, \&
  {Zechmeister}}]{Anglada2016}
{Anglada-Escud{\'e}}, G., {Amado}, P.~J., {Barnes}, J., {et~al.} 2016, \nat,
  536, 437, \dodoi{10.1038/nature19106}

\bibitem[{{Angus} {et~al.}(2019{\natexlab{a}}){Angus}, {Morton}, \&
  {Foreman-Mackey}}]{stardate}
{Angus}, R., {Morton}, T., \& {Foreman-Mackey}, D. 2019{\natexlab{a}}, The
  Journal of Open Source Software, 4, 1469, \dodoi{10.21105/joss.01469}

\bibitem[{{Angus} {et~al.}(2019{\natexlab{b}}){Angus}, {Morton},
  {Foreman-Mackey}, {van Saders}, {Curtis}, {Kane}, {Bedell}, {Kiman}, {Hogg},
  \& {Brewer}}]{Angus2019}
{Angus}, R., {Morton}, T.~D., {Foreman-Mackey}, D., {et~al.}
  2019{\natexlab{b}}, \aj, 158, 173, \dodoi{10.3847/1538-3881/ab3c53}

\bibitem[{{Aschwanden}(1994)}]{Aschwanden1994}
{Aschwanden}, M.~J. 1994, \solphys, 152, 53, \dodoi{10.1007/BF01473183}

\bibitem[{{Astropy Collaboration} {et~al.}(2013){Astropy Collaboration},
  {Robitaille}, {Tollerud}, {Greenfield}, {Droettboom}, {Bray}, {Aldcroft},
  {Davis}, {Ginsburg}, {Price-Whelan}, {Kerzendorf}, {Conley}, {Crighton},
  {Barbary}, {Muna}, {Ferguson}, {Grollier}, {Parikh}, {Nair}, {Unther},
  {Deil}, {Woillez}, {Conseil}, {Kramer}, {Turner}, {Singer}, {Fox}, {Weaver},
  {Zabalza}, {Edwards}, {Azalee Bostroem}, {Burke}, {Casey}, {Crawford},
  {Dencheva}, {Ely}, {Jenness}, {Labrie}, {Lim}, {Pierfederici}, {Pontzen},
  {Ptak}, {Refsdal}, {Servillat}, \& {Streicher}}]{exoplanet:astropy13}
{Astropy Collaboration}, {Robitaille}, T.~P., {Tollerud}, E.~J., {et~al.} 2013,
  \aap, 558, A33, \dodoi{10.1051/0004-6361/201322068}

\bibitem[{{Astropy Collaboration} {et~al.}(2018){Astropy Collaboration},
  {Price-Whelan}, {Sip{\H o}cz}, {G{\"u}nther}, {Lim}, {Crawford}, {Conseil},
  {Shupe}, {Craig}, {Dencheva}, {Ginsburg}, {VanderPlas}, {Bradley},
  {P{\'e}rez-Su{\'a}rez}, {de Val-Borro}, {Aldcroft}, {Cruz}, {Robitaille},
  {Tollerud}, {Ardelean}, {Babej}, {Bach}, {Bachetti}, {Bakanov}, {Bamford},
  {Barentsen}, {Barmby}, {Baumbach}, {Berry}, {Biscani}, {Boquien}, {Bostroem},
  {Bouma}, {Brammer}, {Bray}, {Breytenbach}, {Buddelmeijer}, {Burke},
  {Calderone}, {Cano Rodr{\'{\i}}guez}, {Cara}, {Cardoso}, {Cheedella},
  {Copin}, {Corrales}, {Crichton}, {D'Avella}, {Deil}, {Depagne}, {Dietrich},
  {Donath}, {Droettboom}, {Earl}, {Erben}, {Fabbro}, {Ferreira}, {Finethy},
  {Fox}, {Garrison}, {Gibbons}, {Goldstein}, {Gommers}, {Greco}, {Greenfield},
  {Groener}, {Grollier}, {Hagen}, {Hirst}, {Homeier}, {Horton}, {Hosseinzadeh},
  {Hu}, {Hunkeler}, {Ivezi{\'c}}, {Jain}, {Jenness}, {Kanarek}, {Kendrew},
  {Kern}, {Kerzendorf}, {Khvalko}, {King}, {Kirkby}, {Kulkarni}, {Kumar},
  {Lee}, {Lenz}, {Littlefair}, {Ma}, {Macleod}, {Mastropietro}, {McCully},
  {Montagnac}, {Morris}, {Mueller}, {Mumford}, {Muna}, {Murphy}, {Nelson},
  {Nguyen}, {Ninan}, {N{\"o}the}, {Ogaz}, {Oh}, {Parejko}, {Parley}, {Pascual},
  {Patil}, {Patil}, {Plunkett}, {Prochaska}, {Rastogi}, {Reddy Janga},
  {Sabater}, {Sakurikar}, {Seifert}, {Sherbert}, {Sherwood-Taylor}, {Shih},
  {Sick}, {Silbiger}, {Singanamalla}, {Singer}, {Sladen}, {Sooley},
  {Sornarajah}, {Streicher}, {Teuben}, {Thomas}, {Tremblay}, {Turner},
  {Terr{\'o}n}, {van Kerkwijk}, {de la Vega}, {Watkins}, {Weaver}, {Whitmore},
  {Woillez}, {Zabalza}, \& {Astropy Contributors}}]{exoplanet:astropy18}
{Astropy Collaboration}, {Price-Whelan}, A.~M., {Sip{\H o}cz}, B.~M., {et~al.}
  2018, \aj, 156, 123, \dodoi{10.3847/1538-3881/aabc4f}

\bibitem[{{Bailer-Jones} {et~al.}(2018){Bailer-Jones}, {Rybizki}, {Fouesneau},
  {Mantelet}, \& {Andrae}}]{bailerjones2018}
{Bailer-Jones}, C.~A.~L., {Rybizki}, J., {Fouesneau}, M., {Mantelet}, G., \&
  {Andrae}, R. 2018, \aj, 156, 58, \dodoi{10.3847/1538-3881/aacb21}

\bibitem[{{Bakos} {et~al.}(2013){Bakos}, {Csubry}, {Penev}, {Bayliss},
  {Jord{\'a}n}, {Afonso}, {Hartman}, {Henning}, {Kov{\'a}cs}, {Noyes},
  {B{\'e}ky}, {Suc}, {Cs{\'a}k}, {Rabus}, {L{\'a}z{\'a}r}, {Papp}, {S{\'a}ri},
  {Conroy}, {Zhou}, {Sackett}, {Schmidt}, {Mancini}, {Sasselov}, \&
  {Ueltzhoeffer}}]{bakos:2013:hatsouth}
{Bakos}, G.~{\'A}., {Csubry}, Z., {Penev}, K., {et~al.} 2013, \pasp, 125, 154,
  \dodoi{10.1086/669529}

\bibitem[{{Ballard}(2019)}]{ballard19}
{Ballard}, S. 2019, \aj, 157, 113, \dodoi{10.3847/1538-3881/aaf477}

\bibitem[{{Barclay} {et~al.}(2018){Barclay}, {Pepper}, \&
  {Quintana}}]{Barclay2018}
{Barclay}, T., {Pepper}, J., \& {Quintana}, E.~V. 2018, \apjs, 239, 2,
  \dodoi{10.3847/1538-4365/aae3e9}

\bibitem[{{Becker} \& {Adams}(2017)}]{Becker2017}
{Becker}, J.~C., \& {Adams}, F.~C. 2017, \mnras, 468, 549,
  \dodoi{10.1093/mnras/stx461}

\bibitem[{{Benedict} {et~al.}(2016){Benedict}, {Henry}, {Franz}, {McArthur},
  {Wasserman}, {Jao}, {Cargile}, {Dieterich}, {Bradley}, {Nelan}, \&
  {Whipple}}]{Benedict(2016)}
{Benedict}, G.~F., {Henry}, T.~J., {Franz}, O.~G., {et~al.} 2016, \aj, 152,
  141, \dodoi{10.3847/0004-6256/152/5/141}

\bibitem[{{Bensby} {et~al.}(2014){Bensby}, {Feltzing}, \& {Oey}}]{bensby14}
{Bensby}, T., {Feltzing}, S., \& {Oey}, M.~S. 2014, \aap, 562, A71,
  \dodoi{10.1051/0004-6361/201322631}

\bibitem[{{Bensby} {et~al.}(2010){Bensby}, {Feltzing}, {Johnson}, {Gould},
  {Ad{\'e}n}, {Asplund}, {Mel{\'e}ndez}, {Gal-Yam}, {Lucatello}, {Sana},
  {Sumi}, {Miyake}, {Suzuki}, {Han}, {Bond}, \& {Udalski}}]{bensby2010}
{Bensby}, T., {Feltzing}, S., {Johnson}, J.~A., {et~al.} 2010, \aap, 512, A41,
  \dodoi{10.1051/0004-6361/200913744}

\bibitem[{{Bianchi} {et~al.}(2011){Bianchi}, {Herald}, {Efremova}, {Girardi},
  {Zabot}, {Marigo}, {Conti}, \& {Shiao}}]{bianchi2011}
{Bianchi}, L., {Herald}, J., {Efremova}, B., {et~al.} 2011, \apss, 335, 161,
  \dodoi{10.1007/s10509-010-0581-x}

\bibitem[{{Bolmont} {et~al.}(2017){Bolmont}, {Selsis}, {Owen}, {Ribas},
  {Raymond}, {Leconte}, \& {Gillon}}]{Bolmont2017}
{Bolmont}, E., {Selsis}, F., {Owen}, J.~E., {et~al.} 2017, \mnras, 464, 3728,
  \dodoi{10.1093/mnras/stw2578}

\bibitem[{{Bonfils} {et~al.}(2013){Bonfils}, {Delfosse}, {Udry}, {Forveille},
  {Mayor}, {Perrier}, {Bouchy}, {Gillon}, {Lovis}, {Pepe}, {Queloz}, {Santos},
  {S{\'e}gransan}, \& {Bertaux}}]{Bonfils2013}
{Bonfils}, X., {Delfosse}, X., {Udry}, S., {et~al.} 2013, \aap, 549, A109,
  \dodoi{10.1051/0004-6361/201014704}

\bibitem[{{Borucki} {et~al.}(2010){Borucki}, {Koch}, {Basri}, {Batalha},
  {Brown}, {Caldwell}, {Caldwell}, {Christensen-Dalsgaard}, {Cochran},
  {DeVore}, {Dunham}, {Dupree}, {Gautier}, {Geary}, {Gilliland}, {Gould},
  {Howell}, {Jenkins}, {Kondo}, {Latham}, {Marcy}, {Meibom}, {Kjeldsen},
  {Lissauer}, {Monet}, {Morrison}, {Sasselov}, {Tarter}, {Boss}, {Brownlee},
  {Owen}, {Buzasi}, {Charbonneau}, {Doyle}, {Fortney}, {Ford}, {Holman},
  {Seager}, {Steffen}, {Welsh}, {Rowe}, {Anderson}, {Buchhave}, {Ciardi},
  {Walkowicz}, {Sherry}, {Horch}, {Isaacson}, {Everett}, {Fischer}, {Torres},
  {Johnson}, {Endl}, {MacQueen}, {Bryson}, {Dotson}, {Haas}, {Kolodziejczak},
  {Van Cleve}, {Chandrasekaran}, {Twicken}, {Quintana}, {Clarke}, {Allen},
  {Li}, {Wu}, {Tenenbaum}, {Verner}, {Bruhweiler}, {Barnes}, \&
  {Prsa}}]{borucki10}
{Borucki}, W.~J., {Koch}, D., {Basri}, G., {et~al.} 2010, Science, 327, 977,
  \dodoi{10.1126/science.1185402}

\bibitem[{{Brown} {et~al.}(2013){Brown}, {Baliber}, {Bianco}, {Bowman},
  {Burleson}, {Conway}, {Crellin}, {Depagne}, {De Vera}, {Dilday}, {Dragomir},
  {Dubberley}, {Eastman}, {Elphick}, {Falarski}, {Foale}, {Ford}, {Fulton},
  {Garza}, {Gomez}, {Graham}, {Greene}, {Haldeman}, {Hawkins}, {Haworth},
  {Haynes}, {Hidas}, {Hjelstrom}, {Howell}, {Hygelund}, {Lister}, {Lobdill},
  {Martinez}, {Mullins}, {Norbury}, {Parrent}, {Paulson}, {Petry}, {Pickles},
  {Posner}, {Rosing}, {Ross}, {Sand}, {Saunders}, {Shobbrook}, {Shporer},
  {Street}, {Thomas}, {Tsapras}, {Tufts}, {Valenti}, {Vander Horst}, {Walker},
  {White}, \& {Willis}}]{Brown:2013}
{Brown}, T.~M., {Baliber}, N., {Bianco}, F.~B., {et~al.} 2013, Publications of
  the Astronomical Society of the Pacific, 125, 1031, \dodoi{10.1086/673168}

\bibitem[{{Burke} {et~al.}(2019){Burke}, {Mullally}, {Thompson}, {Coughlin}, \&
  {Rowe}}]{Burke2019}
{Burke}, C.~J., {Mullally}, F., {Thompson}, S.~E., {Coughlin}, J.~L., \&
  {Rowe}, J.~F. 2019, \aj, 157, 143, \dodoi{10.3847/1538-3881/aafb79}

\bibitem[{{Carter} {et~al.}(2012){Carter}, {Agol}, {Chaplin}, {Basu},
  {Bedding}, {Buchhave}, {Christensen-Dalsgaard}, {Deck}, {Elsworth},
  {Fabrycky}, {Ford}, {Fortney}, {Hale}, {Handberg}, {Hekker}, {Holman},
  {Huber}, {Karoff}, {Kawaler}, {Kjeldsen}, {Lissauer}, {Lopez}, {Lund},
  {Lundkvist}, {Metcalfe}, {Miglio}, {Rogers}, {Stello}, {Borucki}, {Bryson},
  {Christiansen}, {Cochran}, {Geary}, {Gilliland}, {Haas}, {Hall}, {Howard},
  {Jenkins}, {Klaus}, {Koch}, {Latham}, {MacQueen}, {Sasselov}, {Steffen},
  {Twicken}, \& {Winn}}]{Carter2012}
{Carter}, J.~A., {Agol}, E., {Chaplin}, W.~J., {et~al.} 2012, Science, 337,
  556, \dodoi{10.1126/science.1223269}

\bibitem[{{Chambers}(1999)}]{chambers99}
{Chambers}, J.~E. 1999, \mnras, 304, 793,
  \dodoi{10.1046/j.1365-8711.1999.02379.x}

\bibitem[{{Chen} \& {Kipping}(2017)}]{forecaster}
{Chen}, J., \& {Kipping}, D. 2017, \apj, 834, 17,
  \dodoi{10.3847/1538-4357/834/1/17}

\bibitem[{{Ciardi} {et~al.}(2015){Ciardi}, {Beichman}, {Horch}, \&
  {Howell}}]{Ciardi2015}
{Ciardi}, D.~R., {Beichman}, C.~A., {Horch}, E.~P., \& {Howell}, S.~B. 2015,
  \apj, 805, 16, \dodoi{10.1088/0004-637X/805/1/16}

\bibitem[{{Claire}(2008)}]{Claire2008}
{Claire}, M.~W. 2008, PhD thesis, University of Washington

\bibitem[{{Clemens} {et~al.}(2004){Clemens}, {Crain}, \& {Anderson}}]{Goodman}
{Clemens}, J.~C., {Crain}, J.~A., \& {Anderson}, R. 2004, in \procspie, Vol.
  5492, Ground-based Instrumentation for Astronomy, ed. A.~F.~M. {Moorwood} \&
  M.~{Iye}, 331--340

\bibitem[{{Cloutier} \& {Menou}(2019)}]{Cloutier2019b}
{Cloutier}, R., \& {Menou}, K. 2019, arXiv e-prints, arXiv:1912.02170.
\newblock \doarXiv{1912.02170}

\bibitem[{{Cloutier} {et~al.}(2019){Cloutier}, {Astudillo-Defru}, {Bonfils},
  {Jenkins}, {Berdi{\~n}as}, {Ricker}, {Vand erspek}, {Latham}, {Seager},
  {Winn}, {Jenkins}, {Almenara}, {Bouchy}, {Delfosse}, {D{\'\i}az},
  {D{\'\i}az}, {Doyon}, {Figueira}, {Forveille}, {Kurtovic}, {Lovis}, {Mayor},
  {Menou}, {Morgan}, {Morris}, {Muirhead}, {Murgas}, {Pepe}, {Santos},
  {S{\'e}gransan}, {Smith}, {Tenenbaum}, {Torres}, {Udry}, {Vezie}, \&
  {Villasenor}}]{cloutier19}
{Cloutier}, R., {Astudillo-Defru}, N., {Bonfils}, X., {et~al.} 2019, \aap, 629,
  A111, \dodoi{10.1051/0004-6361/201935957}

\bibitem[{{Cloutier} {et~al.}(2020){Cloutier}, {Eastman}, {Rodriguez},
  {Astudillo-Defru}, {Bonfils}, {Mortier}, {Watson}, {Stalport}, {Pinamonti},
  {Lienhard}, {Harutyunyan}, {Damasso}, {Latham}, {Collins}, {Massey}, {Irwin},
  {Winters}, {Charbonneau}, {Ziegler}, {Matthews}, {Crossfield}, {Kreidberg},
  {Quinn}, {Ricker}, {Vanderspek}, {Seager}, {Winn}, {Jenkins}, {Vezie},
  {Udry}, {Twicken}, {Tenenbaum}, {Sozzetti}, {S{\'e}gransan}, {Schlieder},
  {Sasselov}, {Santos}, {Rice}, {Rackham}, {Poretti}, {Piotto}, {Phillips},
  {Pepe}, {Molinari}, {Mignon}, {Micela}, {Melo}, {de Medeiros}, {Mayor},
  {Matson}, {Martinez Fiorenzano}, {Mann}, {Magazz{\'u}}, {Lovis},
  {L{\'o}pez-Morales}, {Lopez}, {Lissauer}, {L{\'e}pine}, {Law}, {Kielkopf},
  {Johnson}, {Jensen}, {Howell}, {Gonzales}, {Ghedina}, {Forveille},
  {Figueira}, {Dumusque}, {Dressing}, {Doyon}, {D{\'\i}az}, {Di Fabrizio},
  {Delfosse}, {Cosentino}, {Conti}, {Collins}, {Collier Cameron}, {Ciardi},
  {Caldwell}, {Burke}, {Buchhave}, {Brice{\~n}o}, {Boyd}, {Bouchy}, {Beichman},
  {Artigau}, \& {Almenara}}]{cloutier2020}
{Cloutier}, R., {Eastman}, J.~D., {Rodriguez}, J.~E., {et~al.} 2020, arXiv
  e-prints, arXiv:2003.01136.
\newblock \doarXiv{2003.01136}

\bibitem[{{Cohen} {et~al.}(2014){Cohen}, {Drake}, {Glocer}, {Garraffo},
  {Poppenhaeger}, {Bell}, {Ridley}, \& {Gombosi}}]{Cohen2014}
{Cohen}, O., {Drake}, J.~J., {Glocer}, A., {et~al.} 2014, \apj, 790, 57,
  \dodoi{10.1088/0004-637X/790/1/57}

\bibitem[{Cohen {et~al.}(2020)Cohen, Garraffo, Moschou, Drake, Alvarado-Gomez,
  Glocer, \& Fraschetti}]{cohen2020}
Cohen, O., Garraffo, C., Moschou, S., {et~al.} 2020, The Space Environment and
  Atmospheric Joule Heating of the Habitable Zone Exoplanet TOI700-d.
\newblock \doarXiv{2005.11587}

\bibitem[{{Cohen} {et~al.}(2018){Cohen}, {Glocer}, {Garraffo}, {Drake}, \&
  {Bell}}]{Cohen2018}
{Cohen}, O., {Glocer}, A., {Garraffo}, C., {Drake}, J.~J., \& {Bell}, J.~M.
  2018, \apjl, 856, L11, \dodoi{10.3847/2041-8213/aab5b5}

\bibitem[{{Cohen} {et~al.}(2015){Cohen}, {Ma}, {Drake}, {Glocer}, {Garraffo},
  {Bell}, \& {Gombosi}}]{Cohen2015}
{Cohen}, O., {Ma}, Y., {Drake}, J.~J., {et~al.} 2015, \apj, 806, 41,
  \dodoi{10.1088/0004-637X/806/1/41}

\bibitem[{{Collins} {et~al.}(2017){Collins}, {Kielkopf}, {Stassun}, \&
  {Hessman}}]{karen2017}
{Collins}, K.~A., {Kielkopf}, J.~F., {Stassun}, K.~G., \& {Hessman}, F.~V.
  2017, \aj, 153, 77, \dodoi{10.3847/1538-3881/153/2/77}

\bibitem[{{Co{\textcommabelow s}kuno{\v{g}}lu}
  {et~al.}(2011){Co{\textcommabelow s}kuno{\v{g}}lu}, {Ak}, {Bilir}, {Karaali},
  {Yaz}, {Gilmore}, {Seabroke}, {Bienaym{\'e}}, {Bland-Hawthorn}, {Campbell},
  {Freeman}, {Gibson}, {Grebel}, {Munari}, {Navarro}, {Parker}, {Siebert},
  {Siviero}, {Steinmetz}, {Watson}, {Wyse}, \& {Zwitter}}]{coskunoglu2011}
{Co{\textcommabelow s}kuno{\v{g}}lu}, B., {Ak}, S., {Bilir}, S., {et~al.} 2011,
  \mnras, 412, 1237, \dodoi{10.1111/j.1365-2966.2010.17983.x}

\bibitem[{{Coughlin} {et~al.}(2014){Coughlin}, {Thompson}, {Bryson}, {Burke},
  {Caldwell}, {Christiansen}, {Haas}, {Howell}, {Jenkins}, {Kolodziejczak},
  {Mullally}, \& {Rowe}}]{Coughlin2014}
{Coughlin}, J.~L., {Thompson}, S.~E., {Bryson}, S.~T., {et~al.} 2014, \aj, 147,
  119, \dodoi{10.1088/0004-6256/147/5/119}

\bibitem[{{Coughlin} {et~al.}(2016){Coughlin}, {Mullally}, {Thompson}, {Rowe},
  {Burke}, {Latham}, {Batalha}, {Ofir}, {Quarles}, {Henze}, {Wolfgang},
  {Caldwell}, {Bryson}, {Shporer}, {Catanzarite}, {Akeson}, {Barclay},
  {Borucki}, {Boyajian}, {Campbell}, {Christiansen}, {Girouard}, {Haas},
  {Howell}, {Huber}, {Jenkins}, {Li}, {Patil-Sabale}, {Quintana}, {Ramirez},
  {Seader}, {Smith}, {Tenenbaum}, {Twicken}, \& {Zamudio}}]{coughlin16}
{Coughlin}, J.~L., {Mullally}, F., {Thompson}, S.~E., {et~al.} 2016, \apjs,
  224, 12, \dodoi{10.3847/0067-0049/224/1/12}

\bibitem[{{Cretignier} {et~al.}(2019){Cretignier}, {Dumusque}, {Allart},
  {Pepe}, \& {Lovis}}]{Cretignier19}
{Cretignier}, M., {Dumusque}, X., {Allart}, R., {Pepe}, F., \& {Lovis}, C.
  2019, arXiv e-prints, arXiv:1912.05192.
\newblock \doarXiv{1912.05192}

\bibitem[{{Crossfield} {et~al.}(2019){Crossfield}, {Waalkes}, {Newton},
  {Narita}, {Muirhead}, {Ment}, {Matthews}, {Kraus}, {Kostov}, {Kosiarek},
  {Kane}, {Isaacson}, {Halverson}, {Gonzales}, {Everett}, {Dragomir},
  {Collins}, {Chontos}, {Berardo}, {Winters}, {Winn}, {Scott}, {Rojas-Ayala},
  {Rizzuto}, {Petigura}, {Peterson}, {Mocnik}, {Mikal-Evans}, {Mehrle},
  {Matson}, {Kuzuhara}, {Irwin}, {Huber}, {Huang}, {Howell}, {Howard},
  {Hirano}, {Fulton}, {Dupuy}, {Dressing}, {Dalba}, {Charbonneau}, {Burt},
  {Berta-Thompson}, {Benneke}, {Watanabe}, {Twicken}, {Tamura}, {Schlieder},
  {Seager}, {Rose}, {Ricker}, {Quintana}, {L{\'e}pine}, {Latham}, {Kotani},
  {Jenkins}, {Hori}, {Colon}, \& {Caldwell}}]{crossfield19}
{Crossfield}, I. J.~M., {Waalkes}, W., {Newton}, E.~R., {et~al.} 2019, \apjl,
  883, L16, \dodoi{10.3847/2041-8213/ab3d30}

\bibitem[{{Cushing} {et~al.}(2005){Cushing}, {Rayner}, \&
  {Vacca}}]{Cushing2005}
{Cushing}, M.~C., {Rayner}, J.~T., \& {Vacca}, W.~D. 2005, \apj, 623, 1115

\bibitem[{{Cutri} {et~al.}(2013){Cutri}, {Wright}, {Conrow}, {Fowler},
  {Eisenhardt}, {Grillmair}, {Kirkpatrick}, {Masci}, {McCallon}, {Wheelock},
  {Fajardo-Acosta}, {Yan}, {Benford}, {Harbut}, {Jarrett}, {Lake}, {Leisawitz},
  {Ressler}, {Stanford}, {Tsai}, {Liu}, {Helou}, {Mainzer}, {Gettings},
  {Gonzalez}, {Hoffman}, {Marsh}, {Padgett}, {Skrutskie}, {Beck}, {Papin}, \&
  {Wittman}}]{Cutri2013}
{Cutri}, R.~M., {Wright}, E.~L., {Conrow}, T., {et~al.} 2013, {Explanatory
  Supplement to the AllWISE Data Release Products}, Tech. rep.

\bibitem[{{Deck} {et~al.}(2014){Deck}, {Agol}, {Holman}, \&
  {Nesvorn{\'y}}}]{Deck2014}
{Deck}, K.~M., {Agol}, E., {Holman}, M.~J., \& {Nesvorn{\'y}}, D. 2014, \apj,
  787, 132, \dodoi{10.1088/0004-637X/787/2/132}

\bibitem[{{Delfosse} {et~al.}(1998){Delfosse}, {Forveille}, {Mayor}, {Perrier},
  {Naef}, \& {Queloz}}]{Delfosse1998}
{Delfosse}, X., {Forveille}, T., {Mayor}, M., {et~al.} 1998, \aap, 338, L67.
\newblock \doarXiv{astro-ph/9808026}

\bibitem[{{Dieterich} {et~al.}(2014){Dieterich}, {Henry}, {Jao}, {Winters},
  {Hosey}, {Riedel}, \& {Subasavage}}]{Dieterich2014}
{Dieterich}, S.~B., {Henry}, T.~J., {Jao}, W.-C., {et~al.} 2014, \aj, 147, 94,
  \dodoi{10.1088/0004-6256/147/5/94}

\bibitem[{{Dong} {et~al.}(2020){Dong}, {Jin}, \& {Lingam}}]{dong20}
{Dong}, C., {Jin}, M., \& {Lingam}, M. 2020, arXiv e-prints, arXiv:2005.13190.
\newblock \doarXiv{2005.13190}

\bibitem[{{Dong} {et~al.}(2018){Dong}, {Jin}, {Lingam}, {Airapetian}, {Ma}, \&
  {van der Holst}}]{Dong2018}
{Dong}, C., {Jin}, M., {Lingam}, M., {et~al.} 2018, Proceedings of the National
  Academy of Science, 115, 260, \dodoi{10.1073/pnas.1708010115}

\bibitem[{{Dong} {et~al.}(2017){Dong}, {Lingam}, {Ma}, \& {Cohen}}]{Dong2017}
{Dong}, C., {Lingam}, M., {Ma}, Y., \& {Cohen}, O. 2017, \apjl, 837, L26,
  \dodoi{10.3847/2041-8213/aa6438}

\bibitem[{{Dorn} {et~al.}(2018){Dorn}, {Mosegaard}, {Grimm}, \&
  {Alibert}}]{Dorn2018}
{Dorn}, C., {Mosegaard}, K., {Grimm}, S.~L., \& {Alibert}, Y. 2018, \apj, 865,
  20, \dodoi{10.3847/1538-4357/aad95d}

\bibitem[{{Dressing} \& {Charbonneau}(2013)}]{dressing13}
{Dressing}, C.~D., \& {Charbonneau}, D. 2013, \apj, 767, 95,
  \dodoi{10.1088/0004-637X/767/1/95}

\bibitem[{{Dressing} \& {Charbonneau}(2015)}]{dressing15}
---. 2015, \apj, 807, 45, \dodoi{10.1088/0004-637X/807/1/45}

\bibitem[{{Dressing} {et~al.}(2017){Dressing}, {Newton}, {Schlieder},
  {Charbonneau}, {Knutson}, {Vanderburg}, \& {Sinukoff}}]{Dressing2017}
{Dressing}, C.~D., {Newton}, E.~R., {Schlieder}, J.~E., {et~al.} 2017, \apj,
  836, 167, \dodoi{10.3847/1538-4357/836/2/167}

\bibitem[{{Dressing} {et~al.}(2019){Dressing}, {Hardegree-Ullman}, {Schlieder},
  {Newton}, {Vand erburg}, {Feinstein}, {Duvvuri}, {Arnold}, {Bristow},
  {Thackeray}, {Schwab Abrahams}, {Ciardi}, {Crossfield}, {Yu}, {Martinez},
  {Christiansen}, {Crepp}, \& {Isaacson}}]{Dressing2019}
{Dressing}, C.~D., {Hardegree-Ullman}, K., {Schlieder}, J.~E., {et~al.} 2019,
  \aj, 158, 87, \dodoi{10.3847/1538-3881/ab2895}

\bibitem[{{Dumusque}(2018)}]{Dumusque18}
{Dumusque}, X. 2018, \aap, 620, A47, \dodoi{10.1051/0004-6361/201833795}

\bibitem[{{Evans} {et~al.}(2018){Evans}, {Riello}, {De Angeli}, {Carrasco},
  {Montegriffo}, {Fabricius}, {Jordi}, {Palaversa}, {Diener}, {Busso},
  {Cacciari}, {van Leeuwen}, {Burgess}, {Davidson}, {Harrison}, {Hodgkin},
  {Pancino}, {Richards}, {Altavilla}, {Balaguer-N{\'u}{\~n}ez}, {Barstow},
  {Bellazzini}, {Brown}, {Castellani}, {Cocozza}, {De Luise}, {Delgado},
  {Ducourant}, {Galleti}, {Gilmore}, {Giuffrida}, {Holl}, {Kewley}, {Koposov},
  {Marinoni}, {Marrese}, {Osborne}, {Piersimoni}, {Portell}, {Pulone},
  {Ragaini}, {Sanna}, {Terrett}, {Walton}, {Wevers}, \&
  {Wyrzykowski}}]{Evans2018}
{Evans}, D.~W., {Riello}, M., {De Angeli}, F., {et~al.} 2018, \aap, 616, A4

\bibitem[{{Faria} {et~al.}(2019){Faria}, {Adibekyan}, {Amazo-G{\'o}mez},
  {Barros}, {Camacho}, {Demangeon}, {Figueira}, {Mortier}, {Oshagh}, {Pepe},
  {Santos}, {Gomes da Silva}, {Costa Silva}, {Sousa}, {Ulmer-Moll}, \&
  {Viana}}]{faria19}
{Faria}, J.~P., {Adibekyan}, V., {Amazo-G{\'o}mez}, E.~M., {et~al.} 2019, arXiv
  e-prints, arXiv:1911.11714.
\newblock \doarXiv{1911.11714}

\bibitem[{Foreman-Mackey(2018)}]{exoplanet:exoplanet}
Foreman-Mackey, D. 2018, exoplanet v0.1.3, \dodoi{10.5281/zenodo.2536576}

\bibitem[{{Foreman-Mackey}(2018)}]{exoplanet:foremanmackey18}
{Foreman-Mackey}, D. 2018, Research Notes of the American Astronomical Society,
  2, 31, \dodoi{10.3847/2515-5172/aaaf6c}

\bibitem[{{Foreman-Mackey} {et~al.}(2017){Foreman-Mackey}, {Agol},
  {Ambikasaran}, \& {Angus}}]{exoplanet:foremanmackey17}
{Foreman-Mackey}, D., {Agol}, E., {Ambikasaran}, S., \& {Angus}, R. 2017, \aj,
  154, 220, \dodoi{10.3847/1538-3881/aa9332}

\bibitem[{{Foreman-Mackey} {et~al.}(2013){Foreman-Mackey}, {Hogg}, {Lang}, \&
  {Goodman}}]{foremanmackey12}
{Foreman-Mackey}, D., {Hogg}, D.~W., {Lang}, D., \& {Goodman}, J. 2013, \pasp,
  125, 306, \dodoi{10.1086/670067}

\bibitem[{{Fulton} {et~al.}(2017){Fulton}, {Petigura}, {Howard}, {Isaacson},
  {Marcy}, {Cargile}, {Hebb}, {Weiss}, {Johnson}, {Morton}, {Sinukoff},
  {Crossfield}, \& {Hirsch}}]{Fulton2017}
{Fulton}, B.~J., {Petigura}, E.~A., {Howard}, A.~W., {et~al.} 2017, \aj, 154,
  109, \dodoi{10.3847/1538-3881/aa80eb}

\bibitem[{{Furlan} {et~al.}(2017){Furlan}, {Ciardi}, {Everett}, {Saylors},
  {Teske}, {Horch}, {Howell}, {van Belle}, {Hirsch}, {Gautier}, {Adams},
  {Barrado}, {Cartier}, {Dressing}, {Dupree}, {Gilliland}, {Lillo-Box},
  {Lucas}, \& {Wang}}]{furlanciardi2017}
{Furlan}, E., {Ciardi}, D.~R., {Everett}, M.~E., {et~al.} 2017, \aj, 153, 71,
  \dodoi{10.3847/1538-3881/153/2/71}

\bibitem[{{Gaia Collaboration} {et~al.}(2018){Gaia Collaboration}, {Brown},
  {Vallenari}, {Prusti}, {de Bruijne}, {Babusiaux}, {Bailer-Jones}, {Biermann},
  {Evans}, {Eyer}, {Jansen}, {Jordi}, {Klioner}, {Lammers}, {Lindegren},
  {Luri}, {Mignard}, {Panem}, {Pourbaix}, {Randich}, {Sartoretti}, {Siddiqui},
  {Soubiran}, {van Leeuwen}, {Walton}, {Arenou}, {Bastian}, {Cropper},
  {Drimmel}, {Katz}, {Lattanzi}, {Bakker}, {Cacciari}, {Casta{\~n}eda},
  {Chaoul}, {Cheek}, {De Angeli}, {Fabricius}, {Guerra}, {Holl}, {Masana},
  {Messineo}, {Mowlavi}, {Nienartowicz}, {Panuzzo}, {Portell}, {Riello},
  {Seabroke}, {Tanga}, {Th{\'e}venin}, {Gracia-Abril}, {Comoretto},
  {Garcia-Reinaldos}, {Teyssier}, {Altmann}, {Andrae}, {Audard},
  {Bellas-Velidis}, {Benson}, {Berthier}, {Blomme}, {Burgess}, {Busso},
  {Carry}, {Cellino}, {Clementini}, {Clotet}, {Creevey}, {Davidson}, {De
  Ridder}, {Delchambre}, {Dell'Oro}, {Ducourant}, {Fern{\'a}ndez-
  Hern{\'a}ndez}, {Fouesneau}, {Fr{\'e}mat}, {Galluccio}, {Garc{\'\i}a-Torres},
  {Gonz{\'a}lez-N{\'u}{\~n}ez}, {Gonz{\'a}lez-Vidal}, {Gosset}, {Guy},
  {Halbwachs}, {Hambly}, {Harrison}, {Hern{\'a}ndez}, {Hestroffer}, {Hodgkin},
  {Hutton}, {Jasniewicz}, {Jean-Antoine-Piccolo}, {Jordan}, {Korn},
  {Krone-Martins}, {Lanzafame}, {Lebzelter}, {L{\"o}ffler}, {Manteiga},
  {Marrese}, {Mart{\'\i}n-Fleitas}, {Moitinho}, {Mora}, {Muinonen}, {Osinde},
  {Pancino}, {Pauwels}, {Petit}, {Recio-Blanco}, {Richards}, {Rimoldini},
  {Robin}, {Sarro}, {Siopis}, {Smith}, {Sozzetti}, {S{\"u}veges}, {Torra}, {van
  Reeven}, {Abbas}, {Abreu Aramburu}, {Accart}, {Aerts}, {Altavilla},
  {{\'A}lvarez}, {Alvarez}, {Alves}, {Anderson}, {Andrei}, {Anglada Varela},
  {Antiche}, {Antoja}, {Arcay}, {Astraatmadja}, {Bach}, {Baker},
  {Balaguer-N{\'u}{\~n}ez}, {Balm}, {Barache}, {Barata}, {Barbato}, {Barblan},
  {Barklem}, {Barrado}, {Barros}, {Barstow}, {Bartholom{\'e} Mu{\~n}oz},
  {Bassilana}, {Becciani}, {Bellazzini}, {Berihuete}, {Bertone}, {Bianchi},
  {Bienaym{\'e}}, {Blanco-Cuaresma}, {Boch}, {Boeche}, {Bombrun}, {Borrachero},
  {Bossini}, {Bouquillon}, {Bourda}, {Bragaglia}, {Bramante}, {Breddels},
  {Bressan}, {Brouillet}, {Br{\"u}semeister}, {Brugaletta}, {Bucciarelli},
  {Burlacu}, {Busonero}, {Butkevich}, {Buzzi}, {Caffau}, {Cancelliere},
  {Cannizzaro}, {Cantat-Gaudin}, {Carballo}, {Carlucci}, {Carrasco},
  {Casamiquela}, {Castellani}, {Castro-Ginard}, {Charlot}, {Chemin},
  {Chiavassa}, {Cocozza}, {Costigan}, {Cowell}, {Crifo}, {Crosta}, {Crowley},
  {Cuypers}, {Dafonte}, {Damerdji}, {Dapergolas}, {David}, {David}, {de
  Laverny}, {De Luise}, {De March}, {de Martino}, {de Souza}, {de Torres},
  {Debosscher}, {del Pozo}, {Delbo}, {Delgado}, {Delgado}, {Di Matteo},
  {Diakite}, {Diener}, {Distefano}, {Dolding}, {Drazinos}, {Dur{\'a}n},
  {Edvardsson}, {Enke}, {Eriksson}, {Esquej}, {Eynard Bontemps}, {Fabre},
  {Fabrizio}, {Faigler}, {Falc{\~a}o}, {Farr{\`a}s Casas}, {Federici},
  {Fedorets}, {Fernique}, {Figueras}, {Filippi}, {Findeisen}, {Fonti},
  {Fraile}, {Fraser}, {Fr{\'e}zouls}, {Gai}, {Galleti}, {Garabato},
  {Garc{\'\i}a-Sedano}, {Garofalo}, {Garralda}, {Gavel}, {Gavras}, {Gerssen},
  {Geyer}, {Giacobbe}, {Gilmore}, {Girona}, {Giuffrida}, {Glass}, {Gomes},
  {Granvik}, {Gueguen}, {Guerrier}, {Guiraud}, {Guti{\'e}rrez-S{\'a}nchez},
  {Haigron}, {Hatzidimitriou}, {Hauser}, {Haywood}, {Heiter}, {Helmi}, {Heu},
  {Hilger}, {Hobbs}, {Hofmann}, {Holland}, {Huckle}, {Hypki}, {Icardi},
  {Jan{\ss}en}, {Jevardat de Fombelle}, {Jonker}, {Juh{\'a}sz}, {Julbe},
  {Karampelas}, {Kewley}, {Klar}, {Kochoska}, {Kohley}, {Kolenberg},
  {Kontizas}, {Kontizas}, {Koposov}, {Kordopatis}, {Kostrzewa-Rutkowska},
  {Koubsky}, {Lambert}, {Lanza}, {Lasne}, {Lavigne}, {Le Fustec}, {Le
  Poncin-Lafitte}, {Lebreton}, {Leccia}, {Leclerc}, {Lecoeur-Taibi},
  {Lenhardt}, {Leroux}, {Liao}, {Licata}, {Lindstr{\o}m}, {Lister}, {Livanou},
  {Lobel}, {L{\'o}pez}, {Managau}, {Mann}, {Mantelet}, {Marchal}, {Marchant},
  {Marconi}, {Marinoni}, {Marschalk{\'o}}, {Marshall}, {Martino}, {Marton},
  {Mary}, {Massari}, {Matijevi{\v{c}}}, {Mazeh}, {McMillan}, {Messina},
  {Michalik}, {Millar}, {Molina}, {Molinaro}, {Moln{\'a}r}, {Montegriffo},
  {Mor}, {Morbidelli}, {Morel}, {Morris}, {Mulone}, {Muraveva}, {Musella},
  {Nelemans}, {Nicastro}, {Noval}, {O'Mullane}, {Ord{\'e}novic},
  {Ord{\'o}{\~n}ez-Blanco}, {Osborne}, {Pagani}, {Pagano}, {Pailler},
  {Palacin}, {Palaversa}, {Panahi}, {Pawlak}, {Piersimoni}, {Pineau}, {Plachy},
  {Plum}, {Poggio}, {Poujoulet}, {Pr{\v{s}}a}, {Pulone}, {Racero}, {Ragaini},
  {Rambaux}, {Ramos-Lerate}, {Regibo}, {Reyl{\'e}}, {Riclet}, {Ripepi}, {Riva},
  {Rivard}, {Rixon}, {Roegiers}, {Roelens}, {Romero-G{\'o}mez}, {Rowell},
  {Royer}, {Ruiz-Dern}, {Sadowski}, {Sagrist{\`a} Sell{\'e}s}, {Sahlmann},
  {Salgado}, {Salguero}, {Sanna}, {Santana- Ros}, {Sarasso}, {Savietto},
  {Schultheis}, {Sciacca}, {Segol}, {Segovia}, {S{\'e}gransan}, {Shih},
  {Siltala}, {Silva}, {Smart}, {Smith}, {Solano}, {Solitro}, {Sordo}, {Soria
  Nieto}, {Souchay}, {Spagna}, {Spoto}, {Stampa}, {Steele},
  {Steidelm{\"u}ller}, {Stephenson}, {Stoev}, {Suess}, {Surdej}, {Szabados},
  {Szegedi-Elek}, {Tapiador}, {Taris}, {Tauran}, {Taylor}, {Teixeira},
  {Terrett}, {Teyssandier}, {Thuillot}, {Titarenko}, {Torra Clotet}, {Turon},
  {Ulla}, {Utrilla}, {Uzzi}, {Vaillant}, {Valentini}, {Valette}, {van Elteren},
  {Van Hemelryck}, {van Leeuwen}, {Vaschetto}, {Vecchiato}, {Veljanoski},
  {Viala}, {Vicente}, {Vogt}, {von Essen}, {Voss}, {Votruba}, {Voutsinas},
  {Walmsley}, {Weiler}, {Wertz}, {Wevers}, {Wyrzykowski}, {Yoldas},
  {{\v{Z}}erjal}, {Ziaeepour}, {Zorec}, {Zschocke}, {Zucker}, {Zurbach}, \&
  {Zwitter}}]{gaia2018}
{Gaia Collaboration}, {Brown}, A.~G.~A., {Vallenari}, A., {et~al.} 2018, \aap,
  616, A1, \dodoi{10.1051/0004-6361/201833051}

\bibitem[{{Gaidos} {et~al.}(2016){Gaidos}, {Mann}, {Kraus}, \&
  {Ireland}}]{Gaidos2016}
{Gaidos}, E., {Mann}, A.~W., {Kraus}, A.~L., \& {Ireland}, M. 2016, \mnras,
  457, 2877, \dodoi{10.1093/mnras/stw097}

\bibitem[{{Gaidos} {et~al.}(2014){Gaidos}, {Mann}, {L{\'e}pine}, {Buccino},
  {James}, {Ansdell}, {Petrucci}, {Mauas}, \& {Hilton}}]{Gaidos2014}
{Gaidos}, E., {Mann}, A.~W., {L{\'e}pine}, S., {et~al.} 2014, \mnras, 443, 2561

\bibitem[{{Garcia-Sage} {et~al.}(2017){Garcia-Sage}, {Glocer}, {Drake},
  {Gronoff}, \& {Cohen}}]{Garcia-Sage2017}
{Garcia-Sage}, K., {Glocer}, A., {Drake}, J.~J., {Gronoff}, G., \& {Cohen}, O.
  2017, \apjl, 844, L13, \dodoi{10.3847/2041-8213/aa7eca}

\bibitem[{{Garraffo} {et~al.}(2017){Garraffo}, {Drake}, {Cohen},
  {Alvarado-G{\'o}mez}, \& {Moschou}}]{Garraffo2017}
{Garraffo}, C., {Drake}, J.~J., {Cohen}, O., {Alvarado-G{\'o}mez}, J.~D., \&
  {Moschou}, S.~P. 2017, \apjl, 843, L33, \dodoi{10.3847/2041-8213/aa79ed}

\bibitem[{Gelman \& Rubin(1992)}]{Gelman1992}
Gelman, A., \& Rubin, D.~B. 1992, Statist. Sci., 7, 457,
  \dodoi{10.1214/ss/1177011136}

\bibitem[{{Gillon} {et~al.}(2017){Gillon}, {Triaud}, {Demory}, {Jehin}, {Agol},
  {Deck}, {Lederer}, {de Wit}, {Burdanov}, {Ingalls}, {Bolmont}, {Leconte},
  {Raymond}, {Selsis}, {Turbet}, {Barkaoui}, {Burgasser}, {Burleigh}, {Carey},
  {Chaushev}, {Copperwheat}, {Delrez}, {Fernandes}, {Holdsworth}, {Kotze}, {Van
  Grootel}, {Almleaky}, {Benkhaldoun}, {Magain}, \& {Queloz}}]{Gillon2017}
{Gillon}, M., {Triaud}, A.~H.~M.~J., {Demory}, B.-O., {et~al.} 2017, \nat, 542,
  456, \dodoi{10.1038/nature21360}

\bibitem[{{Grimm} {et~al.}(2018){Grimm}, {Demory}, {Gillon}, {Dorn}, {Agol},
  {Burdanov}, {Delrez}, {Sestovic}, {Triaud}, {Turbet}, {Bolmont}, {Caldas},
  {de Wit}, {Jehin}, {Leconte}, {Raymond}, {Van Grootel}, {Burgasser}, {Carey},
  {Fabrycky}, {Heng}, {Hernandez}, {Ingalls}, {Lederer}, {Selsis}, \&
  {Queloz}}]{grimm18}
{Grimm}, S.~L., {Demory}, B.-O., {Gillon}, M., {et~al.} 2018, \aap, 613, A68,
  \dodoi{10.1051/0004-6361/201732233}

\bibitem[{{G{\"u}nther} {et~al.}(2019){G{\"u}nther}, {Pozuelos}, {Dittmann},
  {Dragomir}, {Kane}, {Daylan}, {Feinstein}, {Huang}, {Morton}, {Bonfanti},
  {Bouma}, {Burt}, {Collins}, {Lissauer}, {Matthews}, {Montet}, {Vand erburg},
  {Wang}, {Winters}, {Ricker}, {Vanderspek}, {Latham}, {Seager}, {Winn},
  {Jenkins}, {Armstrong}, {Barkaoui}, {Batalha}, {Bean}, {Caldwell}, {Ciardi},
  {Collins}, {Crossfield}, {Fausnaugh}, {Furesz}, {Gan}, {Gillon}, {Guerrero},
  {Horne}, {Howell}, {Ireland }, {Isopi}, {Jehin}, {Kielkopf}, {Lepine},
  {Mallia}, {Matson}, {Myers}, {Palle}, {Quinn}, {Relles}, {Rojas-Ayala},
  {Schlieder}, {Sefako}, {Shporer}, {Su{\'a}rez}, {Tan}, {Ting}, {Twicken}, \&
  {Waite}}]{gunther19}
{G{\"u}nther}, M.~N., {Pozuelos}, F.~J., {Dittmann}, J.~A., {et~al.} 2019,
  Nature Astronomy, 420, \dodoi{10.1038/s41550-019-0845-5}

\bibitem[{{Hadden} {et~al.}(2018){Hadden}, {Barclay}, {Payne}, \&
  {Holman}}]{Hadden2018}
{Hadden}, S., {Barclay}, T., {Payne}, M.~J., \& {Holman}, M.~J. 2018, arXiv
  e-prints.
\newblock \doarXiv{1811.01970}

\bibitem[{{Hardegree-Ullman} {et~al.}(2019){Hardegree-Ullman}, {Cushing},
  {Muirhead}, \& {Christiansen}}]{hardegree2019}
{Hardegree-Ullman}, K.~K., {Cushing}, M.~C., {Muirhead}, P.~S., \&
  {Christiansen}, J.~L. 2019, \aj, 158, 75, \dodoi{10.3847/1538-3881/ab21d2}

\bibitem[{{Hedges} {et~al.}(2019){Hedges}, {Saunders}, {Barentsen}, {Coughlin},
  {Cardoso}, {Kostov}, {Dotson}, \& {Cody}}]{hedges19}
{Hedges}, C., {Saunders}, N., {Barentsen}, G., {et~al.} 2019, \apjl, 880, L5,
  \dodoi{10.3847/2041-8213/ab2a74}

\bibitem[{{Henden} {et~al.}(2012){Henden}, {Levine}, {Terrell}, {Smith}, \&
  {Welch}}]{Henden2012}
{Henden}, A.~A., {Levine}, S.~E., {Terrell}, D., {Smith}, T.~C., \& {Welch}, D.
  2012, Journal of the American Association of Variable Star Observers
  (JAAVSO), 40, 430

\bibitem[{{Henden} {et~al.}(2016){Henden}, {Templeton}, {Terrell}, {Smith},
  {Levine}, \& {Welch}}]{Henden2016}
{Henden}, A.~A., {Templeton}, M., {Terrell}, D., {et~al.} 2016, VizieR Online
  Data Catalog, 2336

\bibitem[{Hoffman \& Gelman(2014)}]{NUTS}
Hoffman, M.~D., \& Gelman, A. 2014, Journal of Machine Learning Research, 15,
  1593.
\newblock \url{http://jmlr.org/papers/v15/hoffman14a.html}

\bibitem[{{Howell} {et~al.}(2016){Howell}, {Everett}, {Horch}, {Winters},
  {Hirsch}, {Nusdeo}, \& {Scott}}]{howell2016}
{Howell}, S.~B., {Everett}, M.~E., {Horch}, E.~P., {et~al.} 2016, \apjl, 829,
  L2, \dodoi{10.3847/2041-8205/829/1/L2}

\bibitem[{{Howell} {et~al.}(2011){Howell}, {Everett}, {Sherry}, {Horch}, \&
  {Ciardi}}]{howell2011}
{Howell}, S.~B., {Everett}, M.~E., {Sherry}, W., {Horch}, E., \& {Ciardi},
  D.~R. 2011, \aj, 142, 19, \dodoi{10.1088/0004-6256/142/1/19}

\bibitem[{{Huber} {et~al.}(2016){Huber}, {Bryson}, {Haas}, {Barclay},
  {Barentsen}, {Howell}, {Sharma}, {Stello}, \& {Thompson}}]{Huber2016}
{Huber}, D., {Bryson}, S.~T., {Haas}, M.~R., {et~al.} 2016, \apjs, 224, 2,
  \dodoi{10.3847/0067-0049/224/1/2}

\bibitem[{Hunter(2007)}]{matplotlib}
Hunter, J.~D. 2007, Computing In Science \& Engineering, 9, 90,
  \dodoi{10.1109/MCSE.2007.55}

\bibitem[{{Ikoma} {et~al.}(2001){Ikoma}, {Emori}, \& {Nakazawa}}]{Ikoma2001}
{Ikoma}, M., {Emori}, H., \& {Nakazawa}, K. 2001, \apj, 553, 999,
  \dodoi{10.1086/320954}

\bibitem[{{Jao} {et~al.}(2003){Jao}, {Henry}, {Subasavage}, {Bean}, {Costa},
  {Ianna}, \& {M{\'e}ndez}}]{Jao2003}
{Jao}, W.-C., {Henry}, T.~J., {Subasavage}, J.~P., {et~al.} 2003, \aj, 125,
  332, \dodoi{10.1086/345515}

\bibitem[{{Jao} {et~al.}(2005){Jao}, {Henry}, {Subasavage}, {Brown}, {Ianna},
  {Bartlett}, {Costa}, \& {M{\'e}ndez}}]{Jao2005}
---. 2005, \aj, 129, 1954, \dodoi{10.1086/428489}

\bibitem[{{Jenkins} {et~al.}(2002){Jenkins}, {Caldwell}, \&
  {Borucki}}]{Jenkins2002}
{Jenkins}, J.~M., {Caldwell}, D.~A., \& {Borucki}, W.~J. 2002, \apj, 564, 495,
  \dodoi{10.1086/324143}

\bibitem[{{Jenkins} {et~al.}(2016){Jenkins}, {Twicken}, {McCauliff},
  {Campbell}, {Sanderfer}, {Lung}, {Mansouri-Samani}, {Girouard}, {Tenenbaum},
  {Klaus}, {Smith}, {Caldwell}, {Chacon}, {Henze}, {Heiges}, {Latham},
  {Morgan}, {Swade}, {Rinehart}, \& {Vanderspek}}]{Jenkins2016}
{Jenkins}, J.~M., {Twicken}, J.~D., {McCauliff}, S., {et~al.} 2016, in
  \procspie, Vol. 9913, Software and Cyberinfrastructure for Astronomy IV,
  99133E, \dodoi{10.1117/12.2233418}

\bibitem[{{Jensen}(2013)}]{Jensen:2013}
{Jensen}, E. 2013, {Tapir: A web interface for transit/eclipse observability},
  Astrophysics Source Code Library.
\newblock \doeprint{1306.007}

\bibitem[{{Johnson} \& {Soderblom}(1987)}]{johnson1987}
{Johnson}, D. R.~H., \& {Soderblom}, D.~R. 1987, \aj, 93, 864,
  \dodoi{10.1086/114370}

\bibitem[{{Kane} {et~al.}(2014){Kane}, {Kopparapu}, \&
  {Domagal-Goldman}}]{Kane2014}
{Kane}, S.~R., {Kopparapu}, R.~K., \& {Domagal-Goldman}, S.~D. 2014, \apjl,
  794, L5, \dodoi{10.1088/2041-8205/794/1/L5}

\bibitem[{{Kempton} {et~al.}(2018){Kempton}, {Bean}, {Louie}, {Deming}, {Koll},
  {Mansfield}, {Christiansen}, {L{\'o}pez-Morales}, {Swain}, {Zellem},
  {Ballard}, {Barclay}, {Barstow}, {Batalha}, {Beatty}, {Berta-Thompson},
  {Birkby}, {Buchhave}, {Charbonneau}, {Cowan}, {Crossfield}, {de Val-Borro},
  {Doyon}, {Dragomir}, {Gaidos}, {Heng}, {Hu}, {Kane}, {Kreidberg}, {Mallonn},
  {Morley}, {Narita}, {Nascimbeni}, {Pall{\'e}}, {Quintana}, {Rauscher},
  {Seager}, {Shkolnik}, {Sing}, {Sozzetti}, {Stassun}, {Valenti}, \& {von
  Essen}}]{Kempton2018}
{Kempton}, E.~M.-R., {Bean}, J.~L., {Louie}, D.~R., {et~al.} 2018, \pasp, 130,
  114401, \dodoi{10.1088/1538-3873/aadf6f}

\bibitem[{{Khodachenko} {et~al.}(2007){Khodachenko}, {Ribas}, {Lammer},
  {Grie{\ss}meier}, {Leitner}, {Selsis}, {Eiroa}, {Hanslmeier}, {Biernat},
  {Farrugia}, \& {Rucker}}]{Khodachenko2007}
{Khodachenko}, M.~L., {Ribas}, I., {Lammer}, H., {et~al.} 2007, Astrobiology,
  7, 167, \dodoi{10.1089/ast.2006.0127}

\bibitem[{{Kipping}(2013{\natexlab{a}})}]{exoplanet:kipping13}
{Kipping}, D.~M. 2013{\natexlab{a}}, \mnras, 435, 2152,
  \dodoi{10.1093/mnras/stt1435}

\bibitem[{{Kipping}(2013{\natexlab{b}})}]{Kipping2013}
---. 2013{\natexlab{b}}, \mnras, 434, L51, \dodoi{10.1093/mnrasl/slt075}

\bibitem[{{Kiraga} \& {Stepien}(2007)}]{Kiraga2007}
{Kiraga}, M., \& {Stepien}, K. 2007, \actaa, 57, 149.
\newblock \doarXiv{0707.2577}

\bibitem[{{Kirkpatrick} {et~al.}(1991){Kirkpatrick}, {Henry}, \&
  {McCarthy}}]{Kirkpatrick1991}
{Kirkpatrick}, J.~D., {Henry}, T.~J., \& {McCarthy}, Donald~W., J. 1991, \apjs,
  77, 417, \dodoi{10.1086/191611}

\bibitem[{Kluyver {et~al.}(2016)Kluyver, Ragan-Kelley, P{\'e}rez, Granger,
  Bussonnier, Frederic, Kelley, Hamrick, Grout, Corlay, Ivanov, Avila, Abdalla,
  Willing, \& development~team [Unknown]}]{jupyer}
Kluyver, T., Ragan-Kelley, B., P{\'e}rez, F., {et~al.} 2016, in Positioning and
  Power in Academic Publishing: Players, Agents and Agendas, ed. F.~Loizides \&
  B.~Scmidt (IOS Press), 87--90.
\newblock \url{https://eprints.soton.ac.uk/403913/}

\bibitem[{{Kochanek} {et~al.}(2017){Kochanek}, {Shappee}, {Stanek}, {Holoien},
  {Thompson}, {Prieto}, {Dong}, {Shields}, {Will}, {Britt}, {Perzanowski}, \&
  {Pojma{\'n}ski}}]{Kochanek17}
{Kochanek}, C.~S., {Shappee}, B.~J., {Stanek}, K.~Z., {et~al.} 2017, \pasp,
  129, 104502, \dodoi{10.1088/1538-3873/aa80d9}

\bibitem[{{Kokubo} \& {Ida}(1998)}]{Kokubo1998}
{Kokubo}, E., \& {Ida}, S. 1998, \icarus, 131, 171,
  \dodoi{10.1006/icar.1997.5840}

\bibitem[{{Kopparapu} {et~al.}(2013){Kopparapu}, {Ramirez}, {Kasting}, {Eymet},
  {Robinson}, {Mahadevan}, {Terrien}, {Domagal-Goldman}, {Meadows}, \&
  {Deshpande}}]{kopparapu2013}
{Kopparapu}, R.~K., {Ramirez}, R., {Kasting}, J.~F., {et~al.} 2013, \apj, 765,
  131, \dodoi{10.1088/0004-637X/765/2/131}

\bibitem[{{Kostov} {et~al.}(2019{\natexlab{a}}){Kostov}, {Schlieder},
  {Barclay}, {Quintana}, {Col{\'o}n}, {Brand e}, {Collins}, {Feinstein},
  {Hadden}, {Kane}, {Kreidberg}, {Kruse}, {Lam}, {Matthews}, {Montet},
  {Pozuelos}, {Stassun}, {Winters}, {Ricker}, {Vanderspek}, {Latham}, {Seager},
  {Winn}, {Jenkins}, {Afanasev}, {Armstrong}, {Arney}, {Boyd}, {Barentsen},
  {Barkaoui}, {Batalha}, {Beichman}, {Bayliss}, {Burke}, {Burdanov},
  {Cacciapuoti}, {Carson}, {Charbonneau}, {Christiansen}, {Ciardi}, {Clampin},
  {Collins}, {Conti}, {Coughlin}, {Covone}, {Crossfield}, {Delrez},
  {Domagal-Goldman}, {Dressing}, {Ducrot}, {Essack}, {Everett}, {Fauchez},
  {Foreman-Mackey}, {Gan}, {Gilbert}, {Gillon}, {Gonzales}, {Hamann}, {Hedges},
  {Hocutt}, {Hoffman}, {Horch}, {Horne}, {Howell}, {Hynes}, {Ireland },
  {Irwin}, {Isopi}, {Jensen}, {Jehin}, {Kaltenegger}, {Kielkopf}, {Kopparapu},
  {Lewis}, {Lopez}, {Lissauer}, {Mann}, {Mallia}, {Mandell}, {Matson}, {Mazeh},
  {Monsue}, {Moran}, {Moran}, {Morley}, {Morris}, {Muirhead}, {Mukai},
  {Mullally}, {Mullally}, {Murray}, {Narita}, {Palle}, {Pidhorodetska},
  {Quinn}, {Relles}, {Rinehart}, {Ritsko}, {Rodriguez}, {Rowden}, {Rowe},
  {Sebastian}, {Sefako}, {Shahaf}, {Shporer}, {Ta{\~n}{\'o}n Reyes},
  {Tenenbaum}, {Ting}, {Twicken}, {van Belle}, {Vega}, {Volosin}, {Walkowicz},
  \& {Youngblood}}]{kostov2019b}
{Kostov}, V.~B., {Schlieder}, J.~E., {Barclay}, T., {et~al.}
  2019{\natexlab{a}}, \aj, 158, 32, \dodoi{10.3847/1538-3881/ab2459}

\bibitem[{{Kostov} {et~al.}(2019{\natexlab{b}}){Kostov}, {Mullally},
  {Quintana}, {Coughlin}, {Mullally}, {Barclay}, {Colon}, {Schlieder},
  {Barentsen}, \& {Burke}}]{Kostov2019}
{Kostov}, V.~B., {Mullally}, S.~E., {Quintana}, E.~V., {et~al.}
  2019{\natexlab{b}}, arXiv e-prints, arXiv:1901.07459.
\newblock \doarXiv{1901.07459}

\bibitem[{{Kreidberg} {et~al.}(2019){Kreidberg}, {Koll}, {Morley}, {Hu},
  {Schaefer}, {Deming}, {Stevenson}, {Dittmann}, {Vanderburg}, {Berardo},
  {Guo}, {Stassun}, {Crossfield}, {Charbonneau}, {Latham}, {Loeb}, {Ricker},
  {Seager}, \& {Vand erspek}}]{Kreidberg2019}
{Kreidberg}, L., {Koll}, D. D.~B., {Morley}, C., {et~al.} 2019, \nat, 573, 87,
  \dodoi{10.1038/s41586-019-1497-4}

\bibitem[{{Krishnamurthy} {et~al.}(2019){Krishnamurthy}, {Villasenor},
  {Seager}, {Ricker}, \& {Vanderspek}}]{Krishnamurthy2019}
{Krishnamurthy}, A., {Villasenor}, J., {Seager}, S., {Ricker}, G., \&
  {Vanderspek}, R. 2019, Acta Astronautica, 160, 46,
  \dodoi{10.1016/j.actaastro.2019.04.016}

\bibitem[{{Kruse} {et~al.}(2019){Kruse}, {Agol}, {Luger}, \&
  {Foreman-Mackey}}]{kruse19}
{Kruse}, E., {Agol}, E., {Luger}, R., \& {Foreman-Mackey}, D. 2019, \apjs, 244,
  11, \dodoi{10.3847/1538-4365/ab346b}

\bibitem[{{Kunder} {et~al.}(2017){Kunder}, {Kordopatis}, {Steinmetz},
  {Zwitter}, {McMillan}, {Casagrande}, {Enke}, {Wojno}, {Valentini},
  {Chiappini}, {Matijevi{\v{c}}}, {Siviero}, {de Laverny}, {Recio-Blanco},
  {Bijaoui}, {Wyse}, {Binney}, {Grebel}, {Helmi}, {Jofre}, {Antoja}, {Gilmore},
  {Siebert}, {Famaey}, {Bienaym{\'e}}, {Gibson}, {Freeman}, {Navarro},
  {Munari}, {Seabroke}, {Anguiano}, {{\v{Z}}erjal}, {Minchev}, {Reid},
  {Bland-Hawthorn}, {Kos}, {Sharma}, {Watson}, {Parker}, {Scholz}, {Burton},
  {Cass}, {Hartley}, {Fiegert}, {Stupar}, {Ritter}, {Hawkins}, {Gerhard},
  {Chaplin}, {Davies}, {Elsworth}, {Lund}, {Miglio}, \&
  {Mosser}}]{Kunder(2017)}
{Kunder}, A., {Kordopatis}, G., {Steinmetz}, M., {et~al.} 2017, \aj, 153, 75,
  \dodoi{10.3847/1538-3881/153/2/75}

\bibitem[{{Lambrechts} \& {Johansen}(2014)}]{Lambrechts2014}
{Lambrechts}, M., \& {Johansen}, A. 2014, \aap, 572, A107,
  \dodoi{10.1051/0004-6361/201424343}

\bibitem[{{Lammer} {et~al.}(2003){Lammer}, {Selsis}, {Ribas}, {Guinan},
  {Bauer}, \& {Weiss}}]{Lammer2003}
{Lammer}, H., {Selsis}, F., {Ribas}, I., {et~al.} 2003, \apjl, 598, L121,
  \dodoi{10.1086/380815}

\bibitem[{{Lammer} {et~al.}(2007){Lammer}, {Lichtenegger}, {Kulikov},
  {Grie{\ss}meier}, {Terada}, {Erkaev}, {Biernat}, {Khodachenko}, {Ribas},
  {Penz}, \& {Selsis}}]{Lammer2007}
{Lammer}, H., {Lichtenegger}, H. I.~M., {Kulikov}, Y.~N., {et~al.} 2007,
  Astrobiology, 7, 185, \dodoi{10.1089/ast.2006.0128}

\bibitem[{{Lanza} {et~al.}(2019){Lanza}, {Collier Cameron}, \&
  {Haywood}}]{lanza19}
{Lanza}, A.~F., {Collier Cameron}, A., \& {Haywood}, R.~D. 2019, \mnras, 486,
  3459, \dodoi{10.1093/mnras/stz1055}

\bibitem[{{Lasker}(1994)}]{lasker1994}
{Lasker}, B.~M. 1994, in IAU Symposium, Vol. 161, Astronomy from Wide-Field
  Imaging, ed. H.~T. {MacGillivray}, 167

\bibitem[{{Lasker} {et~al.}(1990){Lasker}, {Sturch}, {McLean}, {Russell},
  {Jenkner}, \& {Shara}}]{lasker1990}
{Lasker}, B.~M., {Sturch}, C.~R., {McLean}, B.~J., {et~al.} 1990, \aj, 99,
  2019, \dodoi{10.1086/115483}

\bibitem[{{Latham} {et~al.}(2011){Latham}, {Rowe}, {Quinn}, {Batalha},
  {Borucki}, {Brown}, {Bryson}, {Buchhave}, {Caldwell}, {Carter},
  {Christiansen}, {Ciardi}, {Cochran}, {Dunham}, {Fabrycky}, {Ford}, {Gautier},
  {Gilliland}, {Holman}, {Howell}, {Ibrahim}, {Isaacson}, {Jenkins}, {Koch},
  {Lissauer}, {Marcy}, {Quintana}, {Ragozzine}, {Sasselov}, {Shporer},
  {Steffen}, {Welsh}, \& {Wohler}}]{latham2011}
{Latham}, D.~W., {Rowe}, J.~F., {Quinn}, S.~N., {et~al.} 2011, \apjl, 732, L24,
  \dodoi{10.1088/2041-8205/732/2/L24}

\bibitem[{{Li} {et~al.}(2019){Li}, {Tenenbaum}, {Twicken}, {Burke}, {Jenkins},
  {Quintana}, {Rowe}, \& {Seader}}]{Li2019}
{Li}, J., {Tenenbaum}, P., {Twicken}, J.~D., {et~al.} 2019, \pasp, 131, 024506,
  \dodoi{10.1088/1538-3873/aaf44d}

\bibitem[{{Lightkurve Collaboration} {et~al.}(2018){Lightkurve Collaboration},
  {Cardoso}, {Hedges}, {Gully-Santiago}, {Saunders}, {Cody}, {Barclay}, {Hall},
  {Sagear}, {Turtelboom}, {Zhang}, {Tzanidakis}, {Mighell}, {Coughlin}, {Bell},
  {Berta- Thompson}, {Williams}, {Dotson}, \& {Barentsen}}]{lightkurve}
{Lightkurve Collaboration}, {Cardoso}, J. V. d.~M., {Hedges}, C., {et~al.}
  2018, {Lightkurve: Kepler and TESS time series analysis in Python}.
\newblock \doeprint{1812.013}

\bibitem[{{Lindegren} {et~al.}(2018){Lindegren}, {Hern{\'a}ndez}, {Bombrun},
  {Klioner}, {Bastian}, {Ramos-Lerate}, {de Torres}, {Steidelm{\"u}ller},
  {Stephenson}, {Hobbs}, {Lammers}, {Biermann}, {Geyer}, {Hilger}, {Michalik},
  {Stampa}, {McMillan}, {Casta{\~n}eda}, {Clotet}, {Comoretto}, {Davidson},
  {Fabricius}, {Gracia}, {Hambly}, {Hutton}, {Mora}, {Portell}, {van Leeuwen},
  {Abbas}, {Abreu}, {Altmann}, {Andrei}, {Anglada}, {Balaguer-N{\'u}{\~n}ez},
  {Barache}, {Becciani}, {Bertone}, {Bianchi}, {Bouquillon}, {Bourda},
  {Br{\"u}semeister}, {Bucciarelli}, {Busonero}, {Buzzi}, {Cancelliere},
  {Carlucci}, {Charlot}, {Cheek}, {Crosta}, {Crowley}, {de Bruijne}, {de
  Felice}, {Drimmel}, {Esquej}, {Fienga}, {Fraile}, {Gai}, {Garralda},
  {Gonz{\'a}lez-Vidal}, {Guerra}, {Hauser}, {Hofmann}, {Holl}, {Jordan},
  {Lattanzi}, {Lenhardt}, {Liao}, {Licata}, {Lister}, {L{\"o}ffler},
  {Marchant}, {Martin-Fleitas}, {Messineo}, {Mignard}, {Morbidelli}, {Poggio},
  {Riva}, {Rowell}, {Salguero}, {Sarasso}, {Sciacca}, {Siddiqui}, {Smart},
  {Spagna}, {Steele}, {Taris}, {Torra}, {van Elteren}, {van Reeven}, \&
  {Vecchiato}}]{GaiaDr2}
{Lindegren}, L., {Hern{\'a}ndez}, J., {Bombrun}, A., {et~al.} 2018, \aap, 616,
  A2

\bibitem[{{Lissauer}(1987)}]{Lissauer1987}
{Lissauer}, J.~J. 1987, \icarus, 69, 249, \dodoi{10.1016/0019-1035(87)90104-7}

\bibitem[{{Lissauer}(2007)}]{Lissauer2007}
---. 2007, \apjl, 660, L149, \dodoi{10.1086/518121}

\bibitem[{{Lissauer} {et~al.}(2012){Lissauer}, {Marcy}, {Rowe}, {Bryson},
  {Adams}, {Buchhave}, {Ciardi}, {Cochran}, {Fabrycky}, {Ford}, {Fressin},
  {Geary}, {Gilliland}, {Holman}, {Howell}, {Jenkins}, {Kinemuchi}, {Koch},
  {Morehead}, {Ragozzine}, {Seader}, {Tanenbaum}, {Torres}, \&
  {Twicken}}]{Lissauer2012}
{Lissauer}, J.~J., {Marcy}, G.~W., {Rowe}, J.~F., {et~al.} 2012, \apj, 750,
  112, \dodoi{10.1088/0004-637X/750/2/112}

\bibitem[{{Lissauer} {et~al.}(2014){Lissauer}, {Marcy}, {Bryson}, {Rowe},
  {Jontof-Hutter}, {Agol}, {Borucki}, {Carter}, {Ford}, {Gilliland}, {Kolbl},
  {Star}, {Steffen}, \& {Torres}}]{Lissauer2014}
{Lissauer}, J.~J., {Marcy}, G.~W., {Bryson}, S.~T., {et~al.} 2014, \apj, 784,
  44, \dodoi{10.1088/0004-637X/784/1/44}

\bibitem[{{Lithwick} \& {Wu}(2011)}]{lithwick11}
{Lithwick}, Y., \& {Wu}, Y. 2011, \apj, 739, 31,
  \dodoi{10.1088/0004-637X/739/1/31}

\bibitem[{{Lopez}(2017)}]{Lopez2017}
{Lopez}, E.~D. 2017, MNRAS, 472, 245, \dodoi{10.1093/mnras/stx1558}

\bibitem[{{Lopez} \& {Fortney}(2013)}]{Lopez2013}
{Lopez}, E.~D., \& {Fortney}, J.~J. 2013, ApJ, 776, 2,
  \dodoi{10.1088/0004-637X/776/1/2}

\bibitem[{{Lopez} \& {Fortney}(2014)}]{Lopez2014}
---. 2014, ApJ, 792, 1, \dodoi{10.1088/0004-637X/792/1/1}

\bibitem[{{Lopez} {et~al.}(2012){Lopez}, {Fortney}, \& {Miller}}]{Lopez2012}
{Lopez}, E.~D., {Fortney}, J.~J., \& {Miller}, N. 2012, ApJ, 761, 59,
  \dodoi{10.1088/0004-637X/761/1/59}

\bibitem[{{Luger} {et~al.}(2018){Luger}, {Agol}, {Foreman-Mackey}, {Fleming},
  {Lustig-Yaeger}, \& {Deitrick}}]{exoplanet:luger18}
{Luger}, R., {Agol}, E., {Foreman-Mackey}, D., {et~al.} 2018, ArXiv e-prints

\bibitem[{{Luger} {et~al.}(2016){Luger}, {Agol}, {Kruse}, {Barnes}, {Becker},
  {Foreman-Mackey}, \& {Deming}}]{Luger2016}
{Luger}, R., {Agol}, E., {Kruse}, E., {et~al.} 2016, \aj, 152, 100,
  \dodoi{10.3847/0004-6256/152/4/100}

\bibitem[{{Luger} {et~al.}(2017){Luger}, {Sestovic}, {Kruse}, {Grimm},
  {Demory}, {Agol}, {Bolmont}, {Fabrycky}, {Fernandes}, {Van Grootel},
  {Burgasser}, {Gillon}, {Ingalls}, {Jehin}, {Raymond}, {Selsis}, {Triaud},
  {Barclay}, {Barentsen}, {Howell}, {Delrez}, {de Wit}, {Foreman-Mackey},
  {Holdsworth}, {Leconte}, {Lederer}, {Turbet}, {Almleaky}, {Benkhaldoun},
  {Magain}, {Morris}, {Heng}, \& {Queloz}}]{Luger2017}
{Luger}, R., {Sestovic}, M., {Kruse}, E., {et~al.} 2017, Nature Astronomy, 1,
  0129, \dodoi{10.1038/s41550-017-0129}

\bibitem[{{Luque} {et~al.}(2019){Luque}, {Pall{\'e}}, {Kossakowski},
  {Dreizler}, {Kemmer}, {Espinoza}, {Burt}, {Anglada-Escud{\'e}}, {B{\'e}jar},
  {Caballero}, {Collins}, {Collins}, {Cort{\'e}s-Contreras},
  {D{\'\i}ez-Alonso}, {Feng}, {Hatzes}, {Hellier}, {Henning}, {Jeffers},
  {Kaltenegger}, {K{\"u}rster}, {Madden}, {Molaverdikhani}, {Montes}, {Narita},
  {Nowak}, {Ofir}, {Oshagh}, {Parviainen}, {Quirrenbach}, {Reffert}, {Reiners},
  {Rodr{\'\i}guez-L{\'o}pez}, {Schlecker}, {Stock}, {Trifonov}, {Winn},
  {Zapatero Osorio}, {Zechmeister}, {Amado}, {Anderson}, {Batalha}, {Bauer},
  {Bluhm}, {Burke}, {Butler}, {Caldwell}, {Chen}, {Crane}, {Dragomir},
  {Dressing}, {Dynes}, {Jenkins}, {Kaminski}, {Klahr}, {Kotani}, {Lafarga},
  {Latham}, {Lewin}, {McDermott}, {Monta{\~n}{\'e}s-Rodr{\'\i}guez}, {Morales},
  {Murgas}, {Nagel}, {Pedraz}, {Ribas}, {Ricker}, {Rowden}, {Seager},
  {Shectman}, {Tamura}, {Teske}, {Twicken}, {Vanderspeck}, {Wang}, \&
  {Wohler}}]{luque19}
{Luque}, R., {Pall{\'e}}, E., {Kossakowski}, D., {et~al.} 2019, \aap, 628, A39,
  \dodoi{10.1051/0004-6361/201935801}

\bibitem[{{Mann} {et~al.}(2013){Mann}, {Brewer}, {Gaidos}, {L{\'e}pine}, \&
  {Hilton}}]{Mann2013a}
{Mann}, A.~W., {Brewer}, J.~M., {Gaidos}, E., {L{\'e}pine}, S., \& {Hilton},
  E.~J. 2013, \aj, 145, 52

\bibitem[{{Mann} {et~al.}(2015){Mann}, {Feiden}, {Gaidos}, {Boyajian}, \& {von
  Braun}}]{Mann2015}
{Mann}, A.~W., {Feiden}, G.~A., {Gaidos}, E., {Boyajian}, T., \& {von Braun},
  K. 2015, \apj, 804, 64, \dodoi{10.1088/0004-637X/804/1/64}

\bibitem[{{Mann} {et~al.}(2019){Mann}, {Dupuy}, {Kraus}, {Gaidos}, {Ansdell},
  {Ireland}, {Rizzuto}, {Hung}, {Dittmann}, {Factor}, {Feiden}, {Martinez},
  {Ru{\'{\i}}z-Rodr{\'{\i}}guez}, \& {Thao}}]{mann19}
{Mann}, A.~W., {Dupuy}, T., {Kraus}, A.~L., {et~al.} 2019, \apj, 871, 63,
  \dodoi{10.3847/1538-4357/aaf3bc}

\bibitem[{{Marcy} {et~al.}(1998){Marcy}, {Butler}, {Vogt}, {Fischer}, \&
  {Lissauer}}]{Marcy1998}
{Marcy}, G.~W., {Butler}, R.~P., {Vogt}, S.~S., {Fischer}, D., \& {Lissauer},
  J.~J. 1998, \apjl, 505, L147, \dodoi{10.1086/311623}

\bibitem[{{Matson} {et~al.}(2018){Matson}, {Howell}, {Horch}, \&
  {Everett}}]{matson2018}
{Matson}, R.~A., {Howell}, S.~B., {Horch}, E.~P., \& {Everett}, M.~E. 2018,
  \aj, 156, 31, \dodoi{10.3847/1538-3881/aac778}

\bibitem[{{Mayor} {et~al.}(2003){Mayor}, {Pepe}, {Queloz}, {Bouchy},
  {Rupprecht}, {Lo Curto}, {Avila}, {Benz}, {Bertaux}, {Bonfils}, {Dall},
  {Dekker}, {Delabre}, {Eckert}, {Fleury}, {Gilliotte}, {Gojak}, {Guzman},
  {Kohler}, {Lizon}, {Longinotti}, {Lovis}, {Megevand}, {Pasquini}, {Reyes},
  {Sivan}, {Sosnowska}, {Soto}, {Udry}, {van Kesteren}, {Weber}, \&
  {Weilenmann}}]{mayor03}
{Mayor}, M., {Pepe}, F., {Queloz}, D., {et~al.} 2003, The Messenger, 114, 20

\bibitem[{{McDonald} {et~al.}(2019){McDonald}, {Kreidberg}, \&
  {Lopez}}]{McDonald2019}
{McDonald}, G.~D., {Kreidberg}, L., \& {Lopez}, E. 2019, \apj, 876, 22,
  \dodoi{10.3847/1538-4357/ab1095}

\bibitem[{McKinney(2010)}]{pandas}
McKinney, W. 2010, in Proceedings of the 9th Python in Science Conference, ed.
  S.~van~der Walt \& J.~Millman, 51 -- 56

\bibitem[{{Millholland} {et~al.}(2017){Millholland}, {Wang}, \&
  {Laughlin}}]{Millholland2017}
{Millholland}, S., {Wang}, S., \& {Laughlin}, G. 2017, \apjl, 849, L33,
  \dodoi{10.3847/2041-8213/aa9714}

\bibitem[{{Monet} {et~al.}(2003){Monet}, {Levine}, {Canzian}, {Ables}, {Bird},
  {Dahn}, {Guetter}, {Harris}, {Henden}, {Leggett}, {Levison}, {Luginbuhl},
  {Martini}, {Monet}, {Munn}, {Pier}, {Rhodes}, {Riepe}, {Sell}, {Stone},
  {Vrba}, {Walker}, {Westerhout}, {Brucato}, {Reid}, {Schoening}, {Hartley},
  {Read}, \& {Tritton}}]{monet2003}
{Monet}, D.~G., {Levine}, S.~E., {Canzian}, B., {et~al.} 2003, \aj, 125, 984,
  \dodoi{10.1086/345888}

\bibitem[{{Morgan} {et~al.}(1992){Morgan}, {Tritton}, {Savage}, {Hartley}, \&
  {Cannon}}]{morgan1992}
{Morgan}, D.~H., {Tritton}, S.~B., {Savage}, A., {Hartley}, M., \& {Cannon},
  R.~D. 1992, Astrophysics and Space Science Library, Vol. 174, {Current and
  Future Programmes with the UK Schmidt Telescope}, ed. H.~T. {MacGillivray} \&
  E.~B. {Thomson}, 11, \dodoi{10.1007/978-94-011-2472-0_3}

\bibitem[{{Morrissey} {et~al.}(2005){Morrissey}, {Schiminovich}, {Barlow},
  {Martin}, {Blakkolb}, {Conrow}, {Cooke}, {Erickson}, {Fanson}, {Friedman},
  {Grange}, {Jelinsky}, {Lee}, {Liu}, {Mazer}, {McLean}, {Milliard}, {Randall},
  {Schmitigal}, {Sen}, {Siegmund}, {Surber}, {Vaughan}, {Viton}, {Welsh},
  {Bianchi}, {Byun}, {Donas}, {Forster}, {Heckman}, {Lee}, {Madore}, {Malina},
  {Neff}, {Rich}, {Small}, {Szalay}, \& {Wyder}}]{morrissey2005}
{Morrissey}, P., {Schiminovich}, D., {Barlow}, T.~A., {et~al.} 2005, \apjl,
  619, L7, \dodoi{10.1086/424734}

\bibitem[{{Morton}(2012)}]{Morton2012}
{Morton}, T.~D. 2012, \apj, 761, 6, \dodoi{10.1088/0004-637X/761/1/6}

\bibitem[{{Morton}(2015)}]{vespa}
---. 2015, {VESPA: False positive probabilities calculator}, Astrophysics
  Source Code Library.
\newblock \doeprint{1503.011}

\bibitem[{{Nava} {et~al.}(2019){Nava}, {L{\'o}pez-Morales}, {Haywood}, \&
  {Giles}}]{nava19}
{Nava}, C., {L{\'o}pez-Morales}, M., {Haywood}, R.~D., \& {Giles}, H. A.~C.
  2019, arXiv e-prints, arXiv:1911.04106.
\newblock \doarXiv{1911.04106}

\bibitem[{{Neil} \& {Rogers}(2019)}]{Neil2019}
{Neil}, A.~R., \& {Rogers}, L.~A. 2019, arXiv e-prints, arXiv:1911.03582.
\newblock \doarXiv{1911.03582}

\bibitem[{{Newton} {et~al.}(2014){Newton}, {Charbonneau}, {Irwin},
  {Berta-Thompson}, {Rojas-Ayala}, {Covey}, \& {Lloyd}}]{Newton2014}
{Newton}, E.~R., {Charbonneau}, D., {Irwin}, J., {et~al.} 2014, \aj, 147, 20

\bibitem[{{Newton} {et~al.}(2017){Newton}, {Irwin}, {Charbonneau}, {Berlind},
  {Calkins}, \& {Mink}}]{Newton2017}
{Newton}, E.~R., {Irwin}, J., {Charbonneau}, D., {et~al.} 2017, \apj, 834, 85,
  \dodoi{10.3847/1538-4357/834/1/85}

\bibitem[{{Newton} {et~al.}(2016){Newton}, {Irwin}, {Charbonneau},
  {Berta-Thompson}, {Dittmann}, \& {West}}]{newton16}
---. 2016, \apj, 821, 93, \dodoi{10.3847/0004-637X/821/2/93}

\bibitem[{{Nowak} {et~al.}(2020){Nowak}, {Luque}, {Parviainen}, {Pall{\'e}},
  {Molaverdikhani}, {B{\'e}jar}, {Lillo-Box}, {Rodr{\'\i}guez-L{\'o}pez},
  {Caballero}, {Zechmeister}, {Passegger}, {Cifuentes}, {Schweitzer}, {Narita},
  {Cale}, {Espinoza}, {Murgas}, {Zapatero Osorio}, {Pozuelos}, {Aceituno},
  {Amado}, {Barkaoui}, {Barrado}, {Bauer}, {Benkhaldoun}, {Caldwell},
  {Casasayas Barris}, {Chaturvedi}, {Chen}, {Collins}, {Collins},
  {Cort{\'e}s-Contreras}, {Crossfield}, {de Le{\'o}n}, {D{\'\i}ez Alonso},
  {Dreizler}, {El Mufti}, {Esparza-Borges}, {Essack}, {Fukui}, {Gillon},
  {Guerra}, {Hatzes}, {Henning}, {Herrero}, {Hesse}, {Hirano}, {Howell},
  {Jeffers}, {Jehin}, {Jenkins}, {Kaminski}, {Kemmer}, {Kielkopf},
  {Kossakowski}, {Kotani}, {K{\"u}rster}, {Lafarga}, {Latham}, {Law},
  {Lissauer}, {Lodieu}, {Madrigal-Aguado}, {Mann}, {Massey}, {Matson},
  {Matthews}, {Monta{\~n}{\'e}s-Rodr{\'\i}guez}, {Montes}, {Morales}, {Mori},
  {Nagel}, {Oshagh}, {Pedraz}, {Plavchan}, {Pollacco}, {Quirrenbach},
  {Reffert}, {Reiners}, {Ribas}, {Rose}, {Schlecker}, {Schlieder}, {Seager},
  {Stangret}, {Stock}, {Tamura}, {Teske}, {Trifonov}, {Twicken}, {Watanabe},
  {Wittrock}, {Ziegler}, \& {Zohrabi}}]{nowak2020}
{Nowak}, G., {Luque}, R., {Parviainen}, H., {et~al.} 2020, arXiv e-prints,
  arXiv:2003.01140.
\newblock \doarXiv{2003.01140}

\bibitem[{Oliphant(2007)}]{scipy}
Oliphant, T.~E. 2007, Computing in Science Engineering, 9, 10,
  \dodoi{10.1109/MCSE.2007.58}

\bibitem[{{Ormel} {et~al.}(2017){Ormel}, {Liu}, \& {Schoonenberg}}]{Ormel2017}
{Ormel}, C.~W., {Liu}, B., \& {Schoonenberg}, D. 2017, \aap, 604, A1,
  \dodoi{10.1051/0004-6361/201730826}

\bibitem[{{Owen} \& {Jackson}(2012)}]{Owen2012}
{Owen}, J.~E., \& {Jackson}, A.~P. 2012, ArXiv e-prints.
\newblock \doarXiv{1206.2367}

\bibitem[{{Owen} \& {Mohanty}(2016)}]{Owen2016}
{Owen}, J.~E., \& {Mohanty}, S. 2016, MNRAS, 459, 4088,
  \dodoi{10.1093/mnras/stw959}

\bibitem[{{Owen} \& {Wu}(2013)}]{Owen2013}
{Owen}, J.~E., \& {Wu}, Y. 2013, ApJ, 775, 105,
  \dodoi{10.1088/0004-637X/775/2/105}

\bibitem[{{Owen} \& {Wu}(2017)}]{Owen2017}
---. 2017, \apj, 847, 29, \dodoi{10.3847/1538-4357/aa890a}

\bibitem[{{Pecaut} \& {Mamajek}(2013)}]{Pecaut(2013)}
{Pecaut}, M.~J., \& {Mamajek}, E.~E. 2013, \apjs, 208, 9,
  \dodoi{10.1088/0067-0049/208/1/9}

\bibitem[{{Penev} {et~al.}(2013){Penev}, {Bakos}, {Bayliss}, {Jord{\'a}n},
  {Mohler}, {Zhou}, {Suc}, {Rabus}, {Hartman}, {Mancini}, {B{\'e}ky}, {Csubry},
  {Buchhave}, {Henning}, {Nikolov}, {Cs{\'a}k}, {Brahm}, {Espinoza}, {Conroy},
  {Noyes}, {Sasselov}, {Schmidt}, {Wright}, {Tinney}, {Addison},
  {L{\'a}z{\'a}r}, {Papp}, \& {S{\'a}ri}}]{penev:2013:hats1}
{Penev}, K., {Bakos}, G.~{\'A}., {Bayliss}, D., {et~al.} 2013, \aj, 145, 5,
  \dodoi{10.1088/0004-6256/145/1/5}

\bibitem[{{Pepe} {et~al.}(2014){Pepe}, {Molaro}, {Cristiani}, {Rebolo},
  {Santos}, {Dekker}, {M{\'e}gevand}, {Zerbi}, {Cabral}, {Di Marcantonio},
  {Abreu}, {Affolter}, {Aliverti}, {Allende Prieto}, {Amate}, {Avila},
  {Baldini}, {Bristow}, {Broeg}, {Cirami}, {Coelho}, {Conconi}, {Coretti},
  {Cupani}, {D'Odorico}, {De Caprio}, {Delabre}, {Dorn}, {Figueira}, {Fragoso},
  {Galeotta}, {Genolet}, {Gomes}, {Gonz{\'a}lez Hern{\'a}ndez}, {Hughes},
  {Iwert}, {Kerber}, {Landoni}, {Lizon}, {Lovis}, {Maire}, {Mannetta},
  {Martins}, {Monteiro}, {Oliveira}, {Poretti}, {Rasilla}, {Riva}, {Santana
  Tschudi}, {Santos}, {Sosnowska}, {Sousa}, {Span{\'o}}, {Tenegi}, {Toso},
  {Vanzella}, {Viel}, \& {Zapatero Osorio}}]{pepe14}
{Pepe}, F., {Molaro}, P., {Cristiani}, S., {et~al.} 2014, Astronomische
  Nachrichten, 335, 8, \dodoi{12.1002/asna.201312004}

\bibitem[{{Peres} {et~al.}(2000){Peres}, {Orlando}, {Reale}, {Rosner}, \&
  {Hudson}}]{Peres2000}
{Peres}, G., {Orlando}, S., {Reale}, F., {Rosner}, R., \& {Hudson}, H. 2000,
  \apj, 528, 537, \dodoi{10.1086/308136}

\bibitem[{Perez \& Granger(2007)}]{ipython}
Perez, F., \& Granger, B.~E. 2007, Computing in Science Engineering, 9, 21,
  \dodoi{10.1109/MCSE.2007.53}

\bibitem[{{Plavchan}(2006)}]{plavchan2006}
{Plavchan}, Peter~Paul, J. 2006, PhD thesis, University of California, Los
  Angeles

\bibitem[{{Quintana} {et~al.}(2014){Quintana}, {Barclay}, {Raymond}, {Rowe},
  {Bolmont}, {Caldwell}, {Howell}, {Kane}, {Huber}, {Crepp}, {Lissauer},
  {Ciardi}, {Coughlin}, {Everett}, {Henze}, {Horch}, {Isaacson}, {Ford},
  {Adams}, {Still}, {Hunter}, {Quarles}, \& {Selsis}}]{Quintana2014}
{Quintana}, E.~V., {Barclay}, T., {Raymond}, S.~N., {et~al.} 2014, Science,
  344, 277, \dodoi{10.1126/science.1249403}

\bibitem[{{Raymond} {et~al.}(2018){Raymond}, {Boulet}, {Izidoro}, {Esteves}, \&
  {Bitsch}}]{Raymond2018}
{Raymond}, S.~N., {Boulet}, T., {Izidoro}, A., {Esteves}, L., \& {Bitsch}, B.
  2018, \mnras, 479, L81, \dodoi{10.1093/mnrasl/sly100}

\bibitem[{{Ricker} {et~al.}(2015){Ricker}, {Winn}, {Vanderspek}, {Latham},
  {Bakos}, {Bean}, {Berta-Thompson}, {Brown}, {Buchhave}, {Butler}, {Butler},
  {Chaplin}, {Charbonneau}, {Christensen-Dalsgaard}, {Clampin}, {Deming},
  {Doty}, {De Lee}, {Dressing}, {Dunham}, {Endl}, {Fressin}, {Ge}, {Henning},
  {Holman}, {Howard}, {Ida}, {Jenkins}, {Jernigan}, {Johnson}, {Kaltenegger},
  {Kawai}, {Kjeldsen}, {Laughlin}, {Levine}, {Lin}, {Lissauer}, {MacQueen},
  {Marcy}, {McCullough}, {Morton}, {Narita}, {Paegert}, {Palle}, {Pepe},
  {Pepper}, {Quirrenbach}, {Rinehart}, {Sasselov}, {Sato}, {Seager},
  {Sozzetti}, {Stassun}, {Sullivan}, {Szentgyorgyi}, {Torres}, {Udry}, \&
  {Villasenor}}]{Ricker2015}
{Ricker}, G.~R., {Winn}, J.~N., {Vanderspek}, R., {et~al.} 2015, Journal of
  Astronomical Telescopes, Instruments, and Systems, 1, 014003,
  \dodoi{10.1117/1.JATIS.1.1.014003}

\bibitem[{{Rivera} {et~al.}(2005){Rivera}, {Lissauer}, {Butler}, {Marcy},
  {Vogt}, {Fischer}, {Brown}, {Laughlin}, \& {Henry}}]{Rivera2005}
{Rivera}, E.~J., {Lissauer}, J.~J., {Butler}, R.~P., {et~al.} 2005, \apj, 634,
  625, \dodoi{10.1086/491669}

\bibitem[{{Rodriguez} {et~al.}(2020){Rodriguez}, {Vanderburg}, {Zieba},
  {Kreidberg}, {Morley}, {Kane}, {Spencer}, {Quinn}, {Eastman}, {Cloutier},
  {Huang}, {Collins}, {Mann}, {Gilbert}, {Schlieder}, {Quintana}, {Barclay},
  {Suissa}, {Kopparapu}, {Dressing}, {Ricker}, {Vanderspek}, {Latham},
  {Seager}, {Winn}, {Jenkins}, {Berta-Thompson}, {Boyd}, {Charbonneau},
  {Caldwell}, {Chiang}, {Christiansen}, {Ciardi}, {Col{\'o}n}, {Doty}, {Gan},
  {Guerrero}, {G{\"u}nther}, {Lee}, {Levine}, {Lopez}, {Muirhead}, {Newton},
  {Rose}, {Twicken}, \& {Villase{\~n}or}}]{Rodriguez2020}
{Rodriguez}, J.~E., {Vanderburg}, A., {Zieba}, S., {et~al.} 2020, arXiv
  e-prints, arXiv:2001.00954.
\newblock \doarXiv{2001.00954}

\bibitem[{{Rogers}(2015)}]{Rogers2015}
{Rogers}, L.~A. 2015, \apj, 801, 41, \dodoi{10.1088/0004-637X/801/1/41}

\bibitem[{{Rojas-Ayala} {et~al.}(2012){Rojas-Ayala}, {Covey}, {Muirhead}, \&
  {Lloyd}}]{RojasAyala2012}
{Rojas-Ayala}, B., {Covey}, K.~R., {Muirhead}, P.~S., \& {Lloyd}, J.~P. 2012,
  \apj, 748, 93, \dodoi{10.1088/0004-637X/748/2/93}

\bibitem[{{Rowe}(2016)}]{Rowe2016c}
{Rowe}, J. 2016, {Kepler: Kepler Transit Model Codebase Release.}, 1.0,
  Zenodo, \dodoi{10.5281/zenodo.60297}

\bibitem[{{Rowe} {et~al.}(2014){Rowe}, {Bryson}, {Marcy}, {Lissauer},
  {Jontof-Hutter}, {Mullally}, {Gilliland}, {Issacson}, {Ford}, {Howell},
  {Borucki}, {Haas}, {Huber}, {Steffen}, {Thompson}, {Quintana}, {Barclay},
  {Still}, {Fortney}, {Gautier}, {Hunter}, {Caldwell}, {Ciardi}, {Devore},
  {Cochran}, {Jenkins}, {Agol}, {Carter}, \& {Geary}}]{Rowe2014}
{Rowe}, J.~F., {Bryson}, S.~T., {Marcy}, G.~W., {et~al.} 2014, \apj, 784, 45,
  \dodoi{10.1088/0004-637X/784/1/45}

\bibitem[{{Rowe} {et~al.}(2015){Rowe}, {Coughlin}, {Antoci}, {Barclay},
  {Batalha}, {Borucki}, {Burke}, {Bryson}, {Caldwell}, {Campbell},
  {Catanzarite}, {Christiansen}, {Cochran}, {Gilliland}, {Girouard}, {Haas},
  {He{\l}miniak}, {Henze}, {Hoffman}, {Howell}, {Huber}, {Hunter},
  {Jang-Condell}, {Jenkins}, {Klaus}, {Latham}, {Li}, {Lissauer}, {McCauliff},
  {Morris}, {Mullally}, {Ofir}, {Quarles}, {Quintana}, {Sabale}, {Seader},
  {Shporer}, {Smith}, {Steffen}, {Still}, {Tenenbaum}, {Thompson}, {Twicken},
  {Van Laerhoven}, {Wolfgang}, \& {Zamudio}}]{Rowe2015}
{Rowe}, J.~F., {Coughlin}, J.~L., {Antoci}, V., {et~al.} 2015, \apjs, 217, 16,
  \dodoi{10.1088/0067-0049/217/1/16}

\bibitem[{Salvatier {et~al.}(2016)Salvatier, Wiecki, \&
  Fonnesbeck}]{exoplanet:pymc3}
Salvatier, J., Wiecki, T.~V., \& Fonnesbeck, C. 2016, PeerJ Computer Science,
  2, e55

\bibitem[{{Sanz-Forcada} {et~al.}(2011){Sanz-Forcada}, {Micela}, {Ribas},
  {Pollock}, {Eiroa}, {Velasco}, {Solano}, \&
  {Garc{\'\i}a-{\'A}lvarez}}]{Sanz-Forcada2011}
{Sanz-Forcada}, J., {Micela}, G., {Ribas}, I., {et~al.} 2011, \aap, 532, A6,
  \dodoi{10.1051/0004-6361/201116594}

\bibitem[{{Schaefer} \& {Fegley}(2007)}]{Schaefer2007}
{Schaefer}, L., \& {Fegley}, B. 2007, \icarus, 186, 462,
  \dodoi{10.1016/j.icarus.2006.09.002}

\bibitem[{{Schlieder} {et~al.}(2016){Schlieder}, {Crossfield}, {Petigura},
  {Howard}, {Aller}, {Sinukoff}, {Isaacson}, {Fulton}, {Ciardi}, {Bonnefoy},
  {Ziegler}, {Morton}, {L{\'e}pine}, {Obermeier}, {Liu}, {Bailey}, {Baranec},
  {Beichman}, {Defr{\`e}re}, {Henning}, {Hinz}, {Law}, {Riddle}, \&
  {Skemer}}]{schlieder2016}
{Schlieder}, J.~E., {Crossfield}, I. J.~M., {Petigura}, E.~A., {et~al.} 2016,
  \apj, 818, 87, \dodoi{10.3847/0004-637X/818/1/87}

\bibitem[{{Shapley}(1953)}]{Shapley1953}
{Shapley}, H. 1953, {Climatic Change: Evidence, Causes, and Effects}

\bibitem[{{Shappee} {et~al.}(2014){Shappee}, {Prieto}, {Grupe}, {Kochanek},
  {Stanek}, {De Rosa}, {Mathur}, {Zu}, {Peterson}, {Pogge}, {Komossa}, {Im},
  {Jencson}, {Holoien}, {Basu}, {Beacom}, {Szczygie{\l}}, {Brimacombe},
  {Adams}, {Campillay}, {Choi}, {Contreras}, {Dietrich}, {Dubberley},
  {Elphick}, {Foale}, {Giustini}, {Gonzalez}, {Hawkins}, {Howell}, {Hsiao},
  {Koss}, {Leighly}, {Morrell}, {Mudd}, {Mullins}, {Nugent}, {Parrent},
  {Phillips}, {Pojmanski}, {Rosing}, {Ross}, {Sand}, {Terndrup}, {Valenti},
  {Walker}, \& {Yoon}}]{Shappee14}
{Shappee}, B.~J., {Prieto}, J.~L., {Grupe}, D., {et~al.} 2014, \apj, 788, 48,
  \dodoi{10.1088/0004-637X/788/1/48}

\bibitem[{{Skrutskie} {et~al.}(2006){Skrutskie}, {Cutri}, {Stiening},
  {Weinberg}, {Schneider}, {Carpenter}, {Beichman}, {Capps}, {Chester},
  {Elias}, {Huchra}, {Liebert}, {Lonsdale}, {Monet}, {Price}, {Seitzer},
  {Jarrett}, {Kirkpatrick}, {Gizis}, {Howard}, {Evans}, {Fowler}, {Fullmer},
  {Hurt}, {Light}, {Kopan}, {Marsh}, {McCallon}, {Tam}, {Van Dyk}, \&
  {Wheelock}}]{Skrutskie2006}
{Skrutskie}, M.~F., {Cutri}, R.~M., {Stiening}, R., {et~al.} 2006, \aj, 131,
  1163, \dodoi{10.1086/498708}

\bibitem[{{Smith} {et~al.}(2012){Smith}, {Stumpe}, {Van Cleve}, {Jenkins},
  {Barclay}, {Fanelli}, {Girouard}, {Kolodziejczak}, {McCauliff}, {Morris}, \&
  {Twicken}}]{Smith2012}
{Smith}, J.~C., {Stumpe}, M.~C., {Van Cleve}, J.~E., {et~al.} 2012, \pasp, 124,
  1000, \dodoi{10.1086/667697}

\bibitem[{{Stassun} \& {Torres}(2018)}]{Stassun2018}
{Stassun}, K.~G., \& {Torres}, G. 2018, \apj, 862, 61,
  \dodoi{10.3847/1538-4357/aacafc}

\bibitem[{{Stassun} {et~al.}(2018){Stassun}, {Oelkers}, {Pepper}, {Paegert},
  {De Lee}, {Torres}, {Latham}, {Charpinet}, {Dressing}, {Huber}, {Kane},
  {L{\'e}pine}, {Mann}, {Muirhead}, {Rojas-Ayala}, {Silvotti}, {Fleming},
  {Levine}, \& {Plavchan}}]{Stassun_CTL_2018}
{Stassun}, K.~G., {Oelkers}, R.~J., {Pepper}, J., {et~al.} 2018, \aj, 156, 102,
  \dodoi{10.3847/1538-3881/aad050}

\bibitem[{{Stassun} {et~al.}(2019){Stassun}, {Oelkers}, {Paegert}, {Torres},
  {Pepper}, {De Lee}, {Collins}, {Latham}, {Muirhead}, {Chittidi},
  {Rojas-Ayala}, {Fleming}, {Rose}, {Tenenbaum}, {Ting}, {Kane}, {Barclay},
  {Bean}, {Brassuer}, {Charbonneau}, {Lissauer}, {Mann}, {McLean}, {Mulally},
  {Narita}, {Plavchan}, {Ricker}, {Sasselov}, {Seager}, {Sharma}, {Shiao},
  {Sozzetti}, {Stello}, {Vand erspek}, {Wallace}, \& {Winn}}]{stassun2019}
{Stassun}, K.~G., {Oelkers}, R.~J., {Paegert}, M., {et~al.} 2019, arXiv
  e-prints, arXiv:1905.10694.
\newblock \doarXiv{1905.10694}

\bibitem[{{Strughold}(1953)}]{Strughold1953}
{Strughold}, H. 1953, The green and red planet; a physiological study of the
  possibility of life on Mars (Albuquerque, NM: University of New Mexico Press)

\bibitem[{{Stumpe} {et~al.}(2014){Stumpe}, {Smith}, {Catanzarite}, {Van Cleve},
  {Jenkins}, {Twicken}, \& {Girouard}}]{Stumpe2014}
{Stumpe}, M.~C., {Smith}, J.~C., {Catanzarite}, J.~H., {et~al.} 2014, \pasp,
  126, 100, \dodoi{10.1086/674989}

\bibitem[{{Su{\'a}rez Mascare{\~n}o} {et~al.}(2020){Su{\'a}rez Mascare{\~n}o},
  {Faria}, {Figueira}, {Lovis}, {Damasso}, {Gonz{\'a}lez Hern{\'a}ndez},
  {Rebolo}, {Cristiano}, {Pepe}, {Santos}, {Zapatero Osorio}, {Adibekyan},
  {Hojjatpanah}, {Sozzetti}, {Murgas}, {Abreo}, {Affolter}, {Alibert},
  {Aliverti}, {Allart}, {Allende Prieto}, {Alves}, {Amate}, {Avila}, {Baldini},
  {Bandi}, {Barros}, {Bianco}, {Benz}, {Bouchy}, {Broeng}, {Cabral},
  {Calderone}, {Cirami}, {Coelho}, {Conconi}, {Coretti}, {Cumani}, {Cupani},
  {D'Odorico}, {Deiries}, {Delabre}, {Di Marcantonio}, {Dumusque},
  {Ehrenreich}, {Fragoso}, {Genolet}, {Genoni}, {G{\'e}nova Santos}, {Hughes},
  {Iwert}, {Ferber}, {Knusdrtrup}, {Landoni}, {Lavie}, {Lillo-Box}, {Lizon},
  {Lo Curto}, {Maire}, {Manescau}, {Martins}, {M{\'e}gevand}, {Mehner},
  {Micela}, {Modigliani}, {Molaro}, {Monteiro}, {Monteiro}, {Moschetti},
  {Mueller}, {Nunes}, {Oggioni}, {Oliveira}, {Pall{\'e}}, {Pariani},
  {Pasquini}, {Poretti}, {Rasilla}, {Redaelli}, {Riva}, {Santana Tschudi},
  {Santin}, {Santos}, {Segovia}, {Sosnoswska}, {Sousa}, {Span{\`o}}, {Tenegi},
  {Udry}, {Zanutta}, \& {Zerbi}}]{proxcen_espresso}
{Su{\'a}rez Mascare{\~n}o}, A., {Faria}, J.~P., {Figueira}, P., {et~al.} 2020,
  arXiv e-prints, arXiv:2005.12114.
\newblock \doarXiv{2005.12114}

\bibitem[{{Suissa} {et~al.}(2020){Suissa}, {Wolf}, {Kopparapu}, {Villanueva},
  {Fauchez}, {Mandell}, {Arney}, {Gilbert}, {Schlieder}, {Barclay}, {Quintana},
  {Lopez}, {Rodriguez}, \& {Vanderburg}}]{Suissa2020}
{Suissa}, G., {Wolf}, E.~T., {Kopparapu}, R.~k., {et~al.} 2020, arXiv e-prints,
  arXiv:2001.00955.
\newblock \doarXiv{2001.00955}

\bibitem[{{Sullivan} {et~al.}(2015){Sullivan}, {Winn}, {Berta-Thompson},
  {Charbonneau}, {Deming}, {Dressing}, {Latham}, {Levine}, {McCullough},
  {Morton}, {Ricker}, {Vanderspek}, \& {Woods}}]{sullivan15}
{Sullivan}, P.~W., {Winn}, J.~N., {Berta-Thompson}, Z.~K., {et~al.} 2015, \apj,
  809, 77, \dodoi{10.1088/0004-637X/809/1/77}

\bibitem[{{Tal-Or} {et~al.}(2018){Tal-Or}, {Zechmeister}, {Reiners}, {Jeffers},
  {Sch{\"o}fer}, {Quirrenbach}, {Amado}, {Ribas}, {Caballero}, {Aceituno},
  {Bauer}, {B{\'e}jar}, {Czesla}, {Dreizler}, {Fuhrmeister}, {Hatzes},
  {Johnson}, {K{\"u}rster}, {Lafarga}, {Montes}, {Morales}, {Reffert},
  {Sadegi}, {Seifert}, \& {Shulyak}}]{talor18}
{Tal-Or}, L., {Zechmeister}, M., {Reiners}, A., {et~al.} 2018, \aap, 614, A122,
  \dodoi{10.1051/0004-6361/201732362}

\bibitem[{{Teske} {et~al.}(2016){Teske}, {Shectman}, {Vogt}, {D{\'\i}az},
  {Butler}, {Crane}, {Thompson}, \& {Arriagada}}]{teske16}
{Teske}, J.~K., {Shectman}, S.~A., {Vogt}, S.~S., {et~al.} 2016, \aj, 152, 167,
  \dodoi{10.3847/0004-6256/152/6/167}

\bibitem[{{Theano Development Team}(2016)}]{exoplanet:theano}
{Theano Development Team}. 2016, arXiv e-prints, abs/1605.02688.
\newblock \url{http://arxiv.org/abs/1605.02688}

\bibitem[{{Thompson} {et~al.}(2018){Thompson}, {Coughlin}, {Hoffman},
  {Mullally}, {Christiansen}, {Burke}, {Bryson}, {Batalha}, {Haas},
  {Catanzarite}, {Rowe}, {Barentsen}, {Caldwell}, {Clarke}, {Jenkins}, {Li},
  {Latham}, {Lissauer}, {Mathur}, {Morris}, {Seader}, {Smith}, {Klaus},
  {Twicken}, {Van Cleve}, {Wohler}, {Akeson}, {Ciardi}, {Cochran}, {Henze},
  {Howell}, {Huber}, {Pr{\v{s}}a}, {Ram{\'\i}rez}, {Morton}, {Barclay},
  {Campbell}, {Chaplin}, {Charbonneau}, {Christensen-Dalsgaard}, {Dotson},
  {Doyle}, {Dunham}, {Dupree}, {Ford}, {Geary}, {Girouard}, {Isaacson},
  {Kjeldsen}, {Quintana}, {Ragozzine}, {Shabram}, {Shporer}, {Silva Aguirre},
  {Steffen}, {Still}, {Tenenbaum}, {Welsh}, {Wolfgang}, {Zamudio}, {Koch}, \&
  {Borucki}}]{Thompson2018}
{Thompson}, S.~E., {Coughlin}, J.~L., {Hoffman}, K., {et~al.} 2018, \apjs, 235,
  38, \dodoi{10.3847/1538-4365/aab4f9}

\bibitem[{{Tokovinin}(2018)}]{tokovinin18}
{Tokovinin}, A. 2018, \pasp, 130, 035002, \dodoi{10.1088/1538-3873/aaa7d9}

\bibitem[{{Tokovinin} {et~al.}(2013){Tokovinin}, {Fischer}, {Bonati},
  {Giguere}, {Moore}, {Schwab}, {Spronck}, \& {Szymkowiak}}]{Tokovinin2013}
{Tokovinin}, A., {Fischer}, D.~A., {Bonati}, M., {et~al.} 2013, \pasp, 125,
  1336, \dodoi{10.1086/674012}

\bibitem[{{Torres} {et~al.}(2010){Torres}, {Andersen}, \&
  {Gim{\'e}nez}}]{Torres2010}
{Torres}, G., {Andersen}, J., \& {Gim{\'e}nez}, A. 2010, \aapr, 18, 67,
  \dodoi{10.1007/s00159-009-0025-1}

\bibitem[{{Torres} {et~al.}(2015){Torres}, {Kipping}, {Fressin}, {Caldwell},
  {Twicken}, {Ballard}, {Batalha}, {Bryson}, {Ciardi}, {Henze}, {Howell},
  {Isaacson}, {Jenkins}, {Muirhead}, {Newton}, {Petigura}, {Barclay},
  {Borucki}, {Crepp}, {Everett}, {Horch}, {Howard}, {Kolbl}, {Marcy},
  {McCauliff}, \& {Quintana}}]{torres15}
{Torres}, G., {Kipping}, D.~M., {Fressin}, F., {et~al.} 2015, \apj, 800, 99,
  \dodoi{10.1088/0004-637X/800/2/99}

\bibitem[{{Twicken} {et~al.}(2018){Twicken}, {Catanzarite}, {Clarke},
  {Girouard}, {Jenkins}, {Klaus}, {Li}, {McCauliff}, {Seader}, {Tenenbaum},
  {Wohler}, {Bryson}, {Burke}, {Caldwell}, {Haas}, {Henze}, \&
  {Sanderfer}}]{Twicken2018}
{Twicken}, J.~D., {Catanzarite}, J.~H., {Clarke}, B.~D., {et~al.} 2018, \pasp,
  130, 064502, \dodoi{10.1088/1538-3873/aab694}

\bibitem[{van~der Walt {et~al.}(2011)van~der Walt, Colbert, \&
  Varoquaux}]{numpy}
van~der Walt, S., Colbert, S.~C., \& Varoquaux, G. 2011, Computing in Science
  Engineering, 13, 22, \dodoi{10.1109/MCSE.2011.37}

\bibitem[{{Van Eylen} {et~al.}(2018){Van Eylen}, {Agentoft}, {Lundkvist},
  {Kjeldsen}, {Owen}, {Fulton}, {Petigura}, \& {Snellen}}]{vaneylen2018}
{Van Eylen}, V., {Agentoft}, C., {Lundkvist}, M.~S., {et~al.} 2018, \mnras,
  479, 4786, \dodoi{10.1093/mnras/sty1783}

\bibitem[{{Van Eylen} {et~al.}(2019){Van Eylen}, {Albrecht}, {Huang},
  {MacDonald}, {Dawson}, {Cai}, {Foreman-Mackey}, {Lundkvist}, {Silva Aguirre},
  {Snellen}, \& {Winn}}]{exoplanet:vaneylen19}
{Van Eylen}, V., {Albrecht}, S., {Huang}, X., {et~al.} 2019, \aj, 157, 61,
  \dodoi{10.3847/1538-3881/aaf22f}

\bibitem[{{Vanderburg} {et~al.}(2016){Vanderburg}, {Plavchan}, {Johnson},
  {Ciardi}, {Swift}, \& {Kane}}]{vanderburg16}
{Vanderburg}, A., {Plavchan}, P., {Johnson}, J.~A., {et~al.} 2016, \mnras, 459,
  3565, \dodoi{10.1093/mnras/stw863}

\bibitem[{{Vanderspek} {et~al.}(2018){Vanderspek}, {Doty}, {Fausnaugh},
  {et~al.}}]{Vanderspek2018b}
{Vanderspek}, R., {Doty}, J.~P., {Fausnaugh}, M., {et~al.} 2018, {TESS
  Instrument Handbook v0.1}, Tech. rep.

\bibitem[{Veyette \& Muirhead(2018)}]{Veyette_2018}
Veyette, M.~J., \& Muirhead, P.~S. 2018, The Astrophysical Journal, 863, 166,
  \dodoi{10.3847/1538-4357/aad40e}

\bibitem[{{Vida} {et~al.}(2017){Vida}, {K{\H{o}}v{\'a}ri}, {P{\'a}l},
  {Ol{\'a}h}, \& {Kriskovics}}]{Vida2017}
{Vida}, K., {K{\H{o}}v{\'a}ri}, Z., {P{\'a}l}, A., {Ol{\'a}h}, K., \&
  {Kriskovics}, L. 2017, \apj, 841, 124, \dodoi{10.3847/1538-4357/aa6f05}

\bibitem[{{Weiss} {et~al.}(2018){Weiss}, {Marcy}, {Petigura}, {Fulton},
  {Howard}, {Winn}, {Isaacson}, {Morton}, {Hirsch}, {Sinukoff}, {Cumming},
  {Hebb}, \& {Cargile}}]{Weiss2018}
{Weiss}, L.~M., {Marcy}, G.~W., {Petigura}, E.~A., {et~al.} 2018, \aj, 155, 48,
  \dodoi{10.3847/1538-3881/aa9ff6}

\bibitem[{{Wheatley} {et~al.}(2017){Wheatley}, {Louden}, {Bourrier},
  {Ehrenreich}, \& {Gillon}}]{Wheatley2017}
{Wheatley}, P.~J., {Louden}, T., {Bourrier}, V., {Ehrenreich}, D., \& {Gillon},
  M. 2017, \mnras, 465, L74, \dodoi{10.1093/mnrasl/slw192}

\bibitem[{{Winters} {et~al.}(2011){Winters}, {Henry}, {Jao}, {Subasavage},
  {Finch}, \& {Hambly}}]{Winters2011}
{Winters}, J.~G., {Henry}, T.~J., {Jao}, W.-C., {et~al.} 2011, \aj, 141, 21,
  \dodoi{10.1088/0004-6256/141/1/21}

\bibitem[{{Winters} {et~al.}(2018){Winters}, {Irwin}, {Newton}, {Charbonneau},
  {Latham}, {Han}, {Muirhead}, {Berlind}, {Calkins}, \&
  {Esquerdo}}]{Winters(2018)}
{Winters}, J.~G., {Irwin}, J., {Newton}, E.~R., {et~al.} 2018, \aj, 155, 125,
  \dodoi{10.3847/1538-3881/aaaa65}

\bibitem[{{Winters} {et~al.}(2019{\natexlab{a}}){Winters}, {Medina}, {Irwin},
  {Charbonneau}, {Astudillo-Defru}, {Horch}, {Eastman}, {Vrijmoet}, {Henry},
  {Diamond-Lowe}, {Winston}, {Barclay}, {Bonfils}, {Ricker}, {Vand erspek},
  {Latham}, {Seager}, {Winn}, {Jenkins}, {Udry}, {Twicken}, {Teske},
  {Tenenbaum}, {Pepe}, {Murgas}, {Muirhead}, {Mink}, {Lovis}, {Levine},
  {L'epine}, {Jao}, {Henze}, {Fur'esz}, {Forveille}, {Figueira}, {Esquerdo},
  {Dressing}, {D'iaz}, {Delfosse}, {Burke}, {Bouchy}, {Berlind}, \&
  {Almenara}}]{winters2019b}
{Winters}, J.~G., {Medina}, A.~A., {Irwin}, J.~M., {et~al.} 2019{\natexlab{a}},
  arXiv e-prints, arXiv:1906.10147.
\newblock \doarXiv{1906.10147}

\bibitem[{{Winters} {et~al.}(2019{\natexlab{b}}){Winters}, {Henry}, {Jao},
  {Subasavage}, {Chatelain}, {Slatten}, {Riedel}, {Silverstein}, \&
  {Payne}}]{Winters(2019a)}
{Winters}, J.~G., {Henry}, T.~J., {Jao}, W.-C., {et~al.} 2019{\natexlab{b}},
  \aj, 157, 216, \dodoi{10.3847/1538-3881/ab05dc}

\bibitem[{{Wright} {et~al.}(2010){Wright}, {Eisenhardt}, {Mainzer}, {Ressler},
  {Cutri}, {Jarrett}, {Kirkpatrick}, {Padgett}, {McMillan}, {Skrutskie},
  {Stanford}, {Cohen}, {Walker}, {Mather}, {Leisawitz}, {Gautier}, {McLean},
  {Benford}, {Lonsdale}, {Blain}, {Mendez}, {Irace}, {Duval}, {Liu}, {Royer},
  {Heinrichsen}, {Howard}, {Shannon}, {Kendall}, {Walsh}, {Larsen}, {Cardon},
  {Schick}, {Schwalm}, {Abid}, {Fabinsky}, {Naes}, \& {Tsai}}]{Wright2010}
{Wright}, E.~L., {Eisenhardt}, P. R.~M., {Mainzer}, A.~K., {et~al.} 2010, \aj,
  140, 1868, \dodoi{10.1088/0004-6256/140/6/1868}

\bibitem[{{Zahnle} \& {Catling}(2017)}]{Zahnle2017}
{Zahnle}, K.~J., \& {Catling}, D.~C. 2017, \apj, 843, 122,
  \dodoi{10.3847/1538-4357/aa7846}

\bibitem[{{Ziegler} {et~al.}(2019){Ziegler}, {Tokovinin}, {Briceno}, {Mang},
  {Law}, \& {Mann}}]{ziegler19}
{Ziegler}, C., {Tokovinin}, A., {Briceno}, C., {et~al.} 2019, arXiv e-prints,
  arXiv:1908.10871.
\newblock \doarXiv{1908.10871}

\end{thebibliography}
\bibliographystyle{aasjournal}
\suppressAffiliationsfalse
\allauthors
\end{document}